\documentclass[showpacs,preprintnumbers,amsmath,amssymb]{revtex4}

\usepackage{amsmath}
\usepackage{amssymb}
\usepackage{bm}
\usepackage{graphicx}
\usepackage{multirow}
\usepackage{textcomp}

\allowdisplaybreaks[1]

\begin{document}


\title{\boldmath Dispersive analysis of glueball masses}
\author{Hsiang-nan Li}
\affiliation{Institute of Physics, Academia Sinica,
Taipei, Taiwan 115, Republic of China}

\date{\today}

\begin{abstract}
We develop an inverse matrix method to solve for resonance masses 
from a dispersion relation obeyed by a correlation function. 
Given the operator product expansion (OPE) of a correlation function
in the deep Euclidean region, we obtain the nonperturbative spectral density, 
which exhibits resonance structures naturally. The value of the gluon condensate 
in the OPE is fixed by producing the $\rho$ meson mass in the formalism,
and then input into the dispersion relations for the scalar, pseudoscalar and
tensor glueballs. It is shown that the low-energy limit of the correlation function
for the scalar glueball, derived from the spectral density, 
discriminates the lattice estimate for the triple-gluon condensate
from the single-instanton estimate. The spectral 
densities for the scalar and pseudoscalar glueballs reveal a double-peak structure:
the peak located at lower mass implies that the $f_0(500)$ and $f_0(980)$ 
($\eta$ ad $\eta'$) mesons contain small amount of gluonium components, and should
be included into scalar (pseudoscalar) mixing frameworks. Another peak 
determines the scalar (pseudoscalar) glueball mass around 1.50 (1.75) GeV
with a broad width about 200 MeV, suggesting that the $f_0(1370)$, $f_0(1500)$ 
and $f_0(1710)$ ($\eta(1760)$) mesons are the glue-rich states. 
We also predict the topological susceptability $\chi_t^{1/4}=75$-78 MeV, deduced from 
the correlation function for the pseudoscalar glueball at zero momentum. Our 
analysis gives no resonance solution for the tensor glueball, which may be attributed 
to the insufficient nonperturbative condensate information in the currently available OPE.

\end{abstract}


\maketitle

%
%
%

\section{INTRODUCTION}

We proposed to handle QCD sum rules \cite{SVZ} for nonpertrbative observables, like 
the $\rho$ meson mass, as an inverse problem recently \cite{Li:2020ejs}. The spectral density 
on the hadron side of a dispersion relation, including both resonance and continuum contributions, is 
regarded as an unknown. The operator product expansion (OPE) on the quark side is calculated 
in the standard way. The spectral density (similar to a source distribution) is then determined from the 
OPE input (similar to a potential observed outside the distribution). This approach does not involve
a continuum threshold, because the spectral density is supposed to be a smooth distribution, and
may differ from the perturbative one (the quark-hadron duality is likely to be broken). It does 
not require a Borel transformation to suppress the continuum contribution, which will be solved 
from the inverse problem. The suppression on higher power corrections to the OPE can be achieved 
by considering the input in the deep Euclidean region. Once a dispersion relation is solved directly, 
the optional stability criterion \cite{Coriano:1993yx,Coriano:1998ge} for sum rules is not necessary
either. The long-existing concern on the rigorousness and predictive power of QCD sum rules 
\cite{Leinweber:1995fn,Gubler:2010cf} is then resolved. As an example, we demonstrated how to 
extract the masses and decay constants of the series of $\rho$ resonances from the dispersion 
relation obeyed by a two-current correlator \cite{Li:2020ejs}. This new viewpoint, 
based only on the analyticity of physical observables, has been extended to the explanation of the 
$D$ meson mixing parameters \cite{Li:2020xrz} and to the constraint on the hadronic vacuum 
polarization contribution to the muon anomalous magnetic moment \cite{Li:2020fiz}. 

The spectral density was parametrized 
as a sum of a pole contribution, which depends on the $\rho$ meson mass and decay constant, 
and an arbitrary continuum contribution in \cite{Li:2020ejs}. The latter was expanded in terms 
of a set of orthogonal polynomials with unknown coefficients, which, together with the 
mass and decay constant, were deduced from the best fit of the hadron side 
to the OPE on the quark side. The continuum contribution turned out to 
be a ramp function, instead of a step function postulated in sum rules. 
A weakness of the above treatment is that the existence of the bound state, presumed from 
the beginning, is not a direct consequence of solving the inverse problem. In fact, the 
Bayesian approach \cite{Gubler:2010cf} was applied to sum rules with a similar motivation, 
in which the form of a spectral density was not specified, but obtained using the maximum 
entropy method. It was found \cite{Ohtani:2012ps} that the resultant spectral density 
associated with the $\rho$ meson exhibits a resonance peak at low energy, followed by a smooth 
continuum tail at high energy. Unfortunately, this approach is numerically intricate, and the
peak location, i.e., the $\rho$ meson mass, depends on some parameters introduced in the
maximum entropy method. The complexity is expected, since it is notoriously difficult to 
solve an ill-posed inverse problem. 

We will develop a simple inverse matrix method to solve for a spectral density from a dispersion relation 
without assuming the existence of a resonance: the whole spectral density is expanded in terms 
of the generalized Laguerre polynomials, whose coefficients are derived from the OPE 
input directly. The quark-hadron duality for a spectral density 
on the hadron side and the discretionary criteria for balancing perturbative and nonperturbative 
contributions \cite{Huang:1998wj,Harnett:2000fy} are not implemented. That is, unambiguous predictions 
for nonperturbative observables can be acquired in the above dispersive formalism. Besides, the 
precision of our predictions can be improved systematically by adding higher-order and 
higher-power corrections to the OPE input. We first test this proposal by solving for the 
$\rho$ meson mass, and demonstrate that a resonance peak shows up in 
the spectral density naturally, a result the same as from the maximum entropy method 
\cite{Ohtani:2012ps}. It means that our formalism is equivalent to but technically simpler than
the Bayesian approach \cite{Gubler:2010cf}. By producing the $\rho$ meson mass
$m_\rho=0.78$ GeV \cite{PDG} from the peak location, we fix the value of the gluon 
condensate $\langle\alpha_s G^2\rangle=0.08$ GeV$^4$, $\alpha_s$ being the strong coupling 
constant, which is within the range accepted in the literature (see \cite{SN98}, for example). 
Our solution contains only a single peak corresponding to the ground-state $\rho$ meson, manifesting
the difficulty to explore properties of excited states with denser spectra, which is also encountered 
by the Bayesian approach \cite{Ohtani:2012ps}.


The framework with the fixed gluon condensate is then applied to the analyses of the scalar, 
pseudoscalar and tensor glueball masses. Though the instanton background may contribute to the 
OPE \cite{NSVZ}, we will not consider its effect as in \cite{SN98} to avoid the model dependence 
from, e.g., parametrizations for the instanton size distribution. We will show that the 
gluon condensates of various dimensions are enough for establishing the gluonium states, and 
confront our findings with those in \cite{SN98}. For sum-rule investigations on scalar and 
pseudoscalar glueball properties in the instanton background, refer to 
\cite{Forkel:2003mk,Forkel:2000fd,Harnett:2000fy,Wen:2010as,Wen:2010qoe,Wang:2015mla}. The quark-loop 
and quark-condensate corrections, which shift the scalar glueball mass by about 
few percent \cite{Yuan:2009vs}, will be neglected in this work. Other theoretical approaches to the 
exploration of glueball physics have been introduced in the review \cite{Mathieu:2008me}.
Note that there exist several estimates for a crucial higher-power contribution to the OPE, i.e., 
the triple-gluon condensate, such as those from the single-instanton \cite{SVZ,NS80,RRY} and 
lattice \cite{PV90} evaluations, and that determined from heavy quark systems \cite{SN10},
which differ dramatically. Once the spectral density is obtained from the dispersion relation, 
one can compute the correlation function associated with the scalar glueball at zero momentum. 
It will be elaborated that the spectral density corresponding to the lattice estimate respects better the 
low-energy theorem for the correlation function \cite{NSVZ,SVZ80,LS92}. That is, our method 
helps discriminate the different estimates for the triple-gluon condensate.


Appropriate moments of a Borel transformed 
correlator have to be selected to form ratios for the extraction of a glueball mass in sum rules, 
because the stability window in the Borel mass may not exist for ratios of other moments 
\cite{SN98,Huang:1998wj}. On the other hand, the lower (higher) moments are more 
sensitive to light (heavy) resonances \cite{SN98}. The above subtlety is not an issue to our 
formalism, in which full information of the OPE input is utilized to solve a dispersion relation. 
It will be seen that a double-peak structure with both light and heavy resonances is revealed 
by the solution to a spectral density, in support of the observation in \cite{NV89}. 
We emphasize that the positivity constraint on the spectral density plays a crucial role for 
establishing the bound states. The shorter peak located at 0.60 GeV in the spectral density for 
the scalar glueball can be identified as the light scalar meson $f_0(500)$ perhaps with little 
admixture from $f_0(980)$. Another located at 1.50 GeV with a broader width about 200 MeV arises 
from the combined contribution of the $f_0(1370)$, $f_0(1500)$ and $f_0(1700)$ mesons, since 
$f_0(1500)$ with a narrow width cannot accommodate the broad peak alone. Our solution indicates 
that $f_0(500)$ and $f_0(980)$ contain some amount of gluonia, and the $f_0(1370)$, $f_0(1500)$ 
and $f_0(1700)$ mesons are the glue-rich states, in agreement with the multiple-resonance mixing 
picture for scalar mesons widely discussed in the literature 
\cite{Close,Giacosa:2005zt,Vento:2004xx,Fariborz:2006xq,Cheng:2015iaa,Noshad:2018afw,Guo:2020akt}. 
The prediction for the scalar glueball mass around 1.50 GeV is close to those from
sum rules \cite{SN98,Wen:2010qoe,Narison:2021xhc} (but a bit lower than in \cite{Chen:2021bck}) and 
quenched lattice QCD \cite{Bali,Morningstar:1999rf,Chen:2005mg,Athenodorou:2020ani}, and
smaller than from holographic QCD \cite{Zhang:2021itx}. The quenched lattice 
calculation in \cite{Lee:1999kv} favors the interpretation of $f_0(1710)$ as composed mainly of the 
lightest scalar glueball. We mention that a double-peak parametrization for the spectral density 
has been found to yield a fit to the sum rule for the scalar glueball with the instanton 
effect better than a single-peak one \cite{Harnett:2000fy}. 

The spectral density for the pseudoscalar glueball is also featured with a double-peak structure: the 
shorter peak located at 0.71 GeV comes from the combined contribution of the $\eta$ and $\eta'$ mesons 
naturally, which have been known to comprise some gluonium components \cite{Kataev:1981aw}.
Another at 1.75 GeV with a broad width about 200 MeV strongly suggests that the $\eta(1760)$ meson
is a promising candidate for the pseudosclar glueball. This mass is lower than most results 
in the literature, which are above 2 GeV, such as those from sum rules \cite{SN98}, quenched lattice QCD 
\cite{Bali,Morningstar:1999rf,Chen:2005mg,Athenodorou:2020ani}, the Bethe-Salpeter approach 
\cite{Huber:2020ngt,Kaptari:2020qlt} and holographic QCD \cite{Zhang:2021itx}. Nevertheless, when the 
resonance contribution was parametrized by a Breit-Wigner form with a finite width, the pseudoscalar 
glueball mass drops to $1.407\pm 0.162$ GeV in sum rules with the 
instanton effect \cite{Wang:2015mla}. We remind that measurements of $J/\psi$ radiative decays do not 
affirm any glue-rich pseudoscalar resonances with masses above 2 GeV (the quantum numbers of
$X(2370)$ are not certain yet) \cite{PDG}. The $\eta(1760)$ meson was proposed to be the 
pseudoscalar glueball two decades ago \cite{Page:1996ss,Wu:2000yt}, 
and examined experimentally via the decay $J/\psi\to\gamma(\eta(1760)\to)\omega\omega$ 
in \cite{BES:2006nqh}. It is abundantly produced in $J/\psi$ radiative decays, 
but not seen in the $J/\psi\to \gamma\gamma V$ channels \cite{MARK}, $V=\rho$, $\phi$, implying that 
the partial width of $\eta(1760)\to \gamma V$ is tiny, namely, $\eta(1760)$ is likely to be a glueball.
Our solution to the spectral density for the pseudoscalar glueball advocates that $\eta(1760)$ 
should be included into the mixing framework for pseudoscalar mesons 
\cite{Cheng:2008ss,He:2009sb,Tsai:2011dp,Gutsche:2009jh,Qin:2017qes} for a complete scenario.


The topological susceptability $\chi_t$, an important quantity 
characterizing the nonperturbative QCD vacuum, is related to the low-energy limit of 
the correlation function for the pseudoscalar glueball, which has been 
analyzed in sum rules with the pure Yang-Mills theory \cite{SN83}.
Similarly, once the spectral density is ready, one can compute the 
corresponding correlation function at zero momentum. We then predict $\chi_t^{1/4}=75$-78 MeV, 
which is quite precise and matches those from lattice QCD 
\cite{Aoki:2017paw,Alexandrou:2017bzk,Bonati:2015vqz,Dimopoulos:2018xkm,Chiu:2020ppa,Bhattacharya:2021lol}
and the chiral perturbatin theory \cite{Luciano:2018pbj,GrillidiCortona:2015jxo,Gorghetto:2018ocs}.
Viewing that our formalism produces the observed $\rho$ meson mass, the widely accepted scalar glueball mass 
around 1.5 GeV, and the topological susceptability in consistency 
with the expectations from other approaches, we are convinced that the $\eta(1760)$ meson
deserves thorough investigations on whether it is the pseudoscalar glueball in the long-lasting quest.

At last, we find no resonance solution to the spectral density for the tensor glueball. Distinct from 
the scalar and pseudoscalar glueball cases, only a single dimension-eight condensate is available for 
the OPE input. It has been verified \cite{Li:2020ejs} that at least two condensates 
of different dimensions are necessary for establishing the $\rho$ meson state. We thus speculate 
that the absence of a resonance solution for the tensor glueball may be due to the insufficient 
nonperturbative input in the present setup. Nevertheless, a tensor glueball mass 
about 2.0 GeV was extracted from specific moments of the sum rule in \cite{SN98}.
As stated before, full information of the OPE input is utilized in our formalism,
so the condition on the existence of a resonance may be more stringent.
It is urged to calculate more higher-power corrections to the OPE of the correlation 
function, so that the tensor glueball mass can be inferred from
the dispersion relation and compared with the lattice predictions
\cite{Morningstar:1999rf,Lucini:2004my,Bennett:2020hqd}.

The rest of the paper is organized as follows.
In Sec.~II we illustrate the inverse matrix method to solve a dispersion relation
as an inverse problem. The mock data from several test functions are constructed,
and treated as inputs. It is shown that the solutions reproduce the test functions accurately 
in the inverse matrix method. Our approach is then applied to the 
determination of the $\rho$ meson mass, via which the value of the gluon condensate is fixed 
for the analyses of the glueball masses. We explain how boundary conditions of a spectral 
density select the suitable set of generalized Laguerre polynomials for its expansion, and why 
it is difficult to probe excited states. In Sec.~III we solve for the scalar, pseudoscalar and 
tensor glueball masses, suggest the physical states they correspond to, and discuss their impact
on meson mixing scenarios. The correlation functions at zero momentum for the scalar 
and pseudoscalar glueballs are also deduced from the spectral densities. The former serves 
to discriminate the different estimates for the triple-gluon condensate, and the latter 
is used to predict the topological susceptability. Section IV contains the conclusion and outlook.

\section{DISPERSIVE RELATION for $\rho$ RESONANCE}

\subsection{Inverse Matrix Method}

We first demonstrate our inverse matrix method to solve a dispersion relation, which belongs 
to the first kind of Fredholm integral equations, by means of several simple examples.
The goal is to find the unknown function $\rho(y)$ from the integral equation 
\begin{eqnarray}
\int_{0}^\infty dy\frac{\rho(y)}{x-y}= \omega(x),\label{su1}
\label{sum2}
\end{eqnarray}
given the input function $\omega(x)$ of the variable $x$. Suppose that $\rho(y)$ decreases quickly enough 
with the variable $y$, so the major contribution to the integral on the left-hand side originates from a finite
range of $y$. It is then justified to expand the integral into a series in $1/x$ up to some power $N$
for a sufficiently large $|x|$ by inserting
\begin{eqnarray}
\frac{1}{x-y}=\sum_{m=1}^N \frac{y^{m-1}}{x^m},
\label{ep1}
\end{eqnarray}
into Eq.~(\ref{su1}). Also suppose that $\omega(x)$ can be 
expanded into a power series in $1/x$ for a large $|x|$, 
\begin{eqnarray}
\omega(x)=\sum_{n=1}^N \frac{b_n}{x^n}.\label{bb1}
\end{eqnarray}
Note that $|x|$ being large enough is only a formal statement,
and does not have a substantial influence on our calculation. 

There are four types of classical orthogonal polynomials constructed from solutions to 
second-order differential equations: the Jacobi-like polynomials (including
the Gegenbauer polynomials, the Legendre polynomials,...) with the support 
$[-1,1]$; the Laguerre polynomials with the support $[0,\infty)$, the
Bessel polynomials with the support $[0,\infty)$, and the Hermite polynomials
with the support $(-\infty,\infty)$ \cite{KL97}. Since $\rho(y)$ in Eq.~(\ref{su1}) 
(and also spectral densities to be studied later)
is defined in the interval $[0,\infty)$, the Laguerre and Bessel polynomials may be
considered for its expansion. However, the latter are orthogonal with respect to the 
weight $\exp(-2/y)$, which suppresses the small-$y$ region we are interested in.
Hence, the Laguerre polynomials, orthogonal with respect to the 
weight $\exp(-y)$, are the only appropriate choice for the basis functions. 
We thus decompose the unknown into
\begin{eqnarray}
\rho(y)=\sum_{n=1}^N a_ny^\alpha e^{-y}L_{n-1}^{(\alpha)}(y),\label{r1}
\end{eqnarray}
in terms of a set of generalized Laguerre functions $L_n^{(\alpha)}$ up to degree $N-1$,
which satisfies the orthogonality
\begin{eqnarray}
\int_0^\infty y^\alpha e^{-y}L_m^{(\alpha)}(y)L_n^{(\alpha)}(y)dy=\frac{\Gamma(n+\alpha+1)}{n!}\delta_{mn}.
\label{or1}
\end{eqnarray}  
The maximal number $N$ will be fixed later, and the choice of the parameter $\alpha$ depends on the 
behavior of $\rho(y)$ in the limit $y\to 0$.
Substituting Eqs.~(\ref{ep1}), (\ref{bb1}) and (\ref{r1}) into Eq.~(\ref{su1}), and equating the
coefficients of $1/x^n$, we arrive at the matrix equation $Ma=b$ with the matrix elements
\begin{eqnarray}
M_{mn}=\int_{0}^\infty dy y^{m-1+\alpha}e^{-y}L_{n-1}^{(\alpha)}(y),\label{m2}
\end{eqnarray}
with $m$ and $n$ running from 1 to $N$, and the vectors $a=(a_1, a_2,\cdots,a_N)$ and $b=(b_1,b_2,\cdots,b_N)$. 

If the inverse of $M$ exists, one can get a solution of $a$ via $a=M^{-1}b$ with the known input $b$ trivially. 
The existence of $M^{-1}$ thus implies the uniqueness of the solution to $\rho(y)$. In principle, the true
solution can be approached to by increasing the number of polynomials $N$ in Eq.~(\ref{r1}). The difference 
between the true solution and the approximate one produces a power correction $1/x^{N+1}$ to the left-hand 
side of Eq.~(\ref{su1}), because of the orthogonality in Eq.~(\ref{or1}), which is beyond the considered
precision. The orthogonality also leads to $M_{mn}=0$ for $m<n$. Namely, $M$ is a triangular matrix, such 
that the coefficients $a_n$ built up previously are not altered, when an additional higher-degree polynomial 
is added to the expansion in Eq.~(\ref{r1}). Nevertheless, both $m$ and $n$ have to stop at a finite $N$ 
in a practical application, since the determinant of $M$ diminishes with its dimension eventually.  
An approximate solution of $a$ would then deviate from the true solution violently, when a tiny 
fluctuation of the input vector $b$ is present and amplified by the huge elements of $M^{-1}$. 
This is a generic feature of an ill-posed inverse problem. 
Hence, the optimal choice of $N$ is set to its maximal value, above which a solution goes out of control.

\begin{figure}
\includegraphics[scale=0.4]{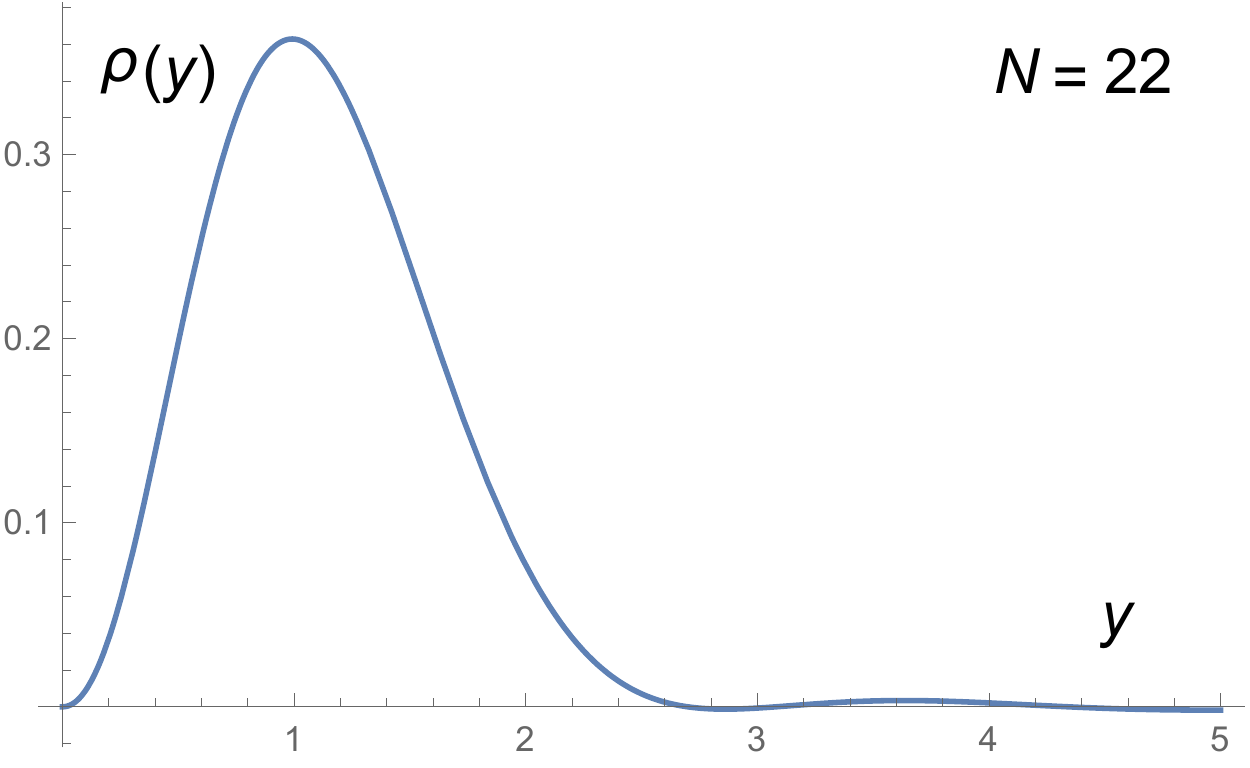}\hspace{0.5cm}
\includegraphics[scale=0.4]{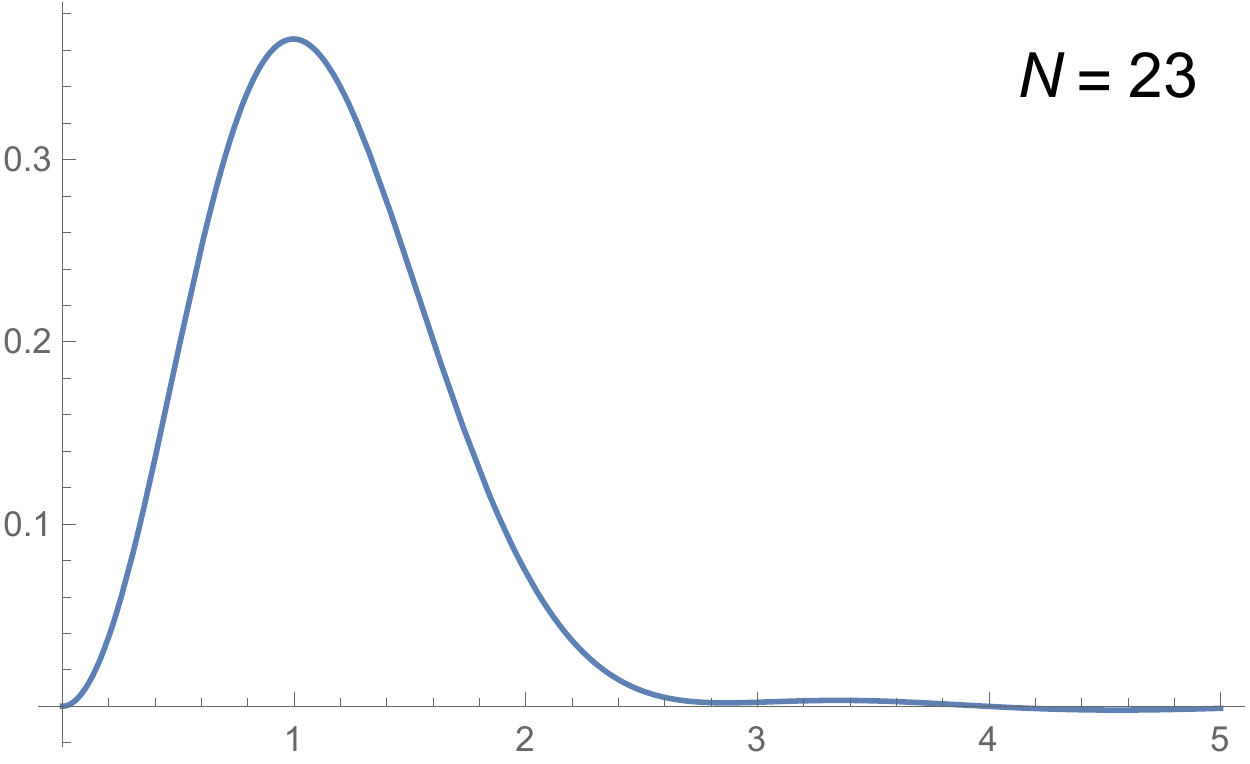}\hspace{0.5cm}
\includegraphics[scale=0.4]{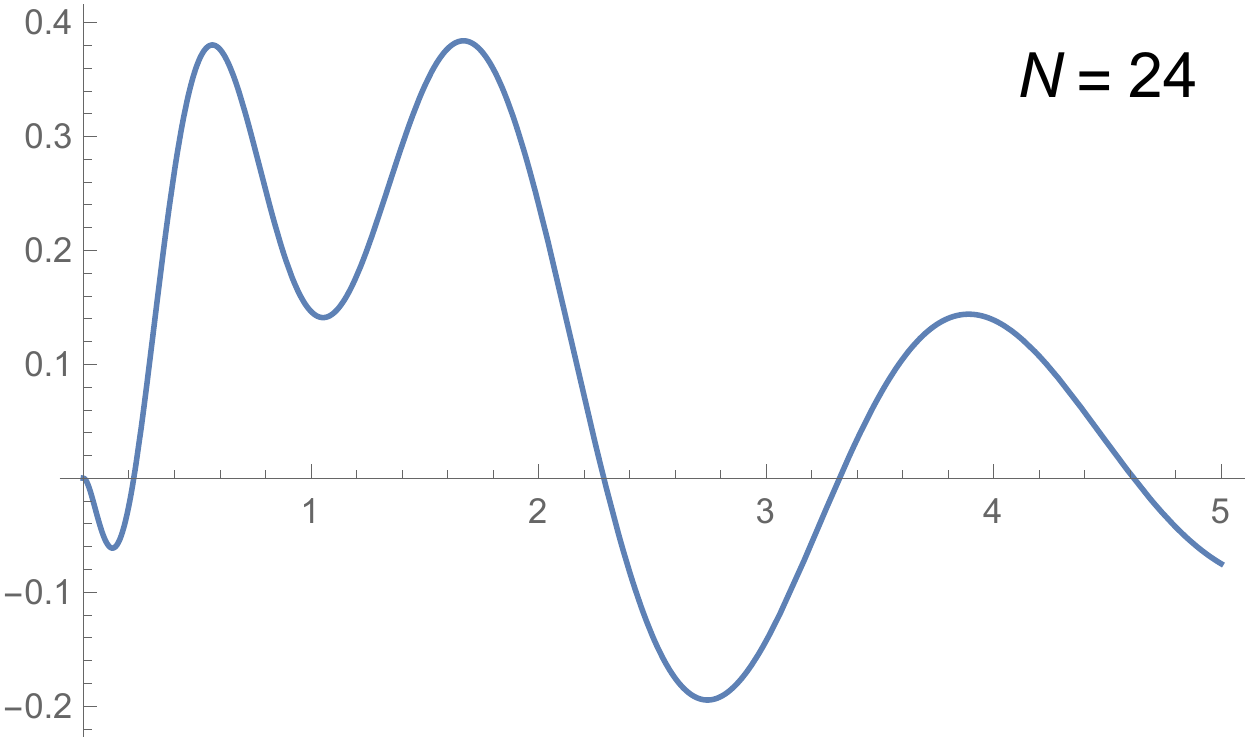}
\caption{\label{fig1}
Solutions to $\rho(y)$ for the input from Eq.~(\ref{bn}) with the expansions up 
to $N=22$, 23 and 24 generalized Laguerre polynomials $L_n^{(2)}(y)$.}
\end{figure}

We test the above inverse matrix method on the following simple examples.
We generate the mock data using a single-peak function
\begin{eqnarray}
\rho(y)=y^2 e^{-y^2},\label{ts1}
\end{eqnarray}
for the input $\omega(x)$ with the coefficients
\begin{eqnarray}
b_n=\int_{0}^\infty dy y^{n-1}y^2e^{-y^2}.\label{bn}
\end{eqnarray}
The factor $e^{-y^2}$ in the test function guarantees that the dominant contribution to the
integral comes from the region with finite $y$, and justifies the power expansion in $1/x$
with a sufficiently large $|x|$.
The matrix $M$ is computed according to Eq.~(\ref{m2}) with $\alpha=2$, motivated by the limit 
$\rho(y)\to y^2$ as $y\to 0$. The inverse $M^{-1}$ gives the solutions for the coefficients 
$a_n$ via $a=M^{-1}b$, and $\rho(y)$ in Eq.~(\ref{r1}), whose behaviors with the 
expansions up to $N=22$, 23 and 24 generalized Laguerre polynomials $L_n^{(2)}(y)$ are displayed in 
Fig.~\ref{fig1}. It is found that the curves labelled by $N=22$ and 23 are almost identical, 
implying the stability of the solutions for a sufficiently large $N$. In fact, the solutions 
have changed little as $N> 20$. It is also seen that the curve
becomes oscillatory, and differs drastically from the single-peak function when
$N$ reaches 24. The big ratio of the last two coefficients $a_{24}/a_{23}\approx 58$,
compared with $a_{22}/a_{21}\approx 1$ for $N=22$ and $a_{23}/a_{22}\approx 2$ for $N=23$,
hints that the inverse $M^{-1}$ is out of control at $N=24$.

\begin{figure}
\includegraphics[scale=0.4]{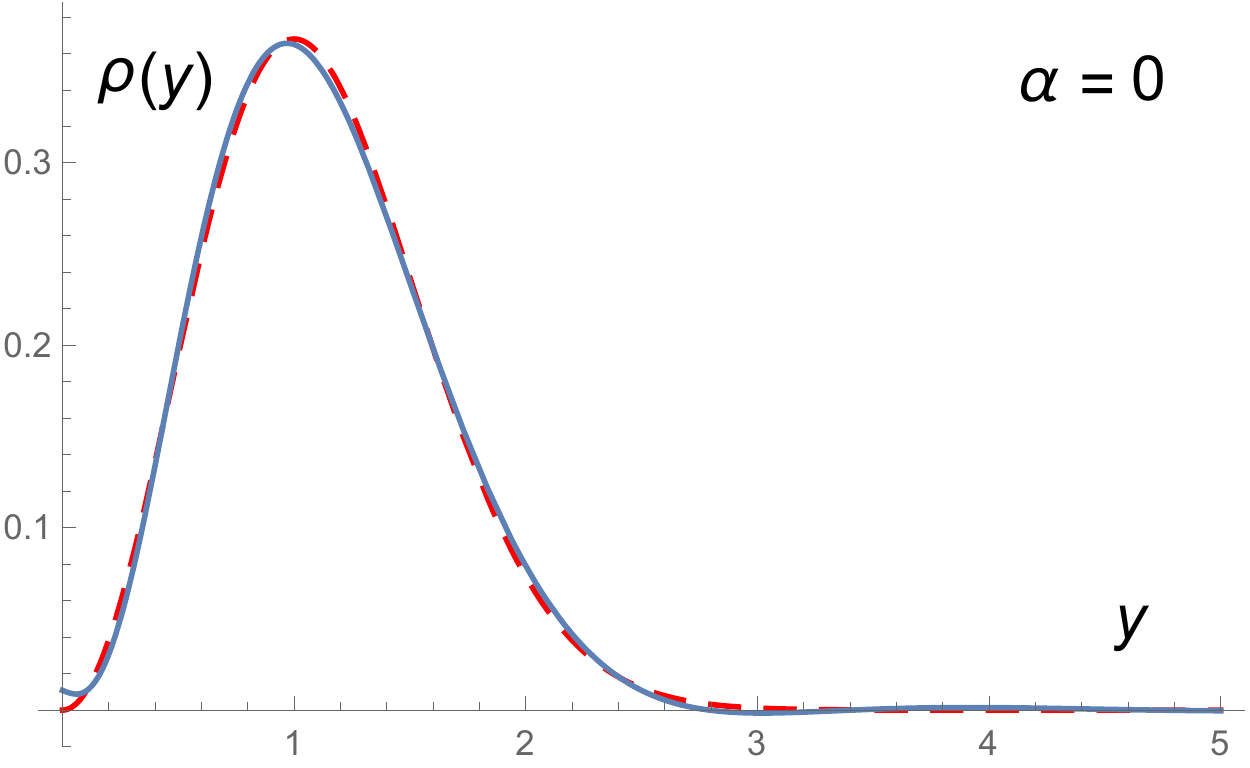}\hspace{0.5cm}
\includegraphics[scale=0.4]{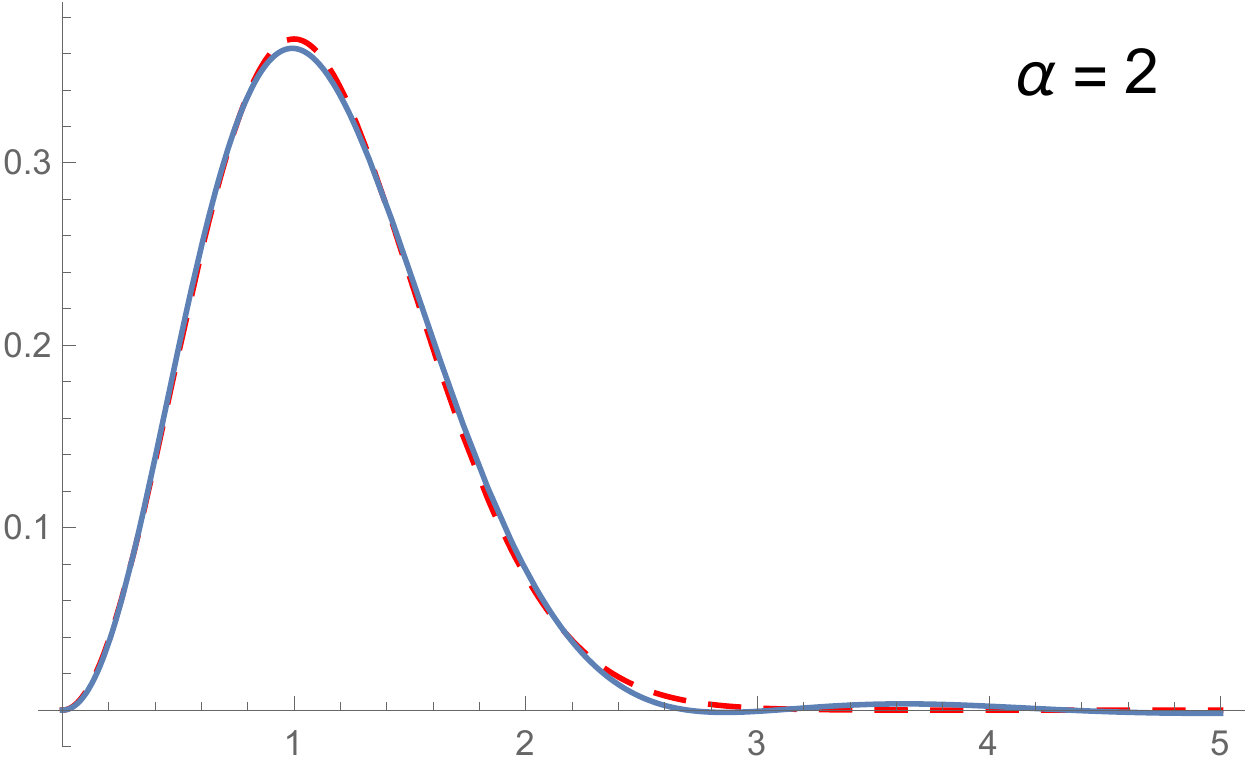}\hspace{0.5cm}
\includegraphics[scale=0.4]{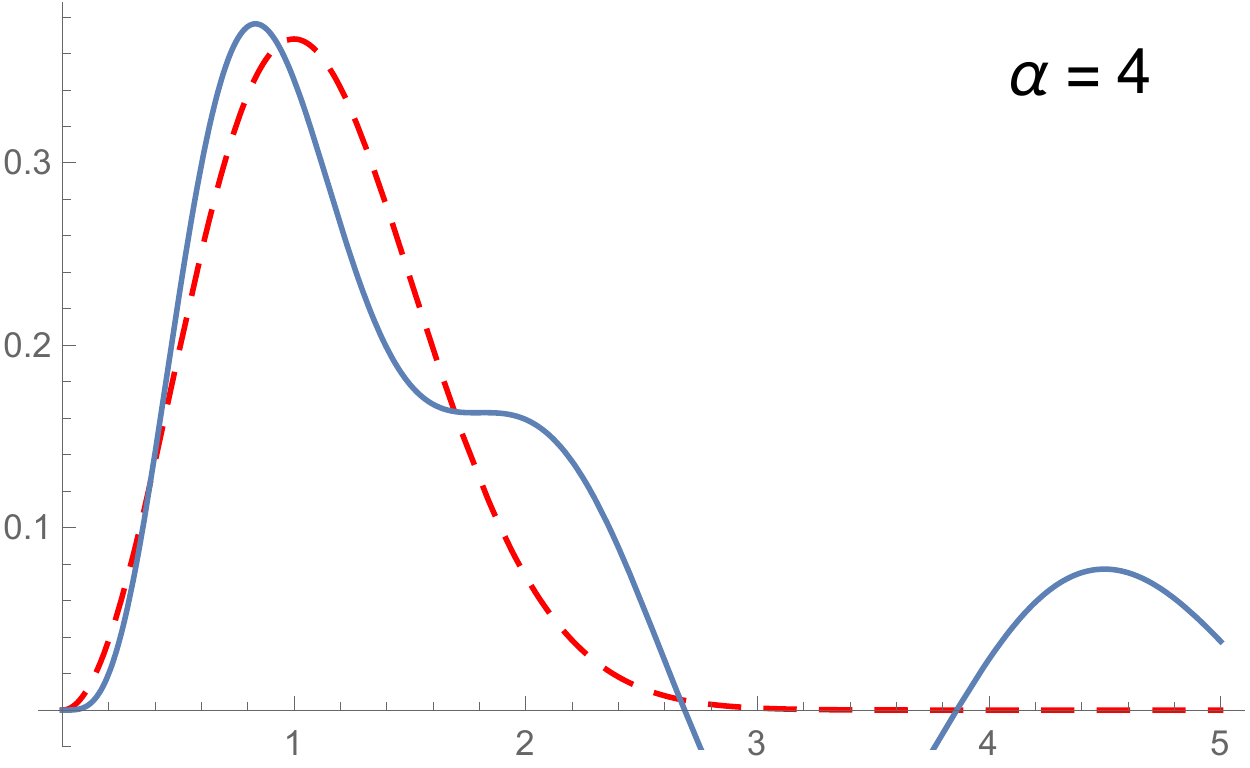}
\caption{\label{fig2}
Solutions to $\rho(y)$ with the expansions up to $22$
generalized Laguerre polynomials $L_n^{(\alpha)}(y)$ for $\alpha=0$, 2 and 4,
compared with the true solution represented by the dashed (red) lines.}
\end{figure}

We fix $N=22$ in the demonstration below, though the choices $N=21$ and $N=23$ also serve the 
purpose. Figure~\ref{fig2} collects the solutions to $\rho(y)$ from the 
parameters $\alpha=0$, 2 and 4 for the generalized Laguerre polynomials $L_n^{(\alpha)}(y)$,
compared with the true solution in Eq.~(\ref{ts1}). The curve corresponding to
$\alpha=2$, namely, the expansion of $\rho(y)$ in Eq.~(\ref{r1}) with the correct behavior 
in the $y\to 0$ limit, matches the true solution most perfectly. The curve labelled by
$\alpha=0$, despite of describing the true solution equally well at finite $y$, shows 
deviation near the origin $y=0$. As $\alpha=4$, the approximate solution differs completely
from the true solution. The above test manifests the importance of the information on
the boundary conditions of the unknown and the choice of the appropriate set of generalized 
Laguerre polynomials for solving Eq.~(\ref{su1}). It will be made explicit in the next sections
that the $\alpha=1$ ($\alpha=2$) set of generalized Laguerre polynomials is selected
for the analysis of the $\rho$ meson (glueball) mass.

\begin{figure}
\includegraphics[scale=0.45]{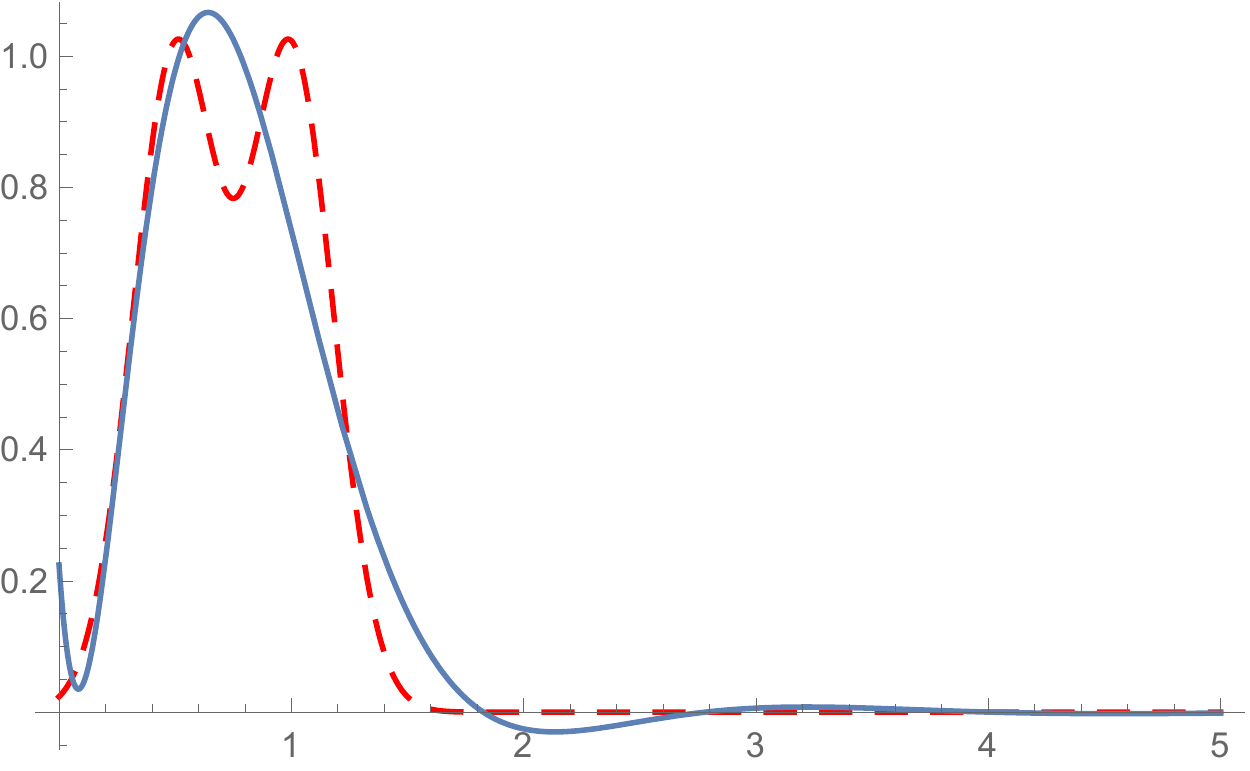}\hspace{1.0cm}
\includegraphics[scale=0.45]{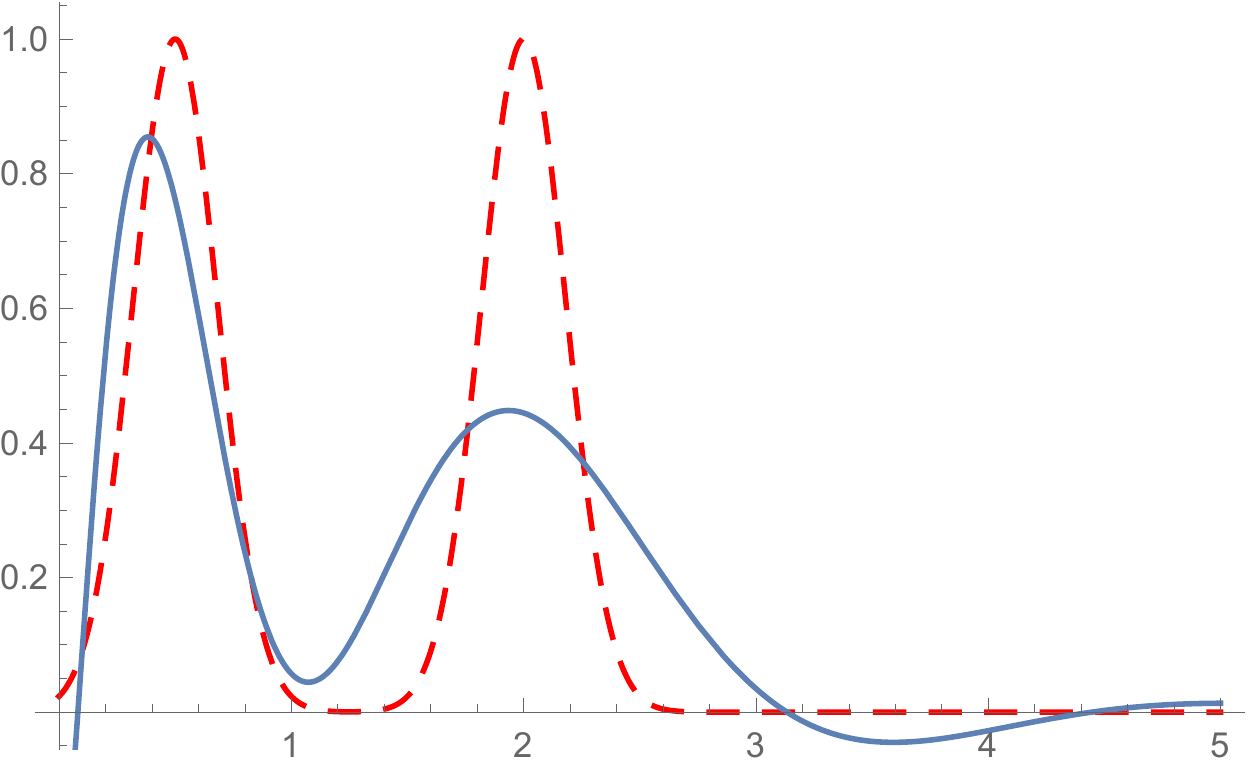}
\caption{\label{fig3}
Solutions to $\rho_1(y)$ (left) and to $\rho_2(y)$ (right) with the expansions up to $22$
generalized Laguerre polynomials $L_n^{(0)}(y)$,
compared with the true solutions represented by the dashed (red) lines.}
\end{figure}

We highlight that the exponential factor $e^{-y}$ in the polynomial expansion in Eq.~(\ref{r1})
characterizes the resolution power $\Delta y\sim 1$ of our method to probe the structure of the 
unknown $\rho(y)$. To elaborate this point, consider the two double-peak functions
\begin{eqnarray}
\rho_1(y)&=&e^{-20(y-0.5)^2}+e^{-20(y-1.0)^2},\nonumber\\
\rho_2(y)&=&e^{-20(y-0.5)^2}+e^{-20(y-2.0)^2},
\end{eqnarray}
in which the two peaks are separated by $\Delta y\sim 0.5$ and $\Delta y \sim 1.5$, respectively.
The large coefficients 20 in the exponents are designed to make sharp peaks for transparent 
illustration. The above two functions are substituted for $y^2e^{-y^2}$ in Eq.~(\ref{bn})
to generate the mock data for the inputs. The same procedure, with the expansion  
up to 22 generalized Laguerre polynomials $L_n^{(0)}(y)$, gives the solutions
to $\rho_1(y)$ and $\rho_2(y)$ in Fig.~\ref{fig3}. It is observed that the approximate solution contains
a single peak in the left plot, since the two peaks of $\rho_1(y)$ are too close to resolve.
On the contrary, the two-peak structure of $\rho_2(y)$, with a separation within the designated resolution,
can be reproduced reasonably. The above test reflects the limitation of our method to probe a physical system
with certain characteristic scales.


\subsection{$\rho$ Meson Mass}

After exploring the aspects of our approach, we apply it to the determination of the
$\rho$ meson mass from the corresponding dispersion relation. We first recapitulate and expand the 
idea of handling QCD sum rules as an inverse problem \cite{Li:2020ejs}, starting with the 
two-point correlator
\begin{eqnarray}
\Pi_{\mu\nu}(q^2)=i\int d^4xe^{iq\cdot x}
\langle 0|T[J_\mu(x)J_\nu(0)]|0\rangle=(q_\mu q_\nu-g_{\mu\nu}q^2)\Pi(q^2),\label{cur}
\end{eqnarray}
for the quark current $J_\mu=(\bar u\gamma_\mu u-\bar d\gamma_\mu d)/\sqrt{2}$.
The vacuum polarization function $\Pi(q^2)$ obeys the identity 
\begin{eqnarray}
\Pi(q^2)=\frac{1}{2\pi i}\oint ds\frac{\Pi(s)}{s-q^2},\label{di1}
\end{eqnarray}
where the contour consists of two pieces of horizontal lines above and below the branch cut along 
the positive real axis on the complex $s$ plane, and a circle of large radius $R$ \cite{Li:2020ejs}.
The OPE of the function $\Pi(q^2)$ in the deep Euclidean region of $q^2$ is 
reliabe, and we have $\Pi^{\rm OPE}(q^2)$ \cite{SVZ} for the left-hand side of Eq.~(\ref{di1}),
\begin{eqnarray}
\Pi^{\rm OPE}(q^2)&=&\Pi^{\rm pert}(q^2)+
\frac{1}{12\pi}\frac{\langle\alpha_sG^2\rangle}{(q^2)^2}+
2\frac{\langle m_q \bar q q\rangle}{(q^2)^2} +\frac{224\pi}{81}
\frac{\kappa \alpha_s\langle \bar q q\rangle^2}{(q^2)^3},\label{ope}\\
\Pi^{\rm pert}(q^2)&=&\frac{1}{4\pi^2}\left(1+\frac{\alpha_s}{\pi}\right)\ln\frac{\mu^2}{-q^2}
\equiv c\ln\frac{\mu^2}{-q^2},\label{ope2}
\end{eqnarray}
up to the dimension-six condensate, ie., to the power correction of $1/(q^2)^3$. The second
expression of the perturbative term $\Pi^{\rm pert}(q^2)$ defines the constant $c$. In Eq.~(\ref{ope}) 
$\langle\alpha_s G^2\rangle\equiv \langle \alpha_s G^a_{\mu\nu}G^{a\mu\nu}\rangle$ is the gluon condensate, 
$m_q$ is a quark mass, and the parameter $\kappa=2$-4 \cite{CDK,SN95,SN09}  
quantifies the violation in the factorization of the four-quark condensate $\langle (\bar q q)^2\rangle$ 
into the square of the quark condensate $\langle \bar q q\rangle$. A regularization-scheme dependent 
constant in $\Pi^{\rm pert}(q^2)$ \cite{Kallen:1955fb} has been dropped,
which is irrelevant to the search for a resonance solution. The fact that this constant can be 
eliminated by the Borel operator in standard sum rules confirms the above statement.

The contour integral on the right-hand side of Eq.~(\ref{di1}) can be written as
\begin{eqnarray}
\frac{1}{2\pi i}\oint ds\frac{\Pi(s)}{s-q^2}=
\frac{1}{\pi}\int_{0}^R ds\frac{{\rm Im}\Pi(s)}{s-q^2}
+\frac{1}{2\pi i}\int_C ds\frac{\Pi^{\rm pert}(s)}{s-q^2},\label{di2}
\end{eqnarray}
in which the lower bound of the first integral on the right-hand side, being of order of 
the pion mass squared, has been set to zero for simplicity, and the imaginary part
${\rm Im}\Pi(s)$, involving nonperturbative dynamics from the low $s$ region, 
will be solved for later. The numerator in the second integral, with $C$ representing the large circle 
of radius $R$, has been replaced by $\Pi^{\rm pert}(s)$, because
the perturbative evaluation of $\Pi(s)$ is reliable for $s$ far away from physical poles, 
in accordance with the employment of the OPE in Eq.~(\ref{ope}). 
We also express the perturbative piece $\Pi^{\rm pert}(q^2)$ by means of an integration along the 
same contour, so Eq.~(\ref{ope}) becomes
\begin{eqnarray}
\Pi^{\rm OPE}(q^2)=\frac{1}{2\pi i}\oint ds\frac{\Pi^{\rm pert}(s)}{s-q^2}+
\frac{1}{12\pi}\frac{\langle\alpha_sG^2\rangle}{(q^2)^2}+
2\frac{\langle m_q \bar q q\rangle}{(q^2)^2} +\frac{224\pi}{81}
\frac{\kappa \alpha_s\langle \bar q q\rangle^2}{(q^2)^3}.\label{di3}
\end{eqnarray}
Equating Eqs.~(\ref{di2}) and (\ref{di3}) according to Eq.~(\ref{di1}), we get
the sum rule
\begin{eqnarray}
\frac{1}{\pi}\int_{0}^R ds\frac{{\rm Im}\Pi(s)}{s-q^2}=
\frac{1}{\pi}\int_{0}^R ds\frac{{\rm Im}\Pi^{\rm pert}(s)}{s-q^2}+
\frac{1}{12\pi}\frac{\langle\alpha_sG^2\rangle}{(q^2)^2}+
2\frac{\langle m_q \bar q q\rangle}{(q^2)^2} +\frac{224\pi}{81}
\frac{\kappa \alpha_s\langle \bar q q\rangle^2}{(q^2)^3},\label{di4}
\end{eqnarray}
where the contributions of $\Pi^{\rm pert}(s)$ along the big circle $C$,
together with the dependence on the renormalization scale $\mu$,
have cancelled from both sides. 

We introduce a subtracted spectral density, which is related to the original 
one $\rho(s)\equiv {\rm Im}\Pi(s)/\pi$ via
\begin{eqnarray}
\Delta\rho(s,\Lambda)=\rho(s)-\frac{1}{\pi}{\rm Im}\Pi^{\rm pert}(s)[1-\exp(-s/\Lambda)].\label{sub}
\end{eqnarray}
The scale $\Lambda$ characterizes the transition of ${\rm Im}\Pi(s)$ to the perturbative expression 
${\rm Im}\Pi^{\rm pert}(s)$. The smooth function $1-\exp(-s/\Lambda)$ 
vanishes like $s$ as $s\to 0$, and approaches to the unity at large $s\gg\Lambda$, such 
that $\Delta\rho(s,\Lambda)$ respects the behavior of $\rho(s)\sim s$ in the limit 
$s\to 0$ \cite{Kwon:2008vq}, and diminishes quickly as $s>\Lambda$. 
Note that $\Delta\rho(s,\Lambda)$ bears the nontrivial resonance structure the same as
$\rho(s)$ for $s<\Lambda$, which is not affected by the perturbative subtraction term.
We have confirmed that other smooth functions with the 
similar limiting behaviors lead to basically identical solutions for $\rho(s)$. 
If one adopts the step function, instead of the smooth function in Eq.~(\ref{sub}), to
define $\Delta\rho(s,\Lambda)$, the resultant $\rho(s)$, as a sum of
the smooth $\Delta\rho(s,\Lambda)$ from solving the dispersion relation and the
discontinuous perturbative contribution caused by the step function, will   
exhibit a sudden jump. The radius $R$ in Eq.~(\ref{di4}) 
can be pushed toward the infinity, when the sum rule is formulated in 
terms of the subtracted spectral density: 
\begin{eqnarray}
\int_{0}^\infty ds\frac{\Delta\rho(s,\Lambda)}{s-q^2}
&=&\int_{0}^\infty ds \frac{c e^{-s/\Lambda}}{s-q^2}
+\frac{1}{12\pi}\frac{\langle\alpha_sG^2\rangle}{(q^2)^2}+
2\frac{\langle m_q \bar q q\rangle}{(q^2)^2} +\frac{224\pi}{81}
\frac{\kappa \alpha_s\langle \bar q q\rangle^2}{(q^2)^3},\label{r20}
\end{eqnarray}
where the constant $c$ has been defined in Eq.~(\ref{ope2}), and the aforementioned
regularization-scheme dependent constant is absent from ${\rm Im}\Pi^{\rm pert}(s)$. 
It is stressed that the quark-hadron duality for the unknown spectral density is not assumed 
at any finite scale $s$ in the above derivation.


Since the subtracted spectral density $\Delta\rho(s,\Lambda)$ is a dimensionless quantity, 
it can be expressed as a function in the form $\Delta\rho(s/\Lambda)$. 
Certainly, the subtracted spectral density may depend on other constant scales, such
as the $\rho$ meson mass $m_\rho$, which, however, appears as a constant ratio $m_\rho/\Lambda$ for a 
given $\Lambda$, and needs not be shown as an explicit argument. 
Equation~(\ref{r20}) then reduces, with the variable changes $x= q^2/\Lambda$ and 
$y= s/\Lambda$, to
\begin{eqnarray}
& &\int_{0}^\infty dy\frac{\Delta\rho(y)}{x-y}
=\int_{0}^\infty dy \frac{c e^{-y}}{x-y}
-\frac{1}{12\pi}\frac{\langle\alpha_sG^2\rangle}{x^2\Lambda^2}-
2\frac{\langle m_q \bar q q\rangle}{x^2\Lambda^2} -\frac{224\pi}{81}
\frac{\kappa \alpha_s\langle \bar q q\rangle^2}{x^3\Lambda^3},
\label{r21}
\end{eqnarray}
where $\Lambda$ in the subtracted spectral density has moved into the condensate terms to make them 
dimensionless. On one hand, the scale $\Lambda$ prefers to be small to enhance the 
resolution of our method. On the other hand, it cannot be too small to spoil the OPE. When 
$\Lambda$ increases from a low scale, the physical $\rho$ meson mass, if generated, corresponds to 
a peak location of $\rho(s)$, which should be insensitive to the change of $\Lambda$. 
As $\Lambda$ further increases, it disappears with the condensate contributions from 
Eq.~(\ref{r21}). Then a solution for $\Delta\rho(y)$, if existent, will imply that $\Delta\rho(s/\Lambda)$
is a solution for an arbitrary $\Lambda$. A peak location of $\rho(s)$, endowed by $\Delta\rho(s/\Lambda)$, thus shifts
with $\Lambda$, such that none of its structure can be interpreted as a physical state. The numerical 
analyses to be performed below verify this tendency of the $\rho$ meson mass $m_\rho$, as well as of 
the glueball masses, with respect to the variation of $\Lambda$: the obtained $m_\rho$ remains stable 
first, and then grows with $\Lambda$ monotonically. In the sense of searching for a stability window 
in which a resonance mass stays constant, $\Lambda$ plays a role similar to the Borel mass in 
conventional sum rules.


Viewing the boundary conditions of $\Delta\rho(y)\sim y$ at $y\to 0$ and $\Delta\rho(y)\to 0$
at $y\to\infty$, we expand $\Delta\rho(y)$ in terms of the generalized Laguerre polynomials $L_n^{(1)}(y)$,
namely, employ Eq.~(\ref{r1}) with the parameter $\alpha=1$. The dependence on the constant ratios 
mentioned before then goes into the coefficients $a_n$: a solution of $a_n$ depends on $\Lambda$,
as indicated by the right-hand side of Eq.~(\ref{r21}). Equation~(\ref{ep1}), which holds for 
$|x|>1$, ie., for $|q^2| > \Lambda$, is inserted into the left-hand side of Eq.~(\ref{r21}) to 
construct the matrix elements $M_{mn}$ in Eq.~(\ref{m2}), and inserted into the integral on the right-hand 
side of Eq.~(\ref{r21}) to compute the coefficients $b_n$ for the input. The coefficients $b_2$ and $b_3$
of the $1/x^2$ and $1/x^3$ terms, respectively, receive additional contributions from the condensates.
The following OPE parameters and the strong coupling $\alpha_s$, evaluated at the scale of 1 GeV 
and within their accepted ranges \cite{CDK,SN95,SN09,Wang:2016sdt,Narison:2014wqa,SVZ,SN98,Forkel:2003mk},
are adopted:
\begin{eqnarray}
& &\langle m_q\bar q q\rangle = 0.007\times(-0.246)^3\;{\rm GeV}^4,\;\;
\langle\alpha_sG^2\rangle=0.08\; {\rm GeV}^4,\nonumber\\
& &\alpha_s\langle \bar q q\rangle^2 = 1.49\times 10^{-4}\;{\rm GeV}^6,\;\;\alpha_s=0.5,\;\; \kappa=2,\label{put}
\end{eqnarray}
which will be shown to produce the observed $\rho$ meson mass 
$m_\rho= 0.78$ GeV \cite{PDG}. Though higher-order corrections to the condensate terms are
available \cite{Wang:2016sdt,ST90}, it may not mean much to include them, because their effects can be 
mimicked by tuning the unknown factorization violation parameter $\kappa$. We have checked that the 
renormalization-group evolutions of $\alpha_s$ and the condensates around the scales 1-2 GeV affect our results
for $m_\rho$ by only few percent, so they will not be taken into account in the numerical study. 




\begin{figure}
\includegraphics[scale=0.4]{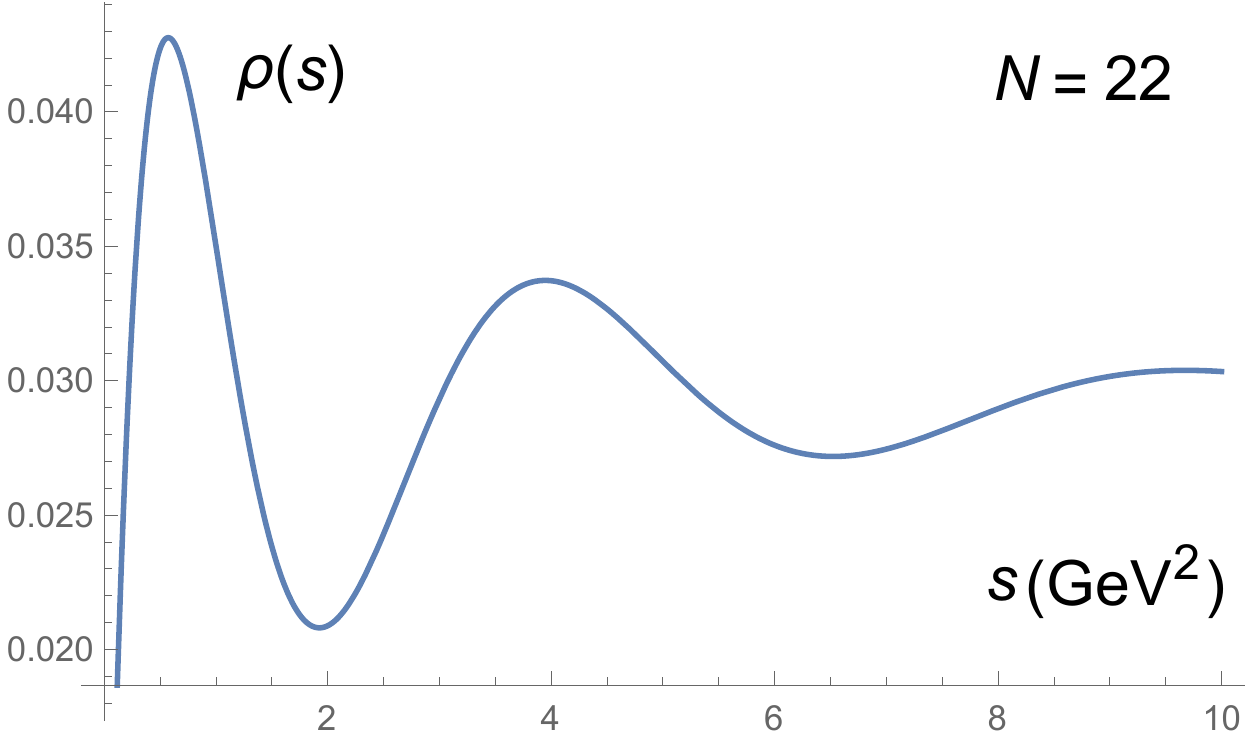}\hspace{0.5cm}
\includegraphics[scale=0.4]{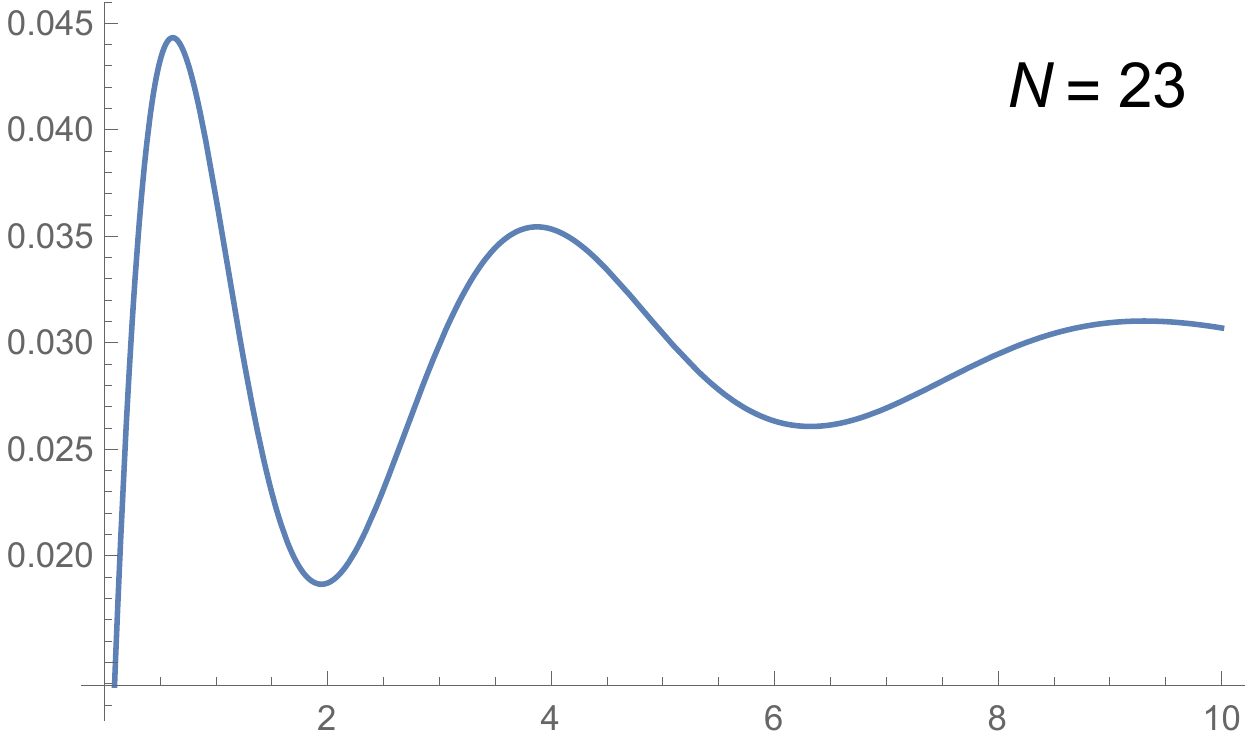}\hspace{0.5cm}
\includegraphics[scale=0.4]{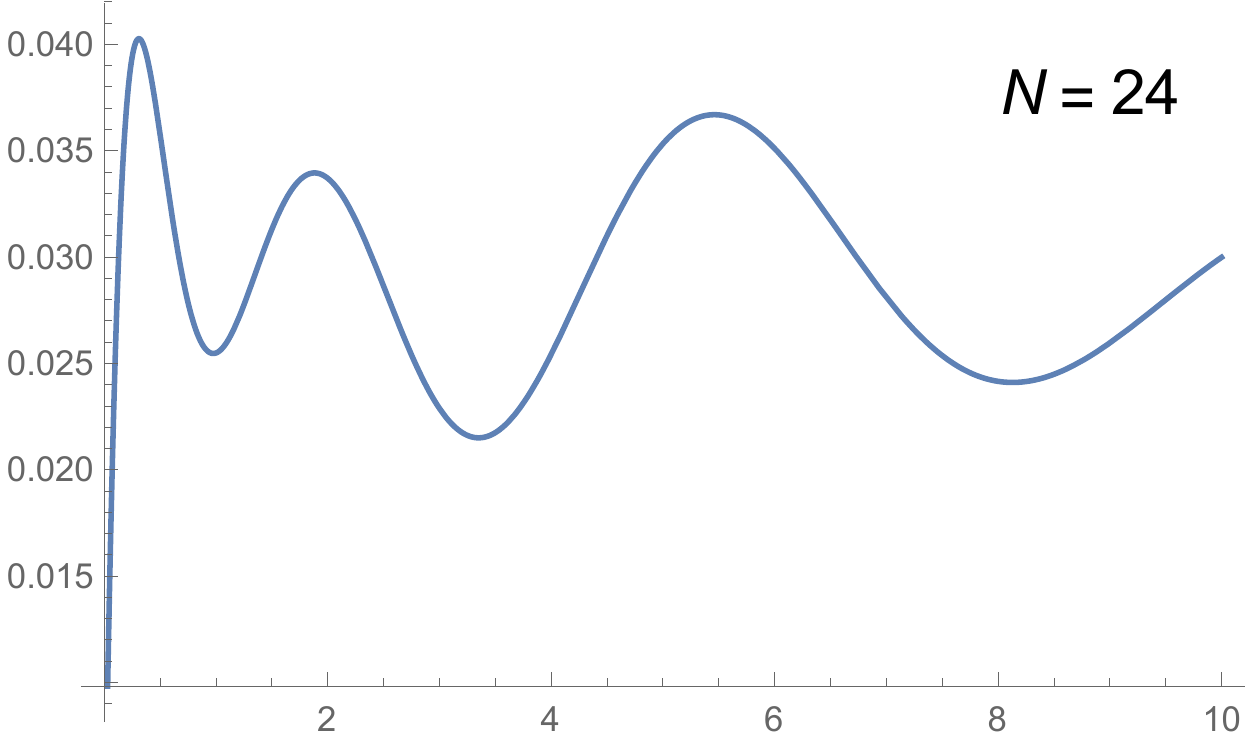}
\caption{\label{fig4}
Solutions to $\rho(s)$ for $\Lambda=2.5$ GeV$^2$ 
with the expansions up to $N=22$, 23 and 24 generalized Laguerre polynomials $L_n^{(1)}(y)$.}
\end{figure}

We derive the inverse matrix $M^{-1}$, the coefficients $a_n$ from the OPE coefficients $b_n$, the
solution to $\Delta \rho(s,\Lambda)$, and then the spectral density $\rho(s)$
from Eq.~(\ref{sub}). The outcomes for the characteristic scale
$\Lambda=2.5$ GeV$^2$ with the expansions up to $N=22$, 23, and 24
generalized Laguerre polynomials $L_n^{(1)}(y)$ are displayed in Fig.~\ref{fig4}.
It is found that the curves labelled by $N=22$ and 23 are very similar, assuring the
stability of the solutions for a sufficiently large $N$, and consistent with those obtained 
in the the maximum entropy method \cite{Ohtani:2012ps}. The curve
becomes oscillatory, and differs significantly from the other two as
$N$ reaches 24. The big ratio of the last two coefficients $a_{24}/a_{23}\approx 7$,
compared with $a_{22}/a_{21}\approx a_{23}/a_{22}< 2$ for $N=22$ and $23$,
indicates that the matrix elements of $M^{-1}$ start to increase rapidly as $N=24$. It is encouraging
that the positivity of the spectral density is satisfied automatically. We read the $\rho$ meson mass
$m_\rho=0.78$ GeV ($m_\rho^2=0.61$ GeV$^2$) off the location of the sharp peak in the plot for $N=23$, 
which agrees with the measured value in \cite{PDG}. The bump located at $s> 2$ GeV$^2$, 
being shorter and broader, may be attributed to the combination of nonresonant contributions 
and resonant ones from excited $\rho$ states. It is obvious that the continuum contribution 
to the spectral density $\rho(s)$, distinct from the perturbative value $c\approx 0.029$, violates the 
local quark-hadron duality, though $\rho(s)$ approaches to $c$ asymptotically.

\begin{figure}
\includegraphics[scale=0.4]{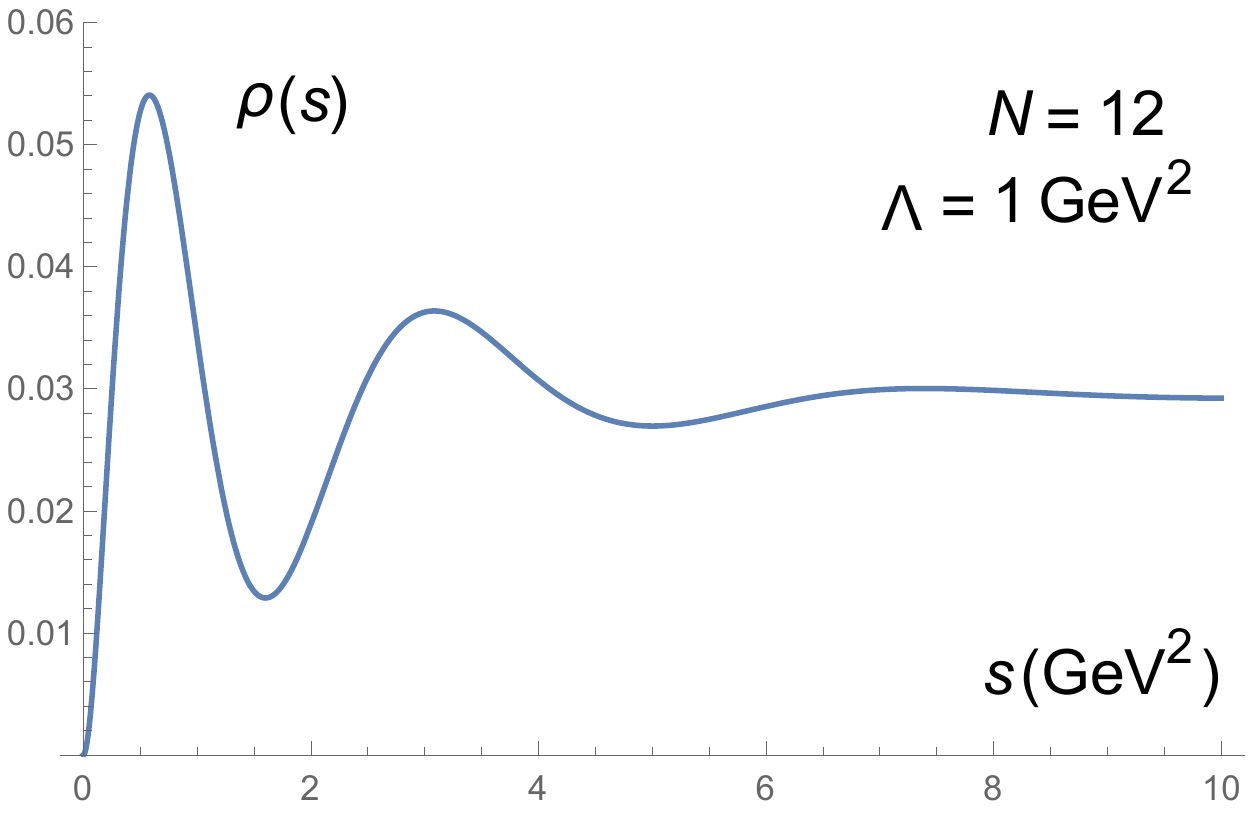}\hspace{0.5cm}
\includegraphics[scale=0.4]{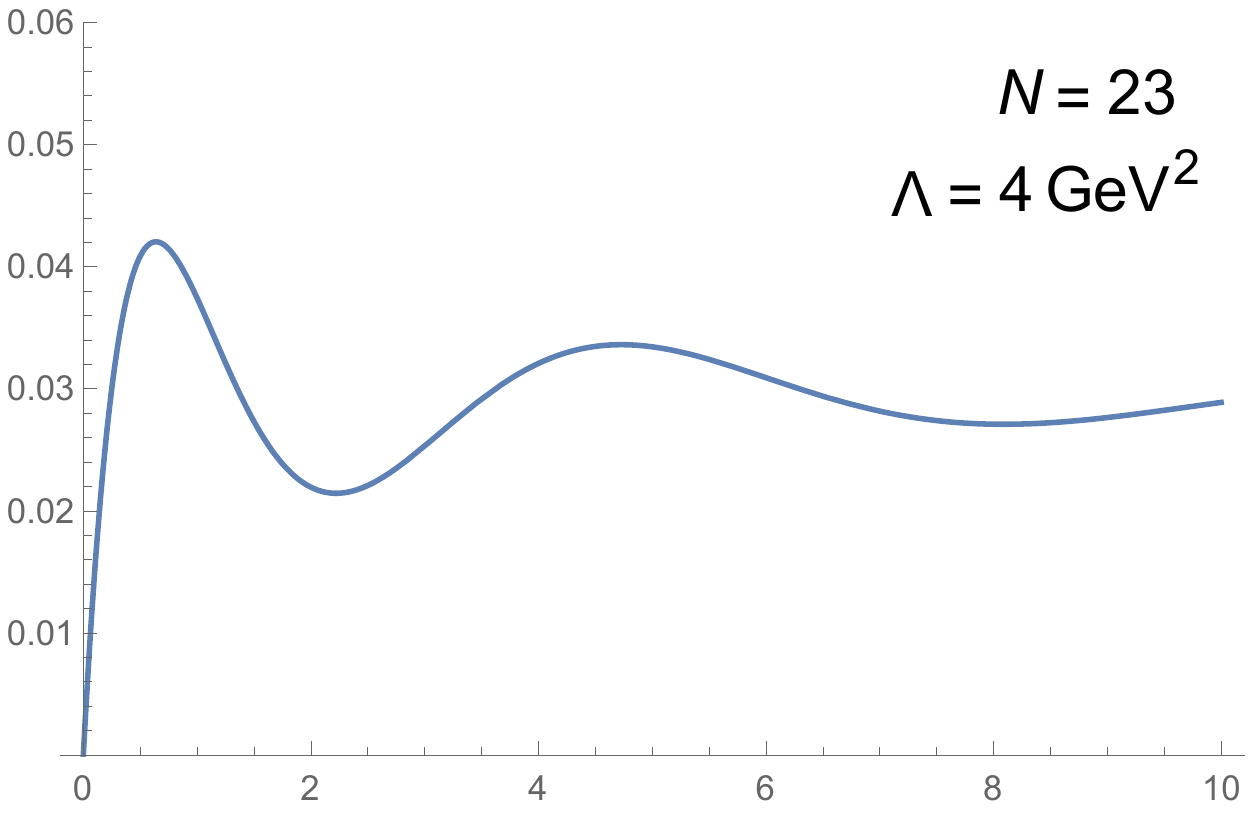}\hspace{0.5cm}
\includegraphics[scale=0.4]{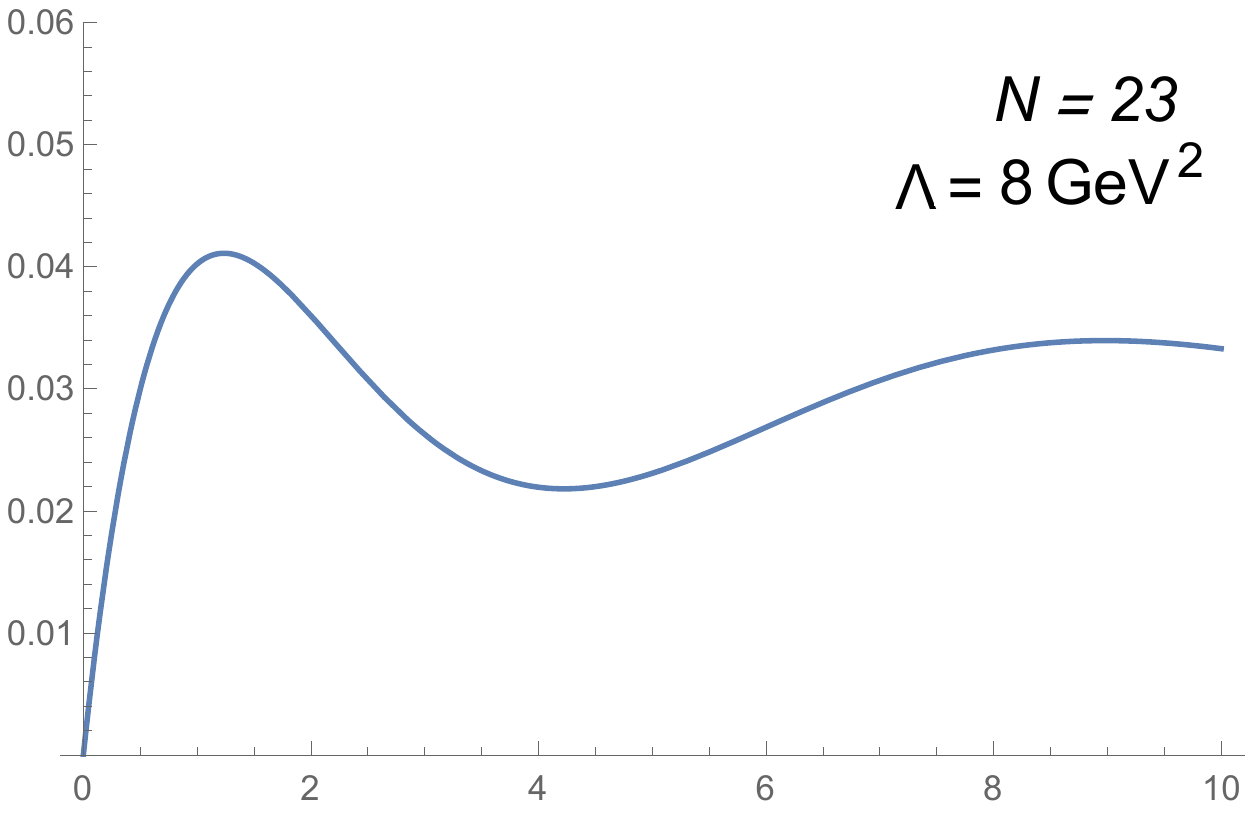}
\caption{\label{fig5}
Solutions to $\rho(s)$ for $\Lambda=1$, 4 and 8 GeV$^2$ 
with the expansions up to $N=12$, 23 and 23 generalized Laguerre polynomials $L_n^{(1)}(y)$, respectively.}
\end{figure}

We then seek the solutions of Eq.~(\ref{r21}) for various characteristic scales $\Lambda=1$-8 GeV$^2$ 
by repeating the above procedure, and present the spectral densities $\rho(s)$ for $\Lambda=1$, 4 
and 8 GeV$^2$ in Fig.~\ref{fig5} for illustration. It is noticed that the expansion in terms of the 
generalized Laguerre polynomials diverges more quickly at lower $\Lambda$, such that solutions
for $\Lambda < 1$ GeV$^2$ with the polynomial numbers roughly smaller than 10 may not be reliable. 
A clear signal for a divergent expansion is that the positivity of the spectral density is lost 
owing to strong oscillation. We have to terminate the expansion at $N=12$ ($N=22$) for $\Lambda=1$ 
GeV$^2$ ($\Lambda=2$ GeV$^2$), and at $N=23$ for higher $\Lambda$. Figure~\ref{fig5} suggests that
the peaks are located at the same $s\approx 0.61$ GeV$^2$ for $\Lambda$ between 1 and 4 GeV$^2$,
which specifies the stability window in $\Lambda$ for the physical solutions. The peak in the
plot for $\Lambda=8$ GeV$^2$ moves toward a bigger $s=1.24$ GeV$^2$, exhibiting the growth of
the mass at large $\Lambda$ with the disappearance of the condensate
effects. We point out that the peak of $\rho(s)$ always exists even without the condensates 
\cite{Li:2020ejs}: the vanishing of the spectral density at $s=0$ gives a contribution to the 
dispersive integral smaller than the constant perturbative piece does, which must be made up 
by a peak at finite $s$ in order to match the OPE input. Therefore, one has to be cautious about 
the identification of a peak in $\rho(s)$ as a physical resonance, for which the stability of the peak location 
inspected above is crucial. Accordingly, the broad bump in Fig.~\ref{fig5} cannot be interpreted as 
a specific bound state, since its location shifts with $\Lambda$ as disclosed in the three plots. 
We also observe that the peak becomes less evident at large $\Lambda$, for the bad
resolution of our method stops probing the structure of $\rho(s)$ at low $s$.

\begin{figure}
\includegraphics[scale=0.5]{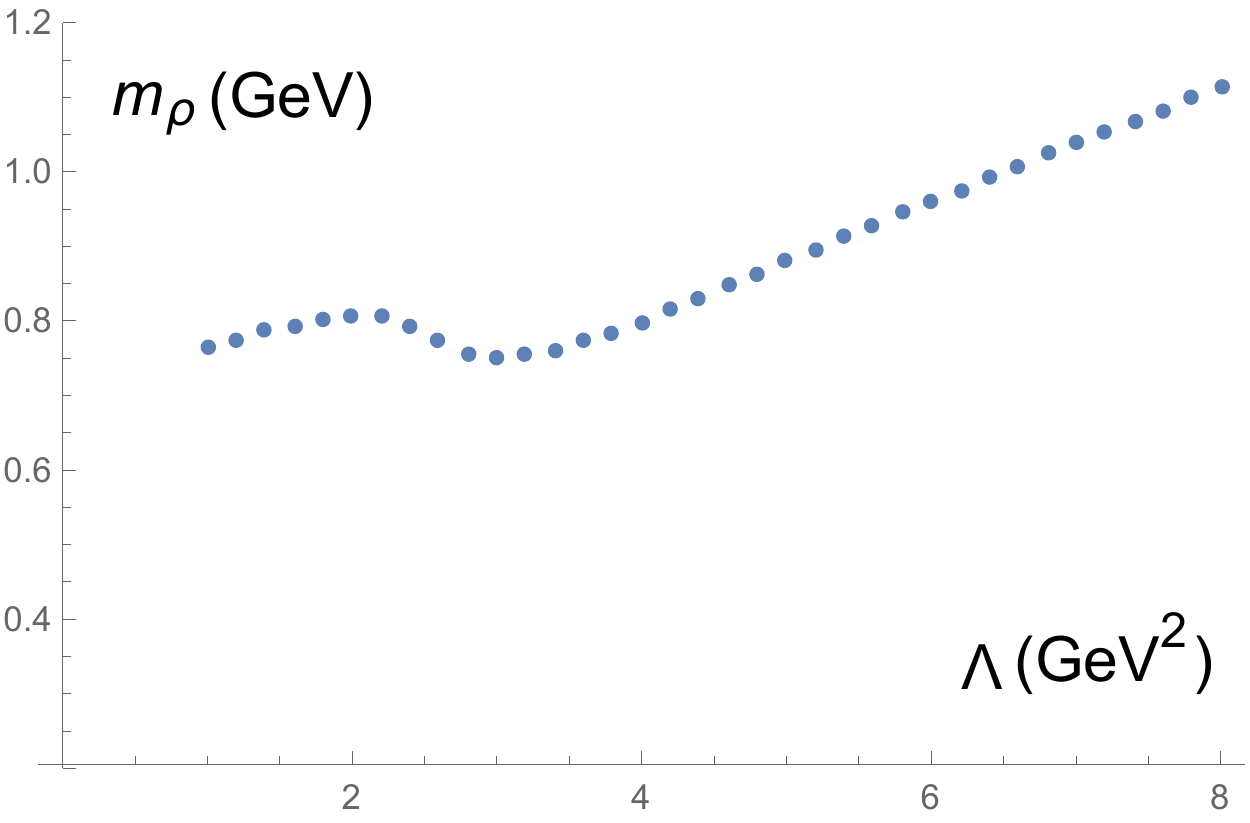}
\caption{\label{fig6}
Dependence of the $\rho$ meson mass $m_\rho$ on the characteristic scale $\Lambda$.}
\end{figure}

Next we scan the $\rho$ meson mass $m_\rho$ in the range 1 GeV$^2<\Lambda <8$ GeV$^2$, and 
depict its dependence on $\Lambda$ in Fig.~\ref{fig6}. The expected features are salient: the 
curve goes up and down around $m_\rho=0.78$ GeV in the interval 1 GeV$^2<\Lambda <3$ GeV$^2$, 
and then ascends monotonically with $\Lambda$ as $\Lambda> 3$ GeV$^2$. 
We assess the theoretical error from our method using the extreme values
at $\Lambda=2.2$ GeV$^2$ and 3.0 GeV$^2$, and get $m_\rho=(0.78\pm 0.03)$ GeV.
To check the sensitivity of our results to the uncertainties of the OPE, 
we vary the gluon condensate $\langle\alpha_sG^2\rangle$ and the factorization violation 
parameter $\kappa$ by $\pm 20\%$ separately. The variation of the former can simulate that of the 
quark condensate $\langle \bar q q\rangle$, which is also of dimension four. The variation of the
latter is equivalent to that of the dimension-six condensate $\langle \bar q q\rangle^2$.
It turns out that the obtained $m_\rho$ changes by only about $\mp 5\%$ and $\pm 7\%$, respectively.
Namely, results in our formalism are insensitive to the OPE uncertainties. Adding the above 
sources of errors in quadrature, we conclude
\begin{eqnarray}
m_\rho=(0.78\pm 0.07)\;{\rm GeV}.
\end{eqnarray} 
The above examination also reveals that $\langle\alpha_sG^2\rangle$ and $\kappa$
must be anti-correlated in order to fix $m_\rho$. Hence, the input value of $\kappa$ 
($\langle\alpha_sG^2\rangle$) in Eq.~(\ref{put}), close to the lower (upper) edge 
of its favored range, is the optimal choice.


It was stated \cite{Yamazaki:2001er} that the area under the resonance peak of the spectral 
density $\rho(s)$ represents the square of the $\rho$ meson decay constant $f_\rho$. In 
our formulation, the resonance peak is appropriately described by the subtracted spectral 
density $\Delta\rho(s,\Lambda)$, where the continuum contribution has been largely removed. 
We thus have 
\begin{eqnarray}
f_\rho^2\approx\int_0^\infty ds\Delta\rho(s,\Lambda),\label{fr}
\end{eqnarray}
with the scale $\Lambda=1.3$ GeV$^2$ corresponding to the $\rho$ meson mass
$m_\rho=0.78$ GeV in Fig.~\ref{fig6}. The above integral leads to
$f_\rho\approx 0.20$ GeV, consistent with the value in \cite{PDG}, which
further supports our approach to the extraction of nonperturbative observables
from dispersion relations. Note that $f_\rho$ derived from Eq.~(\ref{fr})
depends on $\Lambda$: varying $\Lambda$ from 1.0 to 3.0 GeV$^2$,
we find that $f_\rho$ increases from 0.17 to 0.29 GeV. 
This $\Lambda$ dependence of $f_\rho$ is expected, because Eq.~(\ref{fr})
holds better for a narrower resonance, but the $\rho$ meson width is not small:
Eq.~(\ref{fr}) can be understood via a pole parametrization, which originates from the
narrow-width approximation.



At last, a remark is in order. The well-known Weinberg sum rules \cite{SW67} for the difference
$\rho_V(s)-\rho_A(s)$, where $\rho_V(s)$ corresponds to the vector spectral density investigated 
in this subsection and $\rho_A(s)$ denotes the axial-vector spectral density, have been 
analyzed in the literature. These sum rules, according to their derivation, are expected to be 
respected in our formalism, where the spectral densities are solved with the input from the 
OPE of the relevant correlation functions. Certainly, one has to obtain 
$\rho_A(s)$ before verifying the above postulation, which can be a subject of future studies. 
In such an approach, the Weinberg sum rules serve as a check of the theoretical consistency of 
our solutions for the spectral densities. It
differs from the approaches in \cite{Donoghue:1993xb,Kapusta:1993hq}, where the data of $e^+e^-$ 
annihilation and $\tau$ decays were employed to construct $\rho_V(s)$ and $\rho_A(s)$ for examining
whether the sum rules are satisfied.

\section{DISPERSIVE RELATIONS FOR GLUEBALLS}

\subsection{Formalism}

We apply the formalism developed in the previous section to the determination of the scalar 
and pseudoscalar glueball masses. Consider the correlation function for the
glueball channel
\begin{eqnarray}
\Pi_G(q^2)=i\int d^4xe^{iq\cdot x}\langle 0|TO_G(x)O_G(0)|0\rangle,\label{c1}
\end{eqnarray}
where the local composite operators $O_G$ with $G=S$ and $P$ denote the 
gluonic interpolating fields for the scalar $(0^{++})$ and pseudoscalar $(0^{-+})$  
glueballs, respectively. Their explicit definitions with the lowest mass dimension are given by
\begin{eqnarray}
O_S(x)&=&\alpha_s G^a_{\mu\nu}(x)G^{a\mu\nu}(x),\nonumber\\
O_P(x)&=&\alpha_s G^a_{\mu\nu}(x){\tilde G}^{a\mu\nu}(x),
\end{eqnarray}
$\tilde G_{\mu\nu}\equiv i\epsilon_{\mu\nu\rho\sigma}G^{\rho\sigma}/2$ being the dual of 
the gluon field strength. The low-energy theorem demands the zero-momentum limits of the glueball 
correlators \cite{NSVZ,SVZ80}:
\begin{eqnarray}
\Pi_S(q^2=0)&=&\frac{32\pi}{\beta_0}\langle\alpha_s G^2\rangle,\label{le0}\\
\Pi_P(q^2=0)&=&(32\pi^2\alpha_s)^2\chi_t,
\label{le}
\end{eqnarray}
in which $\beta_0=11N_c/3-2N_f/3$, $N_c$ and $N_f$ being the numbers of colors and flavors, respectively, 
is the lowest-order coefficient of the QCD $\beta$-function, and the topological 
susceptibility
\begin{eqnarray}
\chi_t=i\int d^4x\langle 0|Q(x)Q(0)|0\rangle,
\end{eqnarray}
is defined with the topological charge density $Q(x)=G^{a\mu\nu}(x)\tilde G^a_{\mu\nu}(x)/(32\pi^2)$.
The above low-energy limits will be used to test the consistency of our calculations below.


Similarly, we have the OPE of the correlation function $\Pi_G(q^2)$ in the deep Euclidean region of $q^2$ \cite{SVZ},
\begin{eqnarray}
\Pi_G^{\rm OPE}(q^2)&=&q^4 \ln\frac{-q^2}{\mu^2}\left[A_0^{(G)}+A_1^{(G)}\ln\frac{-q^2}{\mu^2}
+A_2^{(G)}\ln^2\frac{-q^2}{\mu^2}\right]\nonumber\\
& &+\left[B_0^{(G)}+B_1^{(G)} \ln\frac{-q^2}{\mu^2}\right]\langle \alpha_s G^2\rangle
-\left[C_0^{(G)}+C_1^{(G)} \ln\frac{-q^2}{\mu^2}\right]\frac{\langle g G^3\rangle}{q^2}+
D_0^{(G)}\frac{\langle \alpha_s^2 G^4\rangle_G}{(q^2)^2},\label{dig3}
\end{eqnarray}
up to the dimension-eight condensate, ie., up to the power correction of $1/(q^2)^2$,
where the various gluon condensates are defined as
\begin{eqnarray}
\langle g G^3\rangle&\equiv&\langle gf^{abc} G^a_{\mu\nu}G^{b\nu}_\rho G^{c\rho\mu}\rangle,\nonumber\\
\langle \alpha_s^2 G^4\rangle_S&\equiv& 14\langle(\alpha_sf^{abc} G^b_{\mu\rho}G^{c\rho}_\nu)^2\rangle
-\langle(\alpha_sf^{abc} G^b_{\mu\nu}G^c_{\rho\lambda})^2\rangle,\nonumber\\
\langle \alpha^2 G^4\rangle_P&\equiv& 2
[10\langle(\alpha_sf^{abc} G^b_{\mu\rho}G^{c\rho}_\nu)^2\rangle
+\langle(\alpha_sf^{abc} G^b_{\mu\nu}G^c_{\rho\lambda})^2\rangle].
\end{eqnarray}
The four-gluon condensates are approximated, under the 
vacuum factorization assumption \cite{NSVZ79,BLP} 
\begin{eqnarray}
\langle (f^{abc}G^b_{\mu\rho}G^{c\rho}_\nu)^2\rangle\approx
\frac{1}{16}\langle G^a_{\mu\nu}G^{a\mu\nu}\rangle^2,\;\;\;\;
\langle (f^{abc}G^b_{\mu\nu}G^c_{\rho\lambda})^2\rangle\approx 
\frac{5}{16}\langle G^a_{\mu\nu}G^{a\mu\nu}\rangle^2,\label{fac}
\end{eqnarray}
by 
\begin{eqnarray}
\langle \alpha_s^2G^4\rangle_S\approx \frac{9}{16}\langle \alpha_sG^2\rangle^2,\;\;\;\;
\langle \alpha_s^2G^4\rangle_P\approx \frac{15}{8}\langle \alpha_sG^2\rangle^2.
\end{eqnarray}
As stated in the introduction, we consider only the condensate contributions to the OPE, 
which are sufficient for establishing the scalar and pseudoscalar glueballs, without the instanton effect.

The coefficients in Eq.~(\ref{dig3}) for the scalar glueball with $N_c = N_f = 3$
read \cite{NSVZ79,Kataev:1981gr,BS90,CKS97,HS01}
\begin{eqnarray}
A_0^{(S)}& =&-2\left(\frac{\alpha_s}{\pi}\right)^2\left[1 +
\frac{659}{36}\frac{\alpha_s}{\pi}+ 247.48\left(\frac{\alpha_s}{\pi}\right)^2\right],\nonumber\\
A_1^{(S)}& =& 2\left(\frac{\alpha_s}{\pi}\right)^3\left(\frac{\beta_0}{4} +
65.781\frac{\alpha_s}{\pi}\right),\;\;\;\;
A_2^{(S)}=-10.125\left(\frac{\alpha_s}{\pi}\right)^4,\nonumber\\
B_0^{(S)}&=& 4\alpha_s\left(1 +\frac{175}{36}\frac{\alpha_s}{\pi}\right),\;\;\;\;
B_1^{(S)}= -\frac{\alpha_s^2}{\pi}\beta_0,\nonumber\\
C_0^{(S)}&=& 8\alpha_s^2,\;\;\;\;
C_1^{(S)}= 0,\;\;\;\;
D_0^{(S)}= 8\pi\alpha_s.
\end{eqnarray}
Those for the pseudoscalar glueball were found to be \cite{NSVZ79,ZS03,AMM92}
\begin{eqnarray}
A_0^{(P)}& =& -2\left(\frac{\alpha_s}{\pi}\right)^2\left[1 +
20.750\frac{\alpha_s}{\pi}+ 305.95\left(\frac{\alpha_s}{\pi}\right)^2\right],\nonumber\\
A_1^{(P)}& =& 2\left(\frac{\alpha_s}{\pi}\right)^3\left(\frac{\beta_0}{4} +
72.531\frac{\alpha_s}{\pi}\right),\;\;\;\;
A_2^{(P)}=-10.125\left(\frac{\alpha_s}{\pi}\right)^4,\nonumber\\
B_0^{(P)}&=& 4\alpha_s,\;\;\;\;
B_1^{(P)}= \frac{\alpha_s^2}{\pi}\beta_0\nonumber\\
C_0^{(P)}&=& -8\alpha_s^2,\;\;\;\;
C_1^{(P)}= 0,\;\;\;\;
D_0^{(P)}= 4\pi\alpha_s.\label{ap}
\end{eqnarray}
Notice $C_1^{(S)}=C_1^{(P)}= 0$, which will not appear in our formulas afterwards.

We extend the derivation of Eqs.~(\ref{di1})-(\ref{di4}) to the construction of the dispersion relations
for the glueball masses, arriving at
\begin{eqnarray}
\frac{1}{\pi}\int_{0}^R ds\frac{{\rm Im}\Pi_G(s)}{s-q^2}
&=& \frac{1}{\pi}\int_{0}^R ds\frac{{\rm Im}\Pi_G^{\rm pert}(s)}{s-q^2}
-C_0^{(G)}\frac{\langle g G^3\rangle}{q^2}+
D_0^{(G)}\frac{\langle \alpha_s^2 G^4\rangle_G}{(q^2)^2}.\label{sg}
\end{eqnarray}
The imaginary part ${\rm Im}\Pi_G^{\rm pert}(s)$ collects the contributions
in Eq.~(\ref{dig3}) without poles at $q^2\to 0$, ie., those which can be
produced by the contour integration of the perturbative piece, like the first term in Eq.~(\ref{di3}):
\begin{eqnarray}
{\rm Im}\Pi^{\rm pert}_G(s)=-\pi\left[A_0^{(G)}s^2+2A_1^{(G)}s^2\ln\frac{s}{\mu^2}
+A_2^{(G)}s^2\left(3\ln^2\frac{s}{\mu^2}-\pi^2\right)
+B_1^{(G)}\langle \alpha_s G^2\rangle\right].\label{per}
\end{eqnarray}
It is seen that the term $B_0^{(G)}\langle \alpha_s G^2\rangle$ in Eq.~(\ref{dig3}) is absent
in the above expression, since it has no discontinuity along the branch cut. As stated before,
a constant piece in the OPE is irrelevant to the search for a resonance solution.
The thresholds in the dispersive integrals on both sides of Eq.~(\ref{sg}) have been set to zero for simplicity.
We remind that ${\rm Im}\Pi_G^{\rm pert}(s)$ contains the nonperturbatve gluon condensate 
$\langle \alpha_s G^2\rangle$ actually.

We introduce the subtracted spectral density
\begin{eqnarray}
\Delta\rho_G(s,\Lambda)&=&\rho_G(s)+s^2\left[A_0^{(G)}+2A_1^{(G)}\ln\frac{s}{\mu^2}
+A_2^{(G)}\left(3\ln^2\frac{s}{\mu^2}-\pi^2\right)\right][1-\exp(-s/\Lambda)]\nonumber\\
& &+B_1^{(G)}\langle \alpha_s G^2\rangle[1-\exp(-s^2/\Lambda^2)],\label{co1}
\end{eqnarray}
in terms of the original one $\rho_G(s)\equiv {\rm Im}\Pi_G(s)/\pi$.
The smooth function $1-\exp(-s/\Lambda)$ vanishes like $s$ and $1-\exp(-s^2/\Lambda^2)$ 
vanishes like $s^2$ as $s\to 0$, so the subtraction terms do not modify the 
behavior of $\rho_G(s)\sim s^2$ in the low-energy limit $s\to 0$ \cite{NSVZ,SVZ80,NS81,MS81}.
Note that $s^2\ln s$ and $s^2\ln^2 s$ are regarded as being much larger than $s^2$
as $s\to 0$, so additional suppression from the first smooth function is necessary. 
Both the smooth functions approach to the unity at large $s\gg\Lambda$, where the subtracted 
spectral density $\Delta\rho(s,\Lambda)$ diminishes quickly, and the radius $R$ can be pushed 
toward the infinity. The scale in the exponent of the second smooth function can differ from 
$\Lambda$ in principle. However, we have verified that our numerical results are insensitive 
to its variation, so it is chosen as $\Lambda$ without loss of generality.
It has been also confirmed that the solutions to $\rho_G$ are basically identical, when different
smooth functions are employed in Eq.~(\ref{co1}). The subtracted spectral density 
$\Delta\rho_G(s,\Lambda)$ exhibits the resonant structure almost the same as
$\rho(s)$ does, which is barely affected by the subtraction term. It will be shown
that the introduction of a subtracted spectral density facilitates 
the evaluation of the correlation function at zero momentum.

Equation~(\ref{sg}) is then converted into 
\begin{eqnarray}
\int_{0}^\infty ds\frac{\Delta\rho_G(s,\Lambda)}{s-q^2}&=&
-\int_{0}^\infty ds \frac{s^2 e^{-s/\Lambda}}{s-q^2}\left[A_0^{(G)}+2A_1^{(G)}\ln\frac{s}{\mu^2}
+A_2^{(G)}\left(3\ln^2\frac{s}{\mu^2}-\pi^2\right)\right]\nonumber\\
& &-\int_{0}^\infty ds \frac{ e^{-s^2/\Lambda^2}}{s-q^2}B_1^{(G)}\langle \alpha_s G^2\rangle
-C_0^{(G)}\frac{\langle g G^3\rangle}{q^2}+
D_0^{(G)}\frac{\langle \alpha_s^2 G^4\rangle_G}{(q^2)^2}.\label{rg20}
\end{eqnarray}
The renormalization scale $\mu$ is usually set to the Borel mass, when the Borel 
transformation is applied to sum rules \cite{Huang:1998wj,SN98,Forkel:2003mk}.
We set $\mu^2$ to the characteristic scale $\Lambda$, 
and apply the variable changes $x= q^2/\Lambda$ and 
$y= s/\Lambda$ to Eq.~(\ref{rg20}), obtaining
\begin{eqnarray}
\int_{0}^\infty dy\frac{\Delta\rho_G(y)}{x-y}&=&
-\int_{0}^\infty dy \frac{y^2 e^{-y}}{x-y}\left[A_0^{(G)}+2A_1^{(G)}\ln y
+A_2^{(G)}\left(3\ln^2y-\pi^2\right)\right]\nonumber\\
& &-\int_{0}^\infty dy \frac{ e^{-y^2}}{x-y}B_1^{(G)}\frac{\langle \alpha_s G^2\rangle}{\Lambda^2}
+C_0^{(G)}\frac{\langle g G^3\rangle}{x\Lambda^3}-
D_0^{(G)}\frac{\langle \alpha_s^2 G^4\rangle_G}{x^2\Lambda^4},\label{rg21}
\end{eqnarray}
where the dimensionless function $\Delta\rho_G(s,\Lambda)/\Lambda^2$ has been replaced by
$\Delta\rho_G(y)$, according to the reasoning in the previous section. The scale $\Lambda$ characterizes 
the region with $y < 1$, from which the dominant nonperturbative contribution to the dispersive 
integral arises. It has been argued that a range of $\Lambda$ exists, in which a glueball 
mass $m_G$, corresponding to a peak location of $\Delta\rho_G(s,\Lambda)$, is stable against the 
variation of $\Lambda$. As $\Lambda$ becomes large enough, it diminishes the nonpertrbative 
condensate effects, and the scaling of a solution with $\Lambda$ appears. When the
scaling occurs, no structure of a solution can be interpreted as a physical state.

Since the subtracted spectral density $\Delta\rho_G(y)$ follows the  
behaviors $\Delta\rho_G(y)\sim y^2$ as $y\to 0$ and $\Delta\rho_G(y)\to 0$ as $y\to\infty$, we expand it
in terms of the generalized Laguerre polynomials $L_n^{(2)}(y)$, namely, adopt Eq.~(\ref{r1}) 
with $\alpha=2$. Equation~(\ref{ep1}) is inserted into the 
left-hand side of Eq.~(\ref{rg21}) to compute the matrix elements $M_{mn}$ in Eq.~(\ref{m2}), 
and inserted into the right-hand side of Eq.~(\ref{rg21})
to gain the coefficients $b_n$ for the input. The coefficients $b_1$ and $b_2$ of the $1/x$ and $1/x^2$
terms, respectively, receive additional contributions from the condensates.
We take the inputs of the gluon condensate $\langle \alpha_s G^2\rangle$ and of the strong coupling 
$\alpha_s$ the same as in Eq.~(\ref{put}) for consistency. 
The triple-gluon condensate is given by 
\begin{eqnarray}
\langle g G^3\rangle &=& 0.27\;{\rm GeV}^2\;\langle\alpha_s G^2\rangle,\label{31}\\
\langle g G^3\rangle &=& -1.5\langle\alpha_s G^2\rangle^{3/2},\label{32}
\end{eqnarray}
from the single-instanton estimate \cite{SVZ,NS80,RRY} and the lattice estimate \cite{PV90}, 
respectively. The value in \cite{SN10} has a positive sign
with magnitude about five times higher than in Eq.~(\ref{31}). They are different 
apparently as having been noticed in \cite{Narison:2018dcr}, and affect numerical outcomes.
We will discriminate the above estimates by the low-energy limit of the correlation function
for the scalar glueball in Eq.~(\ref{le0}). Besides, we do not consider the minor 
renormalization-group effect either. 

As emphasized before, the low-energy limit of the correlation function
can be derived from a solution to the spectral density.
Start with the dispersion relation for the correlation function $\Pi_G(q^2)$ similar to
Eq.~(\ref{di2}), and write
\begin{eqnarray}
\Pi_G(q^2)&=&\int_{0}^R \frac{ds}{s-q^2}\left\{\Delta\rho_G(s,\Lambda)
+s^2\left[A_0^{(G)}+2A_1^{(G)}\ln\frac{s}{\mu^2}
+A_2^{(G)}\left(3\ln^2\frac{s}{\mu^2}-\pi^2\right)\right]\exp(-s/\Lambda)\right.\nonumber\\
& &\left.+B_1^{(G)}\langle \alpha_s G^2\rangle \exp(-s^2/\Lambda^2)\right\}+
\frac{1}{2\pi i}\oint ds\frac{\Pi_G^{\rm perp}(s)}{s-q^2},
\end{eqnarray}
in which Eq.~(\ref{co1}) has been substituted for ${\rm Im}\Pi_G(s)/\pi$, and the pieces in
Eq.~(\ref{per}) have been grouped into the second term on the right-hand side to allow a closed contour. 
Because the dispersive integral is finite, the radius $R$ can be pushed to infinity. We then derive 
the low-energy limit
\begin{eqnarray}
\Pi_G(0)&=&\lim_{\epsilon\to 0}\Pi_G(-\epsilon\Lambda)\nonumber\\
&=&\lim_{\epsilon\to 0}\left[
\int_{0}^\infty \frac{dy}{y+\epsilon}\left\{\Lambda^2\Delta\rho_G(y)
+\Lambda^2\left[A_0^{(G)}+2A_1^{(G)}\ln y
+A_2^{(G)}\left(3\ln^2y-\pi^2\right)\right]y^2e^{-y}\right.\right.\nonumber\\
& &\left.\left.+B_1^{(G)}\langle \alpha_s G^2\rangle e^{-y^2}\right\}
+\left(B_0^{(G)}+B_1^{(G)} \ln\epsilon\right)\langle \alpha_s G^2\rangle\right],
\label{pp}
\end{eqnarray}
where the last term proportional to $\langle \alpha_s G^2\rangle$ is from the contour integral 
for $q^2=0$ and $\mu^2=\Lambda$. The constant $B_0^{(G)}$, despite of being absent in Eq.~(\ref{rg21}),
appears in the above expression. We point out that $\Delta\rho_G(y)$ corresponds exactly to the 
ultraviolet regularized spectral density with the high-frequency contribution being removed, which has been
used to define a correlation function at zero momentum in \cite{Forkel:2003mk}. Equation~(\ref{pp}) 
thus specifies explicitly how to perform the ultraviolet regularization for a dispersive integral. 



\subsection{Scalar Glueball Masses}

\begin{figure}
\includegraphics[scale=0.4]{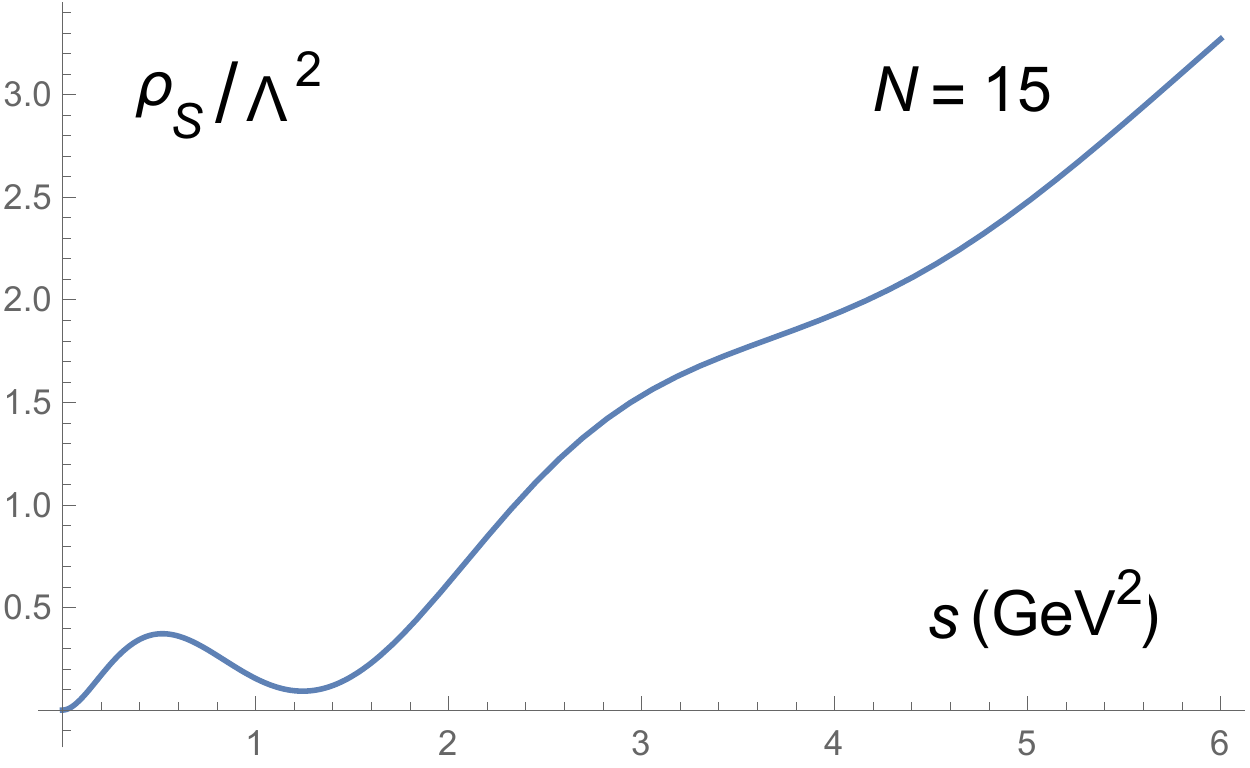}\hspace{0.5cm}
\includegraphics[scale=0.4]{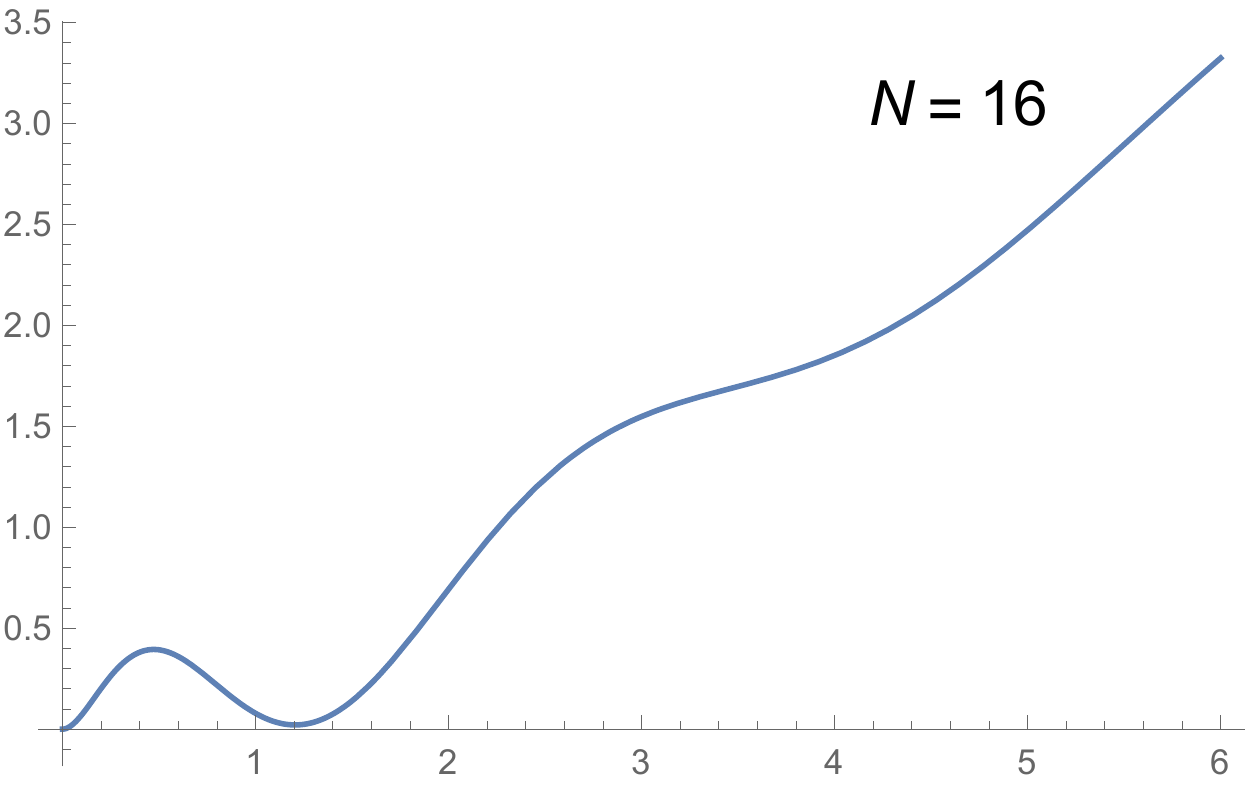}\hspace{0.5cm}
\includegraphics[scale=0.4]{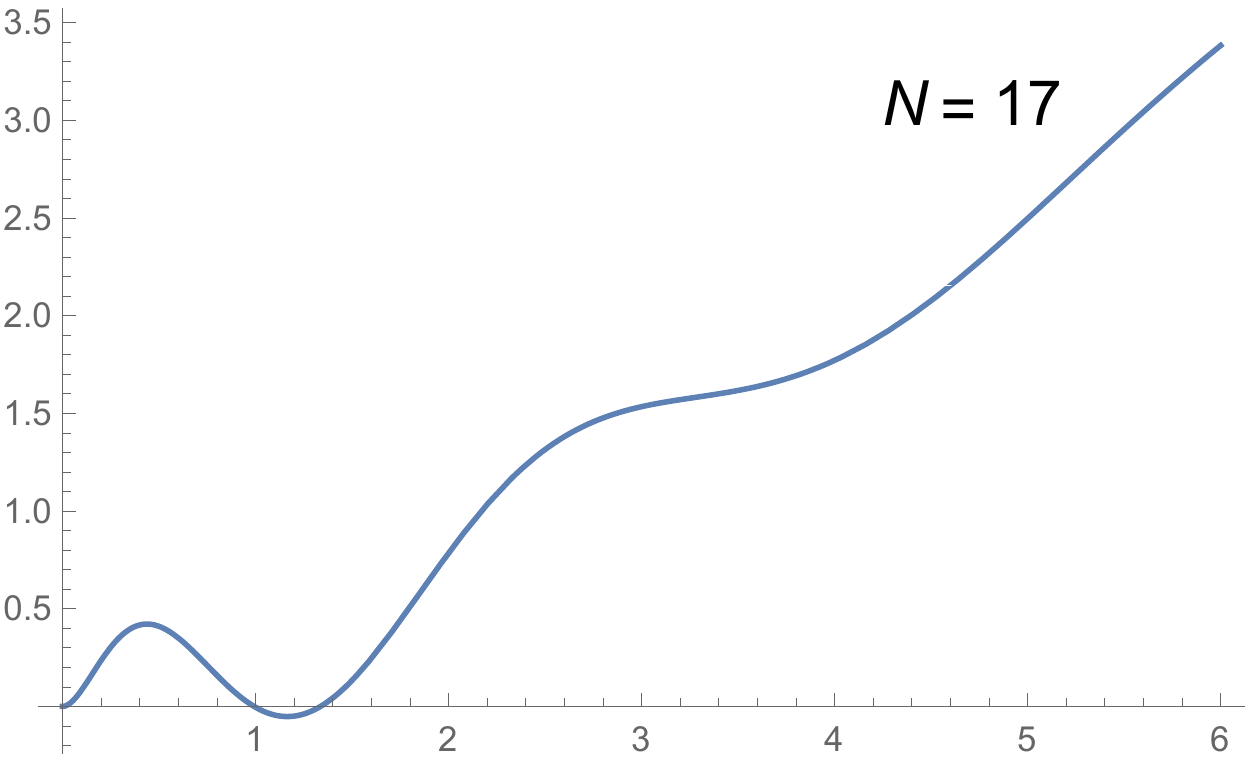}
\caption{\label{fig7}
Solutions to $\rho_S(s)/\Lambda^2$ for $\Lambda=1.5$ GeV$^2$ and the triple-gluon condensate in Eq.~(\ref{31})
with the expansions up to $N=15$, 16 and 17 generalized Laguerre polynomials $L_n^{(2)}(y)$.}
\end{figure}

\begin{figure}
\includegraphics[scale=0.4]{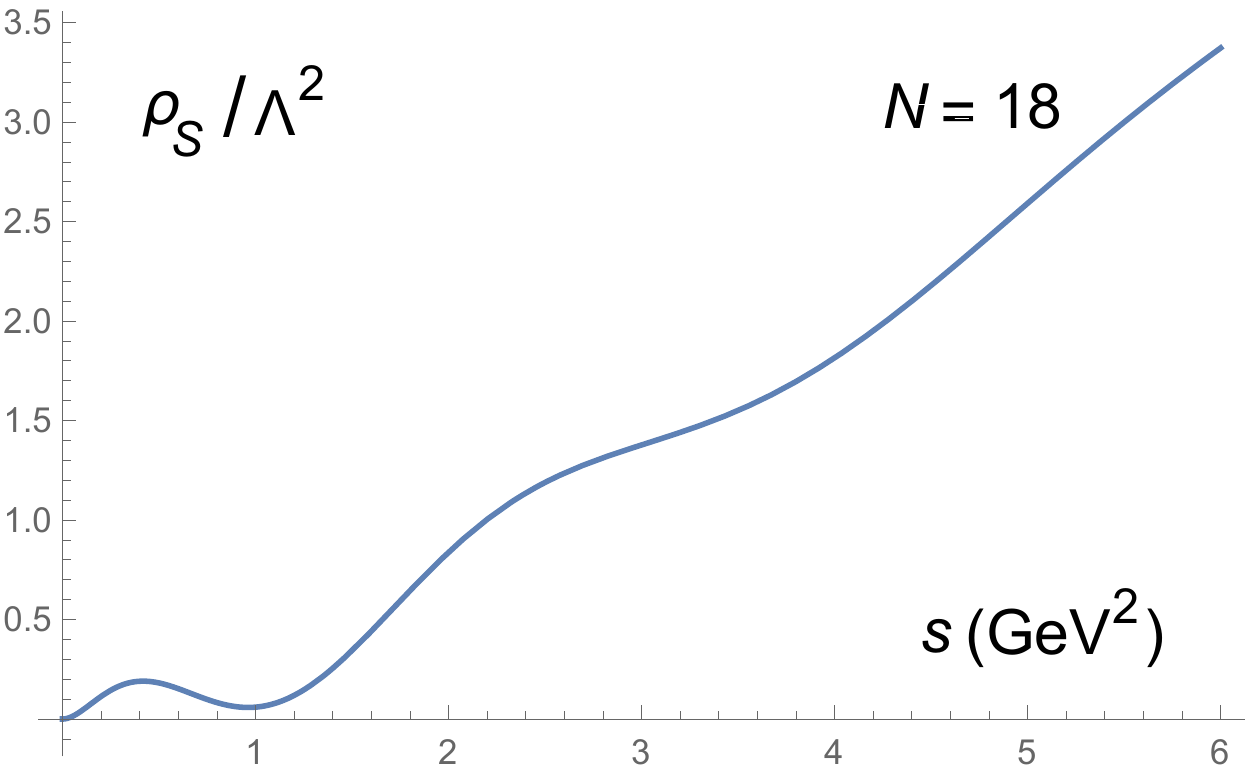}\hspace{0.5cm}
\includegraphics[scale=0.4]{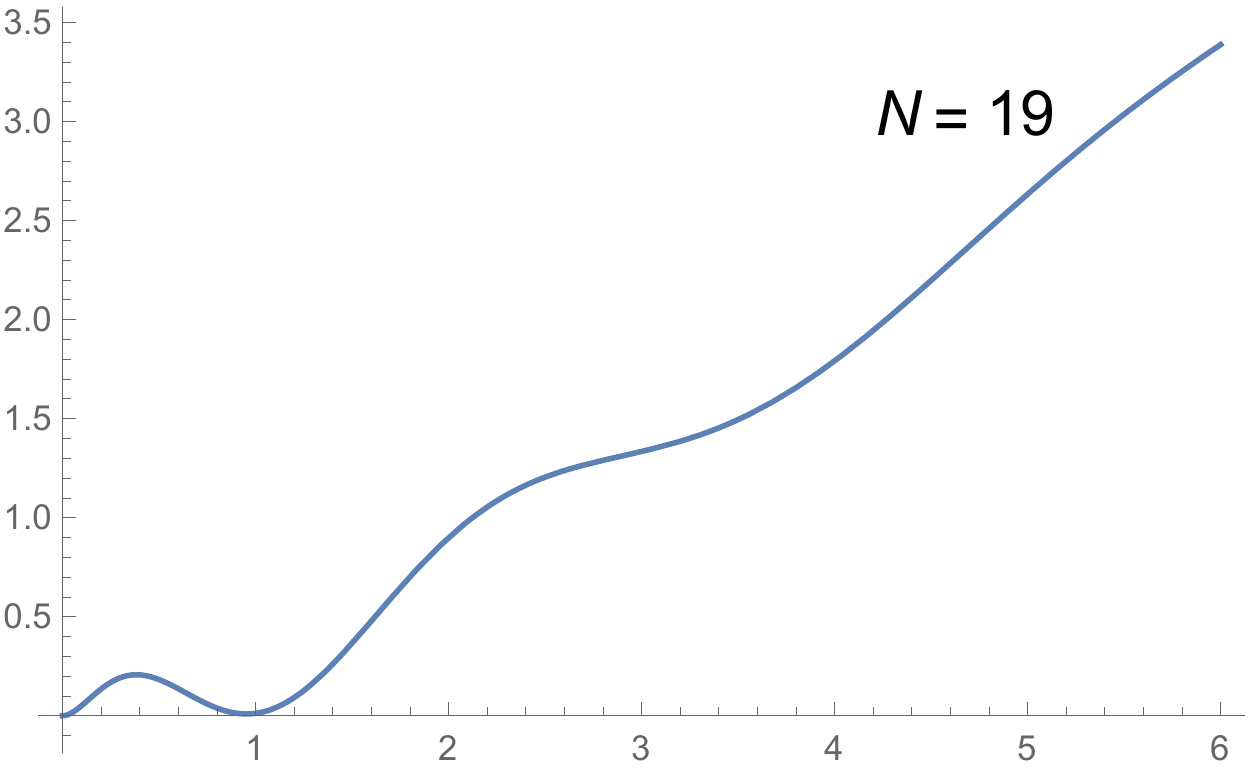}\hspace{0.5cm}
\includegraphics[scale=0.4]{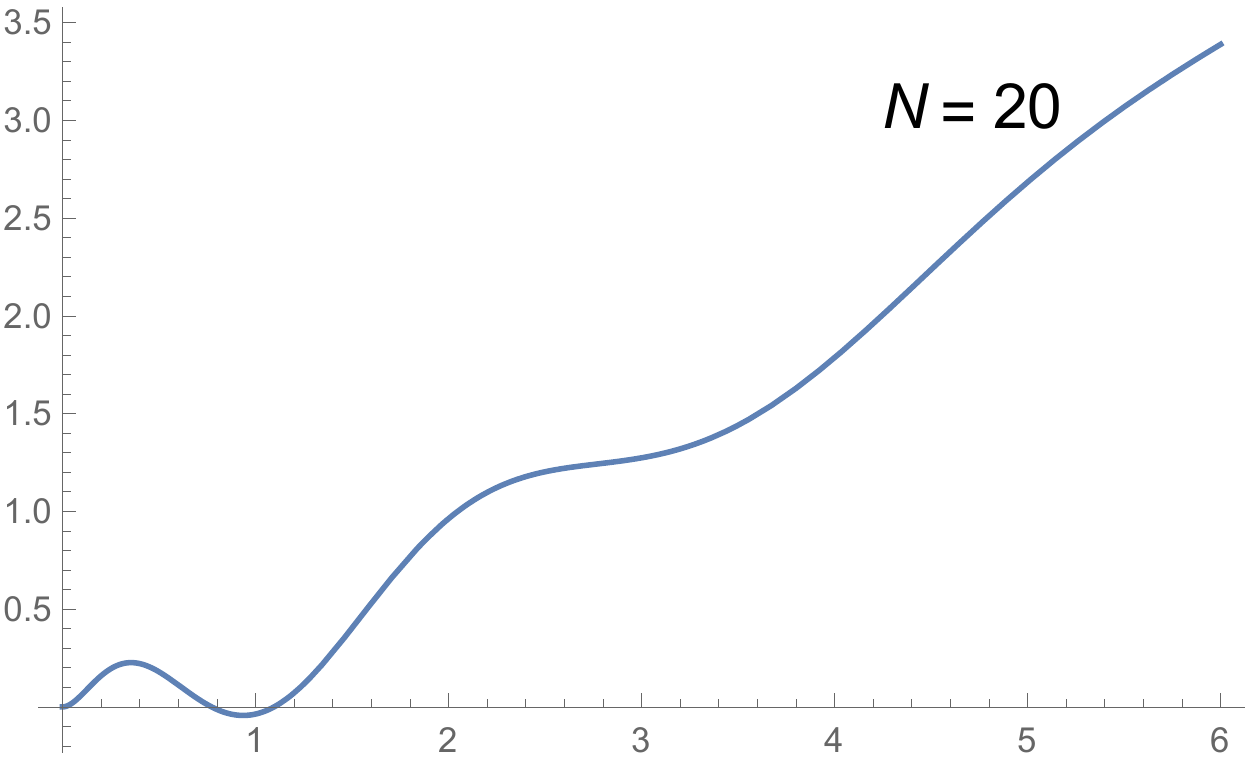}
\caption{\label{fig8}
Solutions to $\rho_S(s)/\Lambda^2$ for $\Lambda=1.5$ GeV$^2$ and the triple-gluon condensate in Eq.~(\ref{32})
with the expansions up to $N=18$, 19 and 20 generalized Laguerre polynomials $L_n^{(2)}(y)$.}
\end{figure}

We extract the scalar glueball mass by solving Eq.~(\ref{rg21}) as an inverse problem first 
with the triple-gluon condensate in Eq.~(\ref{31}),
getting the inverse matrix $M^{-1}$, the coefficients $a_n$ from the OPE coefficients $b_n$, the
solution to $\Delta \rho_S(s,\Lambda)$, and the spectral density $\rho_S(s)$ for various 
characteristic scales $\Lambda$. The results of $\rho_S(s)$ for $\Lambda=1.5$ GeV$^2$ with the 
expansions up to $N=15$, 16 and 17 generalized Laguerre polynomials $L_n^{(2)}(y)$ are displayed in 
Fig.~\ref{fig7}. The coefficients $a_n$ have not yet grown quickly, and the three curves are 
similar, implying the stability of the solutions. We find the ratios of the last two 
coefficients, $a_{15}/a_{14}\approx a_{16}/a_{15}\approx a_{17}/a_{16}\approx 1$ in the three 
cases. However, the spectral density $\rho_S(s)$, supposed to be positive, becomes negative 
around $s\approx 1.2$ GeV$^2$ for $N=17$. The positivity constraint forces us to
terminate the polynomial expansion at $N=16$. That is, the positivity of the spectral
density plays a more important role in the determination of glueball masses than in the
$\rho$ meson case. The same procedure for the triple-gluon condensate
in Eq.~(\ref{32}) yields the solutions for the characteristic scale $\Lambda=1.5$ GeV$^2$ with 
the expansions up to $N=18$, 19 and 20 generalized Laguerre polynomials $L_n^{(2)}(y)$ in Fig.~\ref{fig8}.
The curves in the three plots are also similar, and we select the one with $N=19$ 
as the solution to avoid violating the positivity constraint.


A closer look reveals that the two solutions with $N=16$ in Fig.~\ref{fig7}
and with $N=19$ in Fig.~\ref{fig8} are different: for instance, the first peak of the 
former is taller and located at larger $s\approx 0.5$ GeV$^2$, while the first peak of the 
latter is shorter and located at lower $s\approx 0.4$ GeV$^2$. It means that one has to decide
which triple-gluon condensate is adopted in order to predict glueball masses unambiguously.
The discrimination can be achieved by confronting the solutions with Eq.~(\ref{le0}) from
the low-energy theorem, namely, $\Pi_S(0)=0.78$ GeV$^4$ for the gluon condensate 
$\langle\alpha_sG^2\rangle=0.08$ GeV$^4$ in Eq.~(\ref{put}). Inserting the subtracted 
spectral densities $\Delta \rho_S(s)$ corresponding to Eqs.~(\ref{31}) and (\ref{32}) into Eq.~(\ref{pp}), 
we have $\Pi_S(0)=1.08$ and 0.60 GeV$^4$, respectively. That is, the lattice estimate for the 
triple-gluon condensate in Eq.~(\ref{32}) leads to the zero-momentum correlation function 
more consistent with the low-energy theorem. If the large triple-gluon condensate in \cite{SN10} 
is adopted, we will find $\Pi_S(0)\approx 1.7$ GeV$^4$ for $\Lambda=1.5$ GeV$^2$, and
that the first peak of the spectral density shifts to a higher $s\approx 0.6$ GeV$^2$.
The central value of $\Pi_S(0)$ resulting from the same triple-gluon condensate was also 
found to be larger than that set by the low-energy theorem recently \cite{Narison:2021xhc}. 
It has been ensured by varying the characteristic scale $\Lambda$ that the values $\Pi_S(0)$ 
from the single-instanton estimate in Eq.~(\ref{31}) and from \cite{SN10} are far above
0.78 GeV$^4$ ($\Pi_S(0)$ is not sensitive to the variation of $\Lambda$). Therefore, 
we will pick up the input in Eq.~(\ref{32}) for the rest of numerical investigations.


\begin{figure}
\includegraphics[scale=0.4]{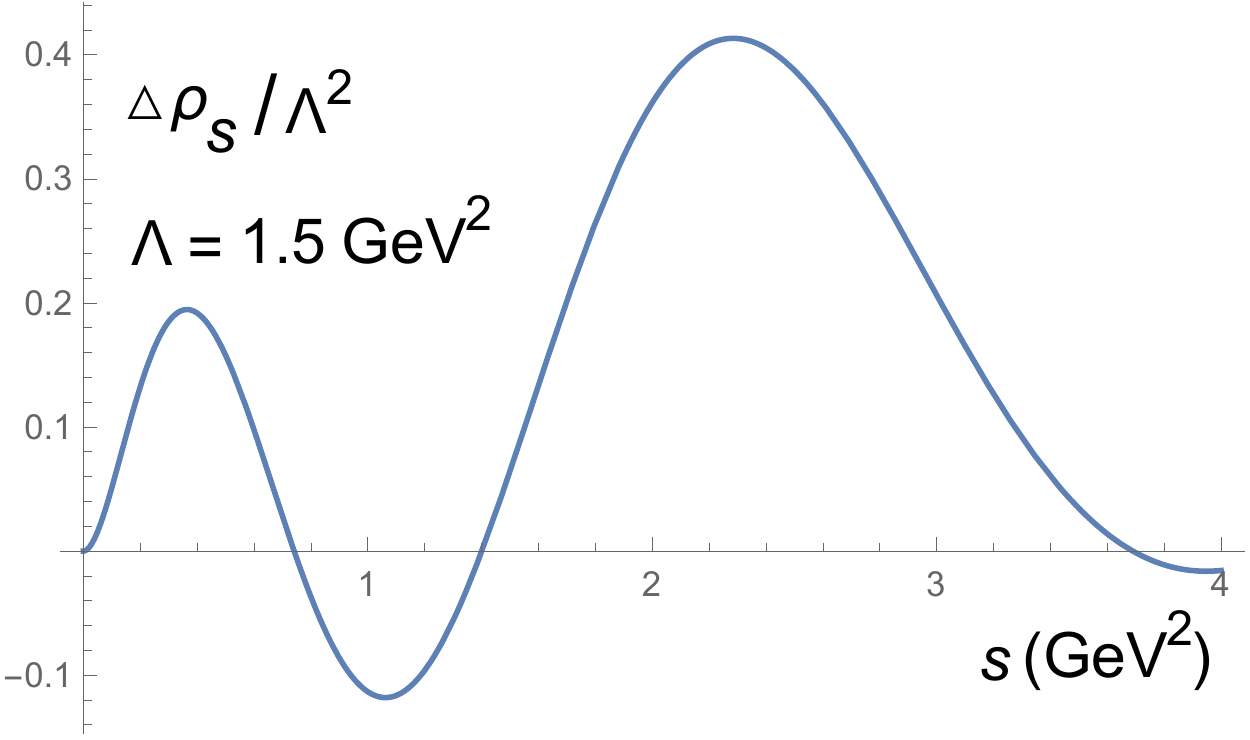}\hspace{0.5cm}
\includegraphics[scale=0.4]{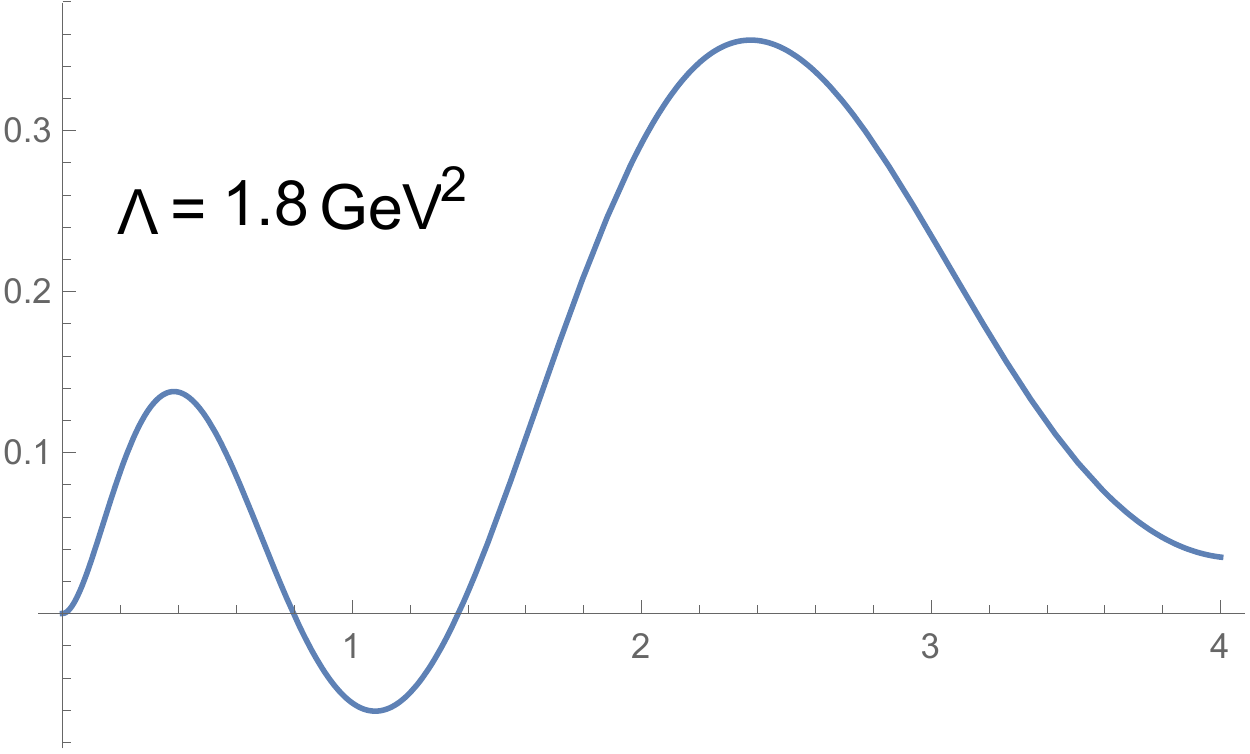}\hspace{0.5cm}
\includegraphics[scale=0.4]{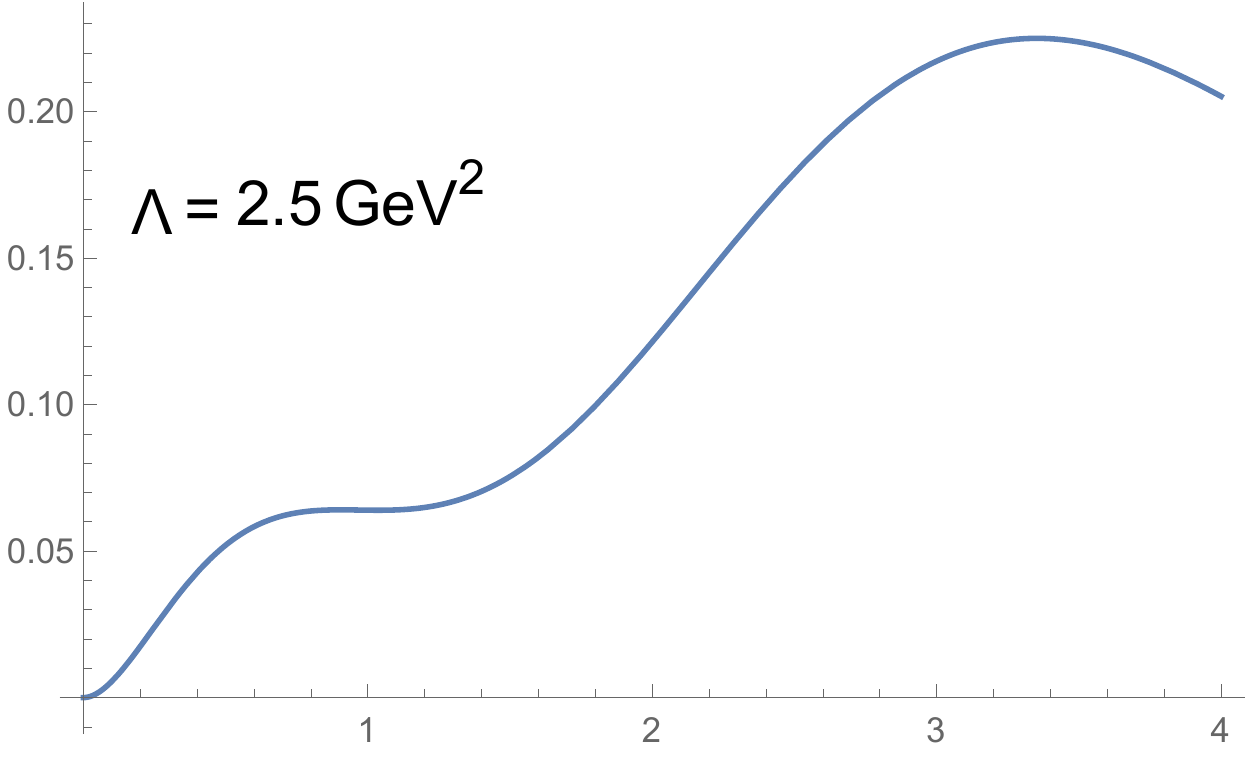}
\caption{\label{fig9}
Solutions to $\Delta\rho_S(s,\Lambda)/\Lambda^2$ for $\Lambda=1.5$, 2.0 and 2.5 GeV$^2$ 
with the expansions up to $N=19$, 22 and 22 generalized Laguerre polynomials $L_n^{(2)}(y)$, respectively.}
\end{figure}

It is interesting to notice in Fig.~\ref{fig8} that two peaks, one at $s\approx 0.4$ GeV$^2$ 
and another at $s\approx 2.3$ GeV$^2$, appear in all the three plots, though the latter seems not 
obvious due to the huge perturbative background from ${\rm Im}\Pi^{\rm perp}_S(s)/\pi$. 
To highlight these two peaks, we turn to the subtracted spectral density $\Delta \rho_S(s,\Lambda)$ 
below. The expansion in terms of the generalized Laguerre polynomials diverges more quickly at 
lower $\Lambda$, such that solutions for $\Lambda < 1$ GeV$^2$ with the polynomial numbers roughly 
smaller than 10 may not be reliable. The maximal number $N$ for the polynomial expansion then
increases with $\Lambda$ under the positivity constraint till $\Lambda=1.6$ GeV$^2$,
above which the spectral density is always positive, but the matrix elements of $M^{-1}$
and the coefficients $a_n$ grow quickly. For example, we find $a_{23}/a_{22}\approx 2.3$ for 
$\Lambda=2.5$ GeV$^2$ and $N=23$, so we stick to $N=22$ for the polynomial expansion in the range 
$\Lambda= 1.7$-2.5 GeV$^2$. The above prescribes how the maximal number of polynomials is fixed for 
an expansion. The behaviors of $\Delta \rho_S(s,\Lambda)$ for $\Lambda=1.5$, 1.8 and 2.5 GeV$^2$ are
shown in Fig.~\ref{fig9}, where the double-peak structure in the first two plots is significant. 
We stress that these two peaks are not a numerical artifact, since we have demonstrated that a 
double-peak structure can be disclosed by our method, as the resolution power, ie., the scale $\Lambda$ is 
appropriately chosen. In the present case $\Lambda\sim O(1)$ GeV$^2$ is able to resolve
the two peaks located at $m_{S_1}^2\approx 0.4$ GeV$^2$ and $m_{S_2}^2\approx 2.3$ GeV$^2$.
The similar shapes in the first two plots of Fig.~\ref{fig9} for $\Lambda=1.5$ and 1.8 GeV$^2$ 
indicate that the solutions for the scalar masses are stable in an interval of $\Lambda$. 
Note that the smooth functions in the subtraction term in Eq.~(\ref{co1}) are close to the unity as $s>4$ 
GeV$^2$ for the considered $\Lambda=1.5$ or 1.8 GeV$^2$. Therefore, the small deviation from zero 
above $s\approx 4$ GeV$^2$ signals the mild violation of the local 
quark-hadron duality of the spectral density $\rho_S(s)$ in the high $s$ region.  It is apparent that the 
third plot in Fig.~\ref{fig9} labelled by $\Lambda=2.5$ GeV$^2$ differs much from the first two: 
the peaks move toward larger $s$, manifesting the scaling behavior of $\Delta \rho_S(s,\Lambda)$ at 
large $\Lambda$ ascribed to the disappearance of the nonperturbative condensate effects. 
The double-peak structure also blurs, since the resolution of our method becomes worse at this 
large $\Lambda$.


We remark that the matrix element $M_{mn}$ in Eq.~(\ref{m2}) corresponds to 
the $m$-th moment of a sum rule for a glueball mass. As postulated in \cite{SN98}, the lower moments of
a sum rule are more sensitive to low-lying resonances, while higher moments are, on the
contrary, more sensitive to heavy resonances: the light masses about 700-900 MeV were 
extracted from the lower moments of the sum rule in \cite{PTN}, and the heavy ones 
about 1.5-1.7 GeV were extracted from the higher moments in \cite{SN98,BS90,Huang:1998wj}.
It is then easy to understand why our formalism, which takes into account more moments than in conventional
sum rules \cite{SN98,BLR,Krasnikov:1982ea}, produces the two peaks simultaneously with the
similar squared masses $m_{S_1}^2\approx 0.4$ GeV$^2$ and $m_{S_2}^2\approx 2.3$ GeV$^2$. 
The double-peak structure in Fig.~\ref{fig9} is consistent with the observation in the sum-rule 
analysis \cite{Harnett:2000fy}, where a double-resonance parametrization for the spectral density was 
shown to give a fit to the OPE side better than a single-resonance one. The mass range 0.8-1.6 GeV for the 
peak locations in \cite{Harnett:2000fy} is similar to ours, but the mass gap between the two resonances, 
about few hundreds of MeV in \cite{Harnett:2000fy}, is smaller than in our solution, which is about 1 GeV. 
Another distinction is that the lighter resonance has a broader width in \cite{Harnett:2000fy}, while 
the heavier one does in our solution, which may be due to the different 
nonperturbative effects included in the theoretical frameworks: they come from the instanton contribution
to the correlation function in \cite{Harnett:2000fy}, but from the gluon condensates in ours.
We mention that the scalar glueball mass around 1.5 GeV, very close to our $m_{S_2}$, 
was derived from sum rules \cite{Wen:2010qoe,Forkel:2000fd} with only the 
instanton effects. It supports our claim that the gluon condensates are sufficient for 
establishing the glueballs.



\begin{figure}
\includegraphics[scale=0.5]{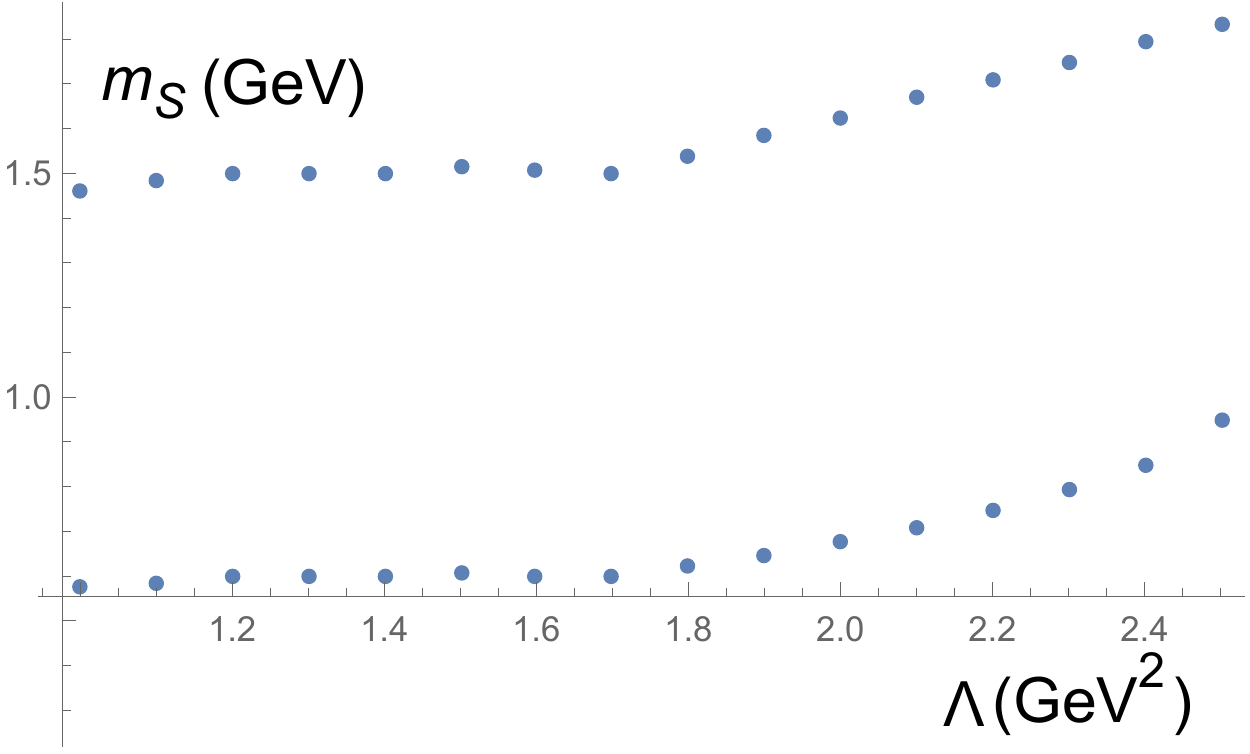}
\caption{\label{fig10}
Dependence of the scalar masses $m_{S_1}$ and $m_{S_2}$ on the characteristic scale $\Lambda$.}
\end{figure}

We scan the scalar masses $m_{S_1}$ and $m_{S_2}$ in the range 1 GeV$^2<\Lambda <2.5$ GeV$^2$, and depict their
dependencies on $\Lambda$ in Fig.~\ref{fig10}, where the two curves describe the two peak locations of 
the subtracted spectral density $\Delta \rho_S(s,\Lambda)$. It is seen that 
the lower and upper curves ascend first from $\Lambda=1$ GeV$^2$, reaching $m_{S_1}=0.60$ GeV and
$m_{S_2}=1.50$ GeV, respectively, become stable in the window 1.2 GeV$^2<\Lambda<1.7$ GeV$^2$, and then go
up again monotonically. These features, having appeared in Fig.~\ref{fig6} for the $\rho$ meson mass, are 
completely anticipated. We then survey the dependence of the correlation function $\Pi_S(0)$ at zero 
momentum on $\Lambda$ according to Eq.~(\ref{pp}), given the solutions for $\Delta\rho_S(s,\Lambda)$, and 
obtain $\Pi_S(0)=0.65$ GeV$^4$ at $\Lambda=1.7$ GeV$^2$. Considering the typical 20\% uncertainty 
around $\Pi_S(0)=0.78$ GeV$^4$ from Eq.~(\ref{le}), we are sure that the low-energy limit is satisfactorily 
respected by our solutions in the stability window. It is legitimate to choose $m_{S_1}=0.60$ 
GeV and $m_{S_2}=1.50$ GeV as the central values, and to estimate the theoretical errors in our method using 
the minimal (maximal) values at $\Lambda=1.4$ (1.5) GeV$^2$. We get $m_{S_1}=(0.60\pm 0.01)$ GeV
and $m_{S_2}=(1.50\pm 0.01)$ GeV, whose tiny errors reflect the remarkable stability of our solutions. 
Because the subtracted spectral densities for other values of $\Lambda$ in the interval 
1.2 GeV$^2<\Lambda<1.7$ GeV$^2$ are very similar to that for $\Lambda=1.5$ GeV$^2$ in Fig.~\ref{fig9},
we do not present them here.


We investigate the theoretical uncertainties arising from the variation of the involved parameters,
which all turn out to be under control. Decreasing the strong coupling $\alpha_s$ from 0.5 to 0.4, which 
corresponds to the scale variation within the stability window roughly, 
we find 8\% enhancement on the determined scalar masses. This check justifies the neglect of the
renormalzation-group effect in our calculation. The typical $\pm 20\%$ change of the gluon condensate 
$\langle\alpha_s G^2\rangle$ causes about $\pm 5\%$ impact. We also examine the sensitivity of the scalar 
masses to choices of the renormalization scale $\mu$, and observe 2\% increase from $\mu^2=2\Lambda$
and $\mu^2=\Lambda/2$. Adding the above sources of errors in quadrature, we conclude
\begin{eqnarray}
m_{S_1}=(0.60\pm 0.06)\;{\rm GeV},\;\;\;\; 
m_{S_2}=(1.50\pm 0.15)\;{\rm GeV}.\label{ms2}
\end{eqnarray}


We interpret the solutions to the spectral density for the scalar glueball, bearing in mind 
that any physical state with a gluonic content can contribute to the considered spectral density. 
The major peak of the subtracted spectral density located at $m_{S_2}= 1.50$ GeV  points to the 
$f_0(1500)$ meson \cite{Close}. Since $f_0(1500)$ has a narrower width 112 MeV \cite{PDG}, 
it cannot accommodate the broad width shown in Fig.~\ref{fig9} alone. Hence, it is likely 
that $f_0(1370)$ and $f_0(1710)$ also have gluonic contents and contribute to the spectral
density, in accordance with the prevailing consensus in the literature \cite{Cheng:2015iaa}. 
That is, all $f_0(1370)$, $f_0(1500)$ and $f_0(1710)$ are glue-rich states (nevertheless,  
a recent phenomenological analysis on $J/\psi$ radiative decays
prefers a higher scalar glueball mass resulting from the mixing with heavier scalar states 
\cite{Sarantsev:2021ein,Klempt:2021ope}).
The shorter peak located at $m_{S_1}\approx 0.60$ GeV might arise from the 
contribution of the $f_0(500)$ meson with a broad width. The smaller area under the peak 
implies a lower $f_0(500)$ decay constant defined via the gluon field, and 
less gluonic content in this light scalar, consistent with the observation in
\cite{Vento:2004xx}. This implication does not depend on the quark structure of the $f_0(500)$ 
meson \cite{Achasov:2020aun}. Note that a little amount of gluonium in the $f_0(980)$ meson
with a narrow width 10 MeV cannot be excluded, as hinted by the spectral densities in Fig.~\ref{fig8}. 
The above interpretation agrees with the conclusion drawn based on a five-state-mixing scenario in 
a nonlinear chiral Lagrangian framework \cite{Fariborz:2006xq,Fariborz:2003uj} and with the analysis 
in \cite{SN98}. The mass gaps between 
$f_0(500)$ and $f_0(980)$, and among $f_0(1370)$, $f_0(1500)$ and $f_0(1710)$ are too small to be 
resolved by our method with the characteristic scale $\Lambda\sim O(1)$ GeV$^2$, so only two peaks are 
revealed in the spectral density.



\subsection{Pseudoscalar Glueball Masses}

\begin{figure}
\includegraphics[scale=0.45]{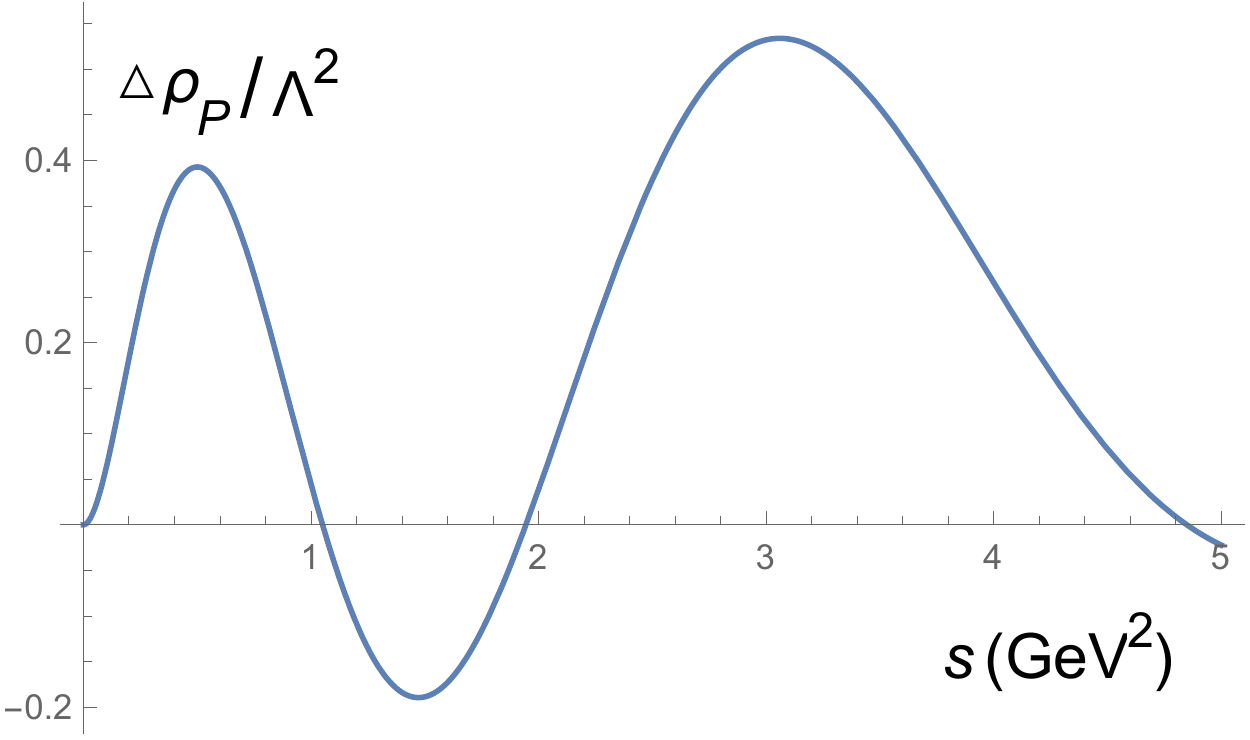}
\caption{\label{fig11}
Solution to $\Delta\rho_P(s,\Lambda)/\Lambda^2$ for $\Lambda=1.5$ GeV$^2$ 
with the expansion up to 14 generalized Laguerre polynomials $L_n^{(2)}(y)$.}
\end{figure}

Next we extract the pseudoscalar glueball mass by solving Eq.~(\ref{rg21}) from the corresponding
set of OPE inputs in Eq.~(\ref{ap}) with the triple-gluon condensate in Eq.~(\ref{32}). 
The prescription for fixing the number $N$ of polynomials 
in an expansion is the same, and the plots for $\rho_P(s)$ are similar to
those in Fig.~\ref{fig8}. The behavior of the subtracted spectral density $\Delta \rho_P(s,\Lambda)$
for $\Lambda=1.5$ GeV$^2$ with the expansion up to $N=14$ generalized Laguerre polynomials 
$L_n^{(2)}(y)$ is exhibited in Fig.~\ref{fig11}, from which we read the pseudoscalar 
masses. The coefficients $a_n$ in this case have not yet grown quickly, but the positivity 
constraint forces us to terminate the polynomial expansion at $N=14$. The maximal number $N$ 
increases with $\Lambda$ under the positivity constraint till $\Lambda=2.3$ GeV$^2$, above which 
the spectral density is always positive, but the matrix elements of $M^{-1}$ and the coefficients 
$a_n$ go out of control as $N>22$. Therefore, we stick to $N=22$ for the polynomial expansion 
in the range $\Lambda= 2.3$-3.5 GeV$^2$. Likewise, we observe the clear double-peak structure
in Fig.~\ref{fig11} with the locations $m_{P_1}^2\approx 0.5$ GeV$^2$ and $m_{P_2}^2=3.1$ GeV$^2$,
which are above the scalar ones $m_{S_1}^2\approx 0.4$ GeV$^2$ and $m_{S_2}^2\approx 2.3$
GeV$^2$ in Fig.~\ref{fig9}, respectively. 



\begin{figure}
\includegraphics[scale=0.5]{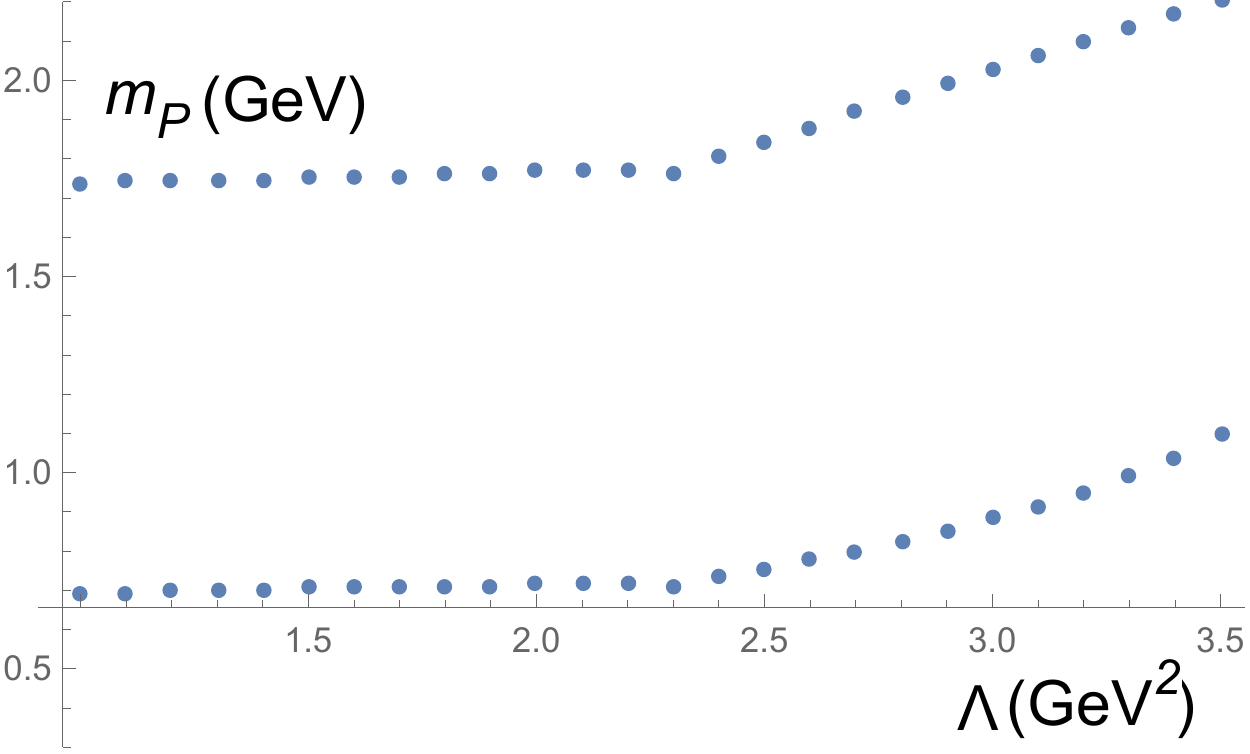}
\caption{\label{fig12}
Dependence of the pseudoscalar masses $m_{P_1}$ and $m_{P_2}$ on the characteristic scale $\Lambda$.}
\end{figure}

We scan the pseudoscalar masses $m_{P_1}$ and $m_{P_2}$ in the range 1 GeV$^2<\Lambda <3.5$ GeV$^2$, and 
display their dependencies on $\Lambda$ in Fig.~\ref{fig12}, where the two curves describe the two peak 
locations of the subtracted spectral density $\Delta\rho_P(s,\Lambda)$. It is found that the lower and 
upper curves almost remain flat in the interval 1.0 GeV$^2<\Lambda<2.3$ GeV$^2$, around $m_{P_1}=0.71$ GeV 
and $m_{P_2}=1.75$ GeV, and then go up monotonically as expected. The stability window is wider than in
the scalar glueball case illustrated in Fig.~\ref{fig10}. We estimate the theoretical errors in our method 
using the extreme values in the window, and get $m_{P_1}=(0.71\pm 0.02)$ GeV and 
$m_{P_2}=(1.75\pm 0.02)$ GeV, whose tiny errors reflect the remarkable stability of our solutions. 
We investigate the theoretical uncertainties from the variation of the involved parameters
in a similar manner. The decrease of the strong coupling $\alpha_s$ from 0.5 to 0.4 yields
only 7\% enhancement on the determined pseudoscalar masses, which justifies the neglect of the
renormalzation-group evolution in our analysis. The typical $\pm 20\%$ change of the gluon condensate 
$\langle\alpha_s G^2\rangle$ causes about $\pm 5\%$ effect. The choice for the renormalization scale 
$\mu^2=2\Lambda$ decreases the pseudoscalar masses by 3\%, and the choice $\mu^2=\Lambda/2$
does not alter the outcomes. Adding the above sources of errors in quadrature, we conclude
\begin{eqnarray}
m_{P_1}=(0.71\pm 0.07)\;{\rm GeV},\;\;\;\; 
m_{P_2}=(1.75\pm 0.16)\;{\rm GeV}.\label{mp2}
\end{eqnarray}
That is, the theoretical uncertainties are under control in our formalism.



The major peak of the subtracted spectral density located at $m_{P_2}\approx 1.75$ GeV may
point to the $\eta(1760)$ meson, whose broad width about 240 MeV \cite{PDG} can
accommodate the width in Fig.~\ref{fig12} alone. This mass is just a bit
higher than the scalar glueball one determined in the previous subsection, as anticipated 
from the minor difference between their OPE inputs and argued in \cite{Faddeev:2003aw}. 
Our result is lower than 
most predictions in the literature, which are above 2 GeV, as stated in the introduction. We 
stress again that measurements of $J/\psi$ radiative decays do not confirm any glue-rich pseudoscalar 
resonances with masses greater than 2 GeV \cite{PDG}. The branching ratio of the $J/\psi\to\gamma X(2370)$ 
decay is so small \cite{PDG}, that $X(2370)$ is unlikely to be the pseudoscalar glueball,
even if it carried the correct quantum numbers. On the contrary, the $\eta(1760)$ 
meson is abundantly produced in $J/\psi$ radiative decays, but not seen in the 
$J/\psi\to \gamma\gamma V$ decays \cite{MARK}. Another nearby pseudoscalar $X(1835)$
is produced less abundantly in $J/\psi$ radiative decays, and seen in the 
$J/\psi\to \gamma\gamma \phi$ decay \cite{BESIII:2018dim}. It should be reminded that
the relevant experimental studies are not yet conclusive. For instance,
it is puzzling that the $J/\psi \to\gamma(\eta(1760)\to) \omega\omega$ branching ratio,
despite of the stronger phase space suppression, is about one order of magnitude larger than 
the $J/\psi \to\gamma(\eta(1760)\to) \rho^0\rho^0$ one \cite{PDG}. Nevertheless,
the currently available data do support that $\eta(1760)$ is a glueball. 

Our solution disfavors the speculation, deduced from a pseudoscalar meson mixing 
formalism based on the anomalous Ward identity \cite{Cheng:2008ss,He:2009sb,Tsai:2011dp}, 
that the $\eta(1405)$ is the lightest pseudoscalar glubeball \cite{MCU}. It should be pointed
out, however, that a pseudoscalar glueball mass as heavy as 1.75 GeV is not excluded 
in \cite{Cheng:2008ss}, when some inputs are allowed to vary. A similar mixing formalism with 
more conservative assumption \cite{Qin:2017qes} shows that the pseudoscalar glueball tends to 
be heavier. Our prediction 1.75 GeV matches their results with a large angle $\phi_G$ for the 
mixing between the pure glueball and the flavor-singlet light quark states. It is fair to allege, 
based on the theoretical uncertainties, that the $\eta(1405/1475)$ 
mesons \footnote{The experimental issues concerning the line shapes and mass shifts between 
$\eta(1405)$ and $\eta(1475)$ were addressed in \cite{Wu:2011yx,Wu:2012pg,Du:2019idk}, where 
the key triangle singularity mechanism was proposed.}  
contain some gluonium components, which, though, are not dominant as found in \cite{SN98}. 
Indeed, the $\eta(1405/1475)$ mesons are produced copiously in $J/\psi$ radiative decays, 
and also seen in the $J/\psi\to \gamma\gamma \rho$ decay \cite{PDG}.

The shorter peak located at $m_{P_1}\approx 0.71$ GeV between the $\eta$ and 
$\eta'$ meson masses comes from the combined contributions of these two states with  
similar weights. A low-lying state with mass around 1 GeV has been also identified and assigned to 
the $\eta'$ meson in the lattice calculation \cite{Sun:2017ipk}, when the topological charge density 
with a strong coupling to flavor-singlet light quark states, the same as in the present 
work, is employed to define the correlation function.
This observation is in accordance with the implementation of the $\eta$-$\eta'$-glueball mixing,
which has been intensively discussed in the literature. The comparison of the relative heights between 
the two peaks in Figs.~\ref{fig9} and \ref{fig11} suggests that $\eta$ and $\eta'$ have more gluonic 
content than $f_0(500)$ and $f_0(980)$ do. Our solutions hint that a more complete mixing
scenario involving $\eta$, $\eta'$, $\eta(1405/1475)$ and $\eta(1760)$ is needed for a
deeper understanding of the pseudoscalar glueball properties. We will examine whether more 
quantitative information on the scalar and pseudoscalar mixings can be drawn from our formalism 
in the future. For the same reason, the mass gap between $\eta$ and $\eta'$ is too 
small to be resolved by our method with the characteristic scale $\Lambda\sim O(1)$ GeV$^2$, although they
have quite narrow widths.

Once the spectral density $\Delta\rho_P(s,\Lambda)$ is ready, we evaluate 
the correlation function $\Pi_P(0)$ at zero momentum in the stability window of $\Lambda$ 
following Eq.~(\ref{pp}), and deduce the topological susceptability from 
Eq.~(\ref{le}). It is seen that the topological susceptability 
increases from $\chi^{1/4}_t=75$ MeV at $\Lambda=1.0$ GeV$^2$ slightly,  
reaches $\chi^{1/4}_t=78$ MeV at $\Lambda= 1.5$ GeV$^2$, and then saturates. 
Namely, the topological susceptability is predicted to be
in the range $\chi^{1/4}_t=75$-78 MeV, which is almost independent of the scale $\Lambda$.
This prediction is compatible with the results $\chi^{1/4}_t=66$-120 MeV from 
recent lattice QCD evaluations involving at least two light flavors of fermions
\cite{Aoki:2017paw,Alexandrou:2017bzk,Bonati:2015vqz,Dimopoulos:2018xkm,Chiu:2020ppa,Bhattacharya:2021lol},
and $\chi^{1/4}\approx 75$ MeV in chiral perturbatin theory 
\cite{Luciano:2018pbj,GrillidiCortona:2015jxo,Gorghetto:2018ocs}.

\subsection{Tensor Glueball Mass}

We extend the above formalism to the study of the tensor glueball  $(2^{++})$, 
for which the correlation function is defined as
\begin{eqnarray}
\Pi_{\mu\nu\rho\sigma}(q^2)&\equiv&i\int d^4xe^{iq\cdot x}\langle 0|T\Theta_{\mu\nu}(x)\Theta_{\rho\sigma}(0)|0\rangle,\nonumber\\
&=&\frac{1}{2}\left(\eta_{\mu\rho}\eta_{\nu\sigma}+\eta_{\mu\sigma}\eta_{\nu\rho}-\frac{2}{3}\eta_{\mu\nu}\eta_{\rho\sigma}
\right)\Pi_T(q^2),
\label{ct}
\end{eqnarray}
with $\eta_{\mu\nu}=g_{\mu\nu}-q_\mu q_\nu/q^2$, and
\begin{eqnarray}
\Theta_{\mu\nu}(x)=-\alpha_sG_{\mu}^{a\alpha} G^a_{\nu\alpha}
+\frac{g_{\mu\nu}}{4}\alpha_s G_{\alpha\beta}^a G^{a\alpha\beta},
\end{eqnarray}
being the energy-momentum stress tensor of QCD. The OPE of the correlation function 
$\Pi_T(q^2)$ in the deep Euclidean region of $q^2$ is given by \cite{NSVZ}
\begin{eqnarray}
\Pi_T^{\rm OPE}(q^2)=-\frac{1}{20}\left(\frac{\alpha_s}{\pi}\right)^2q^4\ln\frac{-q^2}{\mu^2}
+\frac{5}{3}\pi\alpha_s\frac{\langle \alpha_s^2G^4\rangle_T}{(q^2)^2},\label{dt3}
\end{eqnarray}
up to the dimension-eight condensate, ie., to the power correction of $1/(q^2)^2$, which 
can be approximated under the vacuum factorization assumption by
\begin{eqnarray}
\langle \alpha_s^2G^4\rangle_T&=& 
2\langle(\alpha_sf^{abc} G^b_{\mu\rho}G^{c\rho}_\nu)^2\rangle
-\langle(\alpha_sf^{abc} G^b_{\mu\nu}G^c_{\rho\lambda})^2\rangle,\nonumber\\
&\approx &-\frac{3}{16}\langle \alpha_sG^2\rangle^2.
\end{eqnarray}

\begin{figure}
\includegraphics[scale=0.45]{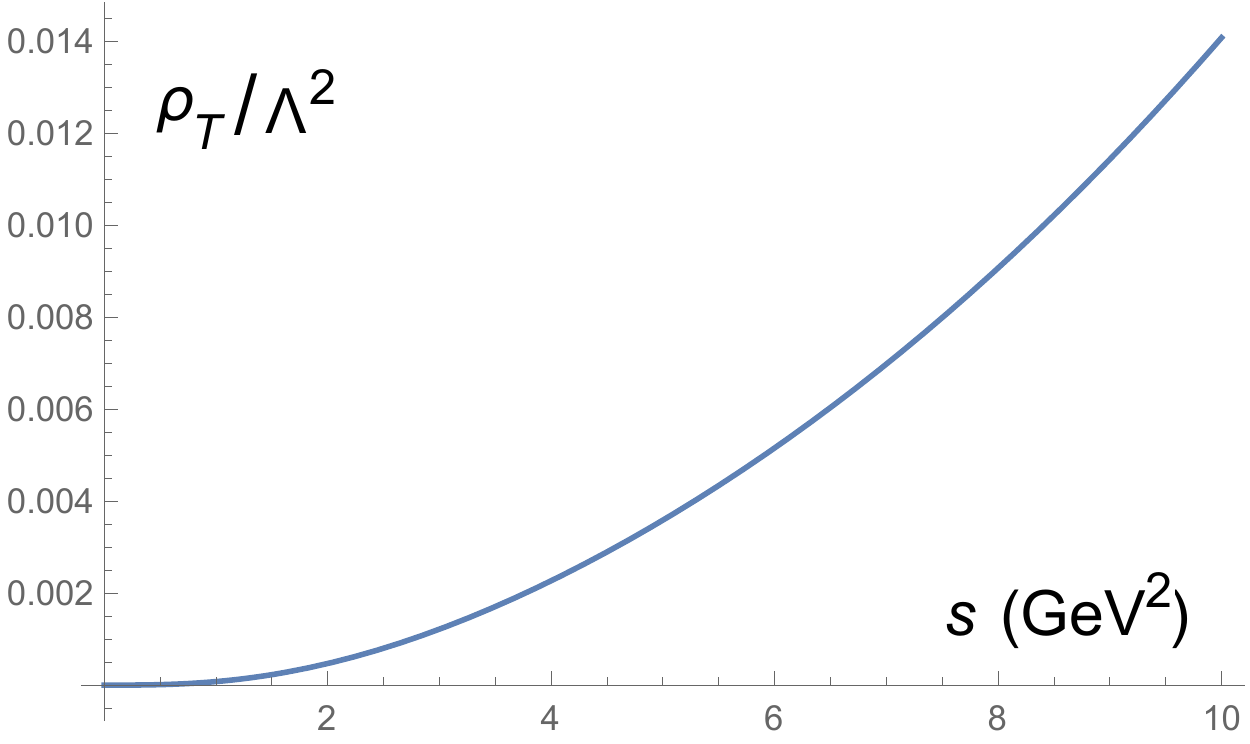}\hspace{1.0cm}
\includegraphics[scale=0.45]{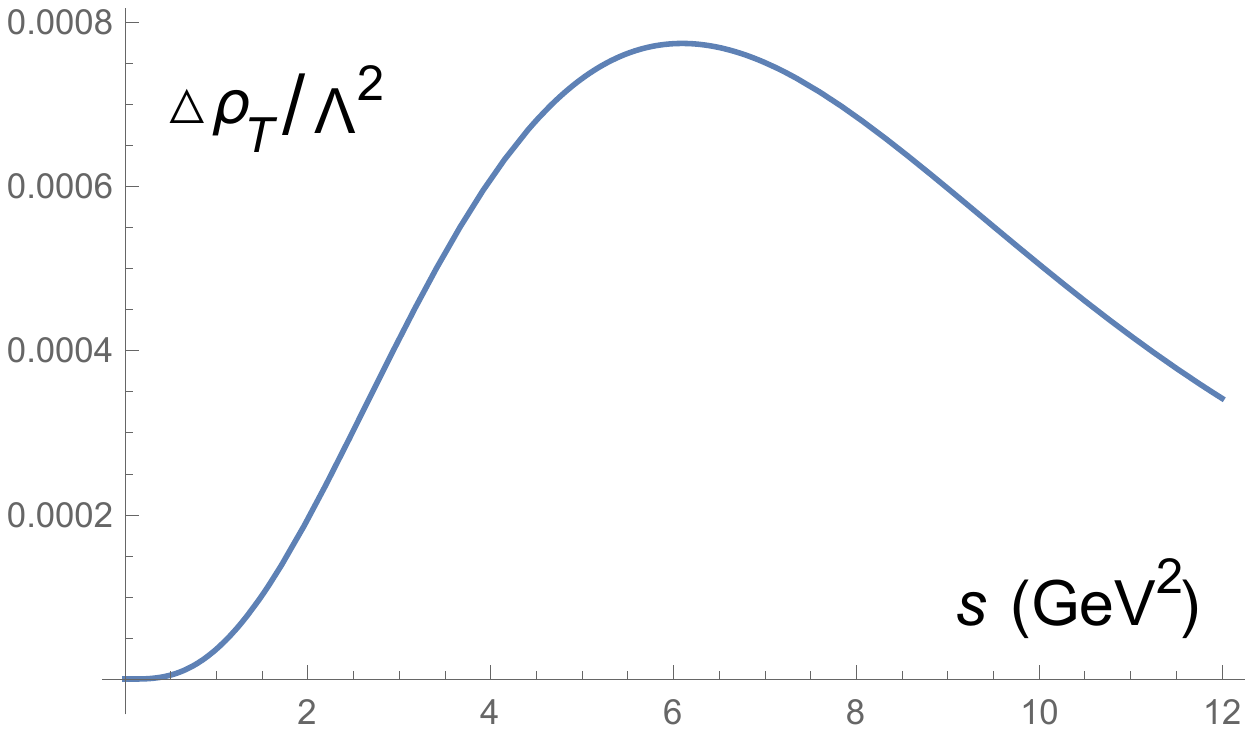}
\caption{\label{fig13}
Solutions to $\rho_T(s)/\Lambda^2$ and $\Delta\rho_T(s,\Lambda)/\Lambda^2$ for $\Lambda=3.0$ GeV$^2$
with the expansion up to $6$ generalized Laguerre polynomials $L_n^{(2)}(y)$.}
\end{figure}

Repeating the same steps, we notice that the tensor glueball case differs much from
those of the scalar and pseudoscalar glueballs. The positivity requirement on the spectral density 
forces us to terminate the expansion in terms of the generalized Laguerre polynomials $L_n^{(2)}(y)$
at very low $N$: the maximal $N$'s are found to be 3, 4, 6 and 8 for 
$\Lambda=1.5$, 2.0, 3.0 and 4.0 GeV$^2$, respectively. The maximal number $N$ for a
polynomial expansion increases with $\Lambda$ under the positivity constraint, but it  
reaches only $N=10$ even when $\Lambda$ is as high as 5.0 GeV$^2$. Hence, we accept the expansions
up to fewer polynomials in the analysis of the tensor glueball mass, and obtain the  
spectral density $\rho_T(s)$ and the subtracted spectral density $\Delta\rho_T(s,\Lambda)$ for 
$\Lambda=3.0$ GeV$^2$ and $N=6$ given in Fig.~\ref{fig13}. The coefficients $a_n$ in the
expansion have not increased rapidly, with $a_6/a_5\approx 1.2$. If $N=7$ is chosen, 
$\rho_T(s)$ will become negative and violate the positivity constraint in the range  $0<s<0.7$ GeV$^2$.
No evident bump like those in Fig.~\ref{fig7} appears in the curve for $\rho_T(s)$, which
always ascends in the range 1.0 GeV$^2<\Lambda<5.0$ GeV$^2$. After the huge perturbative background 
is removed, the subtracted spectral density $\Delta\rho_T(s,\Lambda)$ exhibits a single peak 
with diminishing height (note the scale on the vertical axis in Fig.~\ref{fig13}).
The single-peak structure of $\Delta\rho_T(s,\Lambda)$ persists till $\Lambda=5.0$ GeV$^2$, 
and becomes broader with $\Lambda$. It has been underlined that a peak of the
spectral density cannot be interpreted as a physical state, unless a stability window in $\Lambda$
for its location exists. 


\begin{figure}
\includegraphics[scale=0.5]{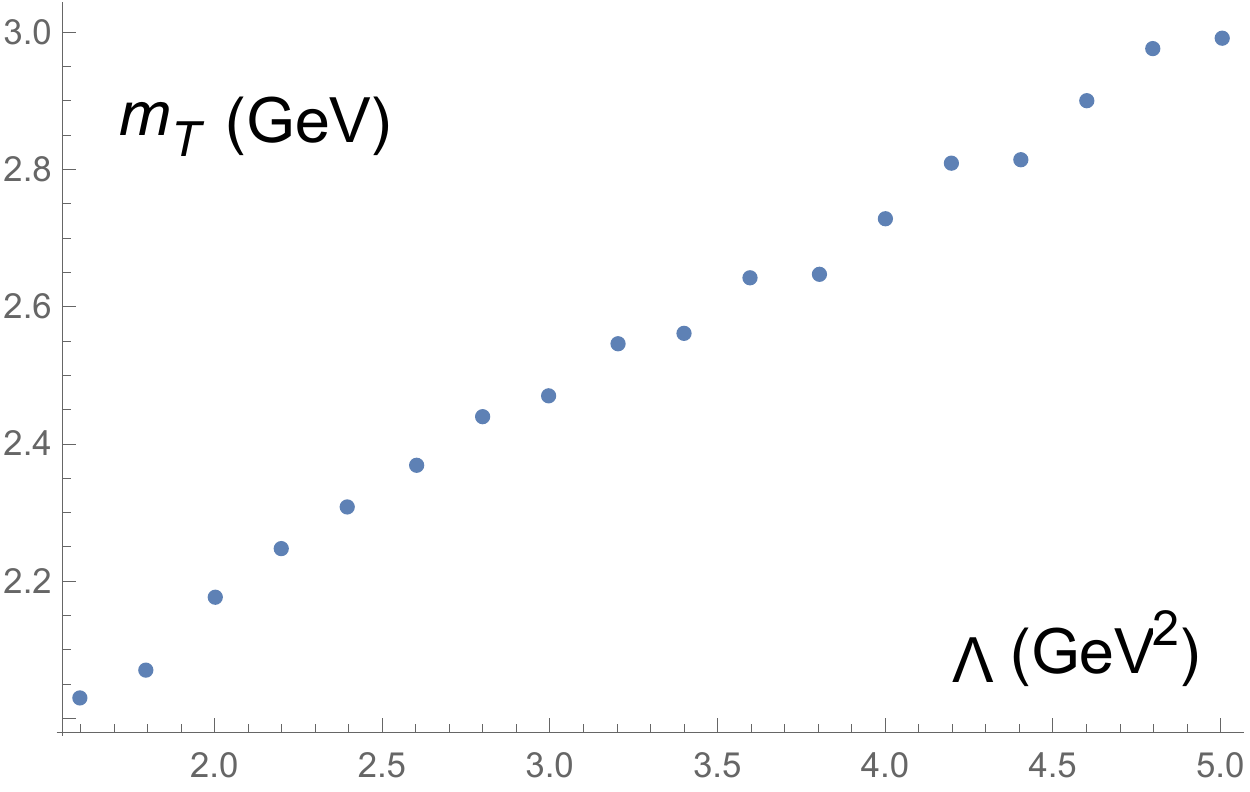}
\caption{\label{fig14}
Dependence of the tensor glueball mass $m_{T}$ on the characteristic scale $\Lambda$.}
\end{figure}

We thus investigate the dependence of the peak location on $\Lambda$ in the range 
1.5 GeV$^2<\Lambda <5.0$ GeV$^2$, and present it in Fig.~\ref{fig14}. It turns out that 
the mass $m_T$ always grows with $\Lambda$ monotonically, namely, 
a stable region for the tensor glubeball mass is not found. The number of the generalized 
Laguerre polynomials is smaller than 3 for $\Lambda<1.5$ GeV$^2$, but the resultant masses,
if depicted in Fig.~\ref{fig14}, follow the same trend. The curve is a bit bumpy, because the
polynomials are few, such that the discontinuity at each increment of $N$ 
is more significant. It has been elaborated \cite{Li:2020ejs} that at least two condensates with 
different dimensions are necessary for establishing the $\rho$ meson state. Hence, we speculate 
that the absence of a resonance solution for the tensor glueball may be ascribed to the insufficient 
nonperturbative input in the present setup: the single four-gluon condensate
may not be able to fix the tensor glueball mass. An OPE, which includes higher-dimensional operators 
compared to Eq.~(\ref{dt3}), is urged. Note that a minimum of $m_T\approx 2.0$ GeV  
was extracted from the ratio of two selected moments by varying the Borel mass in sum 
rules \cite{SN98}. However, no minimum existed as the
the continuum threshold was varied, a situation different from the
scalar and pseudoscalar glueball cases \cite{SN98}. We suspect that  
more moments considered in our formalism may have imposed a stronger constraint
on the existence of the tensor glueball.  This issue is worth more thorough studies.

\section{CONCLUSION}

We have refined our previous proposal for handing QCD sum rules
as an inverse problem by solving dispersion relations with OPE inputs directly. 
A nonperturbative spectral density is expanded in terms of a suitable set of 
generalized Laguerre polynomials up to some degree, according to its boundary 
condition at vanishing energy. A dispersive integral and power corrections to an OPE from 
condensates render possible expansions of both sides of a dispersion relation into power 
series in $1/q^2$. The coefficients in the power series on the hadron side can then be 
derived in the inverse matrix method from the known OPE coefficients on the quark side 
straightforwardly. An additional polynomial in the expansion for the spectral density appears 
as a higher-power correction in $1/q^2$ to the dispersion relation, such that 
the convergence of the OPE guarantees the convergence of the polynomial expansion for 
the spectral density. Certainly, a solution from an ill-posed inverse problem will 
go out of control, as the matrix dimension becomes sufficiently large. However, a reasonable 
solution can be obtained in most cases, as elaborated in Sec.~IIA by means of mock data 
constructed from several test functions. The employment of the generalized Laguerre 
polynomials introduces an arbitrary scale $\Lambda$ into the theoretical framework, 
which characterizes the resolution power of the method. It has been explained that a 
peak in the spectral density can be identified as a physical resonance, if its location 
is stable against the variation of $\Lambda$.

The applications of the above formalism to the determinations of the $\rho$ meson, scalar 
glueball and pseudoscalar glueball masses have been quite successful in the sense that the 
stability windows in $\Lambda$ do exist. By producing the $\rho$ meson mass $m_\rho=0.78$ GeV, 
we fixed the gluon condensate $\langle\alpha_s G^2\rangle$ and the factorization violation 
parameter $\kappa$, which are both within the ranges commonly accepted in the literature. The 
double-peak structures, appearing naturally in the spectral densities for the scalar 
and pseudoscalar glueballs, indicate some gluonium components in the light quark states.
The locations of the major broad peaks, 1.50 and 1.75 GeV, point to the $f_0(1500)$ and $\eta(1760)$ 
as the glue-rich states in the scalar and pseudoscalar sectors, respectively. The former, 
together with $f_0(1370)$ and $f_0(1710)$, have been well recognized and extensively analyzed 
in mixing frameworks. However, the latter, with its mass below most predictions from quenched
lattice QCD and sum rules, has not, and deserves more theoretical and experimental endeavors. 
We stress that we did not find a resonance solution for the spectral density associated with
the tensor glueball. As shown in our previous work, at least two condensates with 
different dimensions in the OPE input are required for establishing the $\rho$ meson state. We have 
speculated that the absence of a solution for the tensor glueball may be due to the insufficient 
nonperturbative input at present. An OPE for the corresponding correlation 
function with higher-dimensional operators is thus in demand.

A merit of our approach is that the correlation functions for the scalar and 
pseudoscalar glueballs at zero momentum can be calculated from the subtracted spectral 
densities introduced in this paper, which realize explicitly the ultraviolet regularization 
required for dispersive integrals in the literature. The former served to 
discriminate the lattice estimate for 
the triple-gluon condensate from the others, and the latter gave rise to
our prediction for the topological susceptability $\chi_t^{1/4}=75$-78 MeV, whose 
range overlaps with that from lattice QCD. Combined with the other findings on the measured $\rho$ 
meson mass and the widely accepted scalar glueball mass around 1.5 GeV in the same 
formalism, and the experimental observations from $J/\psi$ radiative decays, we tend to
advocate that the $\eta(1760)$ meson is a promising candidate for the pseudoscalar glueball.


We have explored various sources of theoretical uncertainties in our method, which are all
under control. The error from the variation of $\Lambda$ in the stability window is tiny, 
as reflected by the flat curves for the resonance masses. The error from the 
renormalization-group evolution is also minor, because it is the relative importance among the 
different terms in an OPE that determines the resonance masses, which is 
not sensitive to the running effect. The variations of the condensates are not crucial either: 
a typical 20\% change of the gluon condensate causes about 5\% effects on the $\rho$ meson and 
glueball masses. Compared to conventional sum rules, outcomes from directly solving dispersive 
relations have less model dependence. Besides, our theoretical framework can be 
improved systematically by taking into account higher-order and higher-power corrections to 
an OPE on the quark side. Note that the dimension-eight condensates are still 
quite uncertain \cite{Blok:1997yd,GPP,BGJ,Dominguez:2014fua}, on which more progress 
is needed.

Our work suggests that the multiple-resonance mixing scenario for the pseudoscaalr mesons should 
include $\eta(1760)$ for completeness. To explore mixing properties in our formalism, one has 
to consider additional off-diagonal correlation functions, in which one of the gluon operator is 
replaced by a quark one \cite{NPP,BBN}. The decay constants of glueballs ought to be 
extracted in mixing frameworks, since their definitions depend on the currents 
adopted. Moreover, the input from the triple-gluon condensate, which affects mixing patterns
\cite{Narison:2021xhc}, should be fixed accurately first.
Though our method is powerful for studying properties of low-lying resonances, it is difficult to 
probe excited states and finer structures, such as the $\rho$-$\omega$ mixing. To attempt the former, 
one may resort to the multiple-pole parametrization 
plus an arbitrary continuum contribution for a spectral density as in conventional sum rules
\cite{Krasnikov:1982ea,SVVZ,NOS,Krasnikov:1981vw,Bakulev:1998pf,Pimikov:2013usa}. However, it is 
likely that we can avoid the ad hoc prescriptions for choosing relevant hadronic parameters, such as 
continuum thresholds \cite{MaiordeSousa:2012vv}. It is worthwhile 
to extend our formalism to investigations of three-gluon states \cite{LPN} and other hadronic
states. There is no doubt that that our proposal will have wide applications to analyses 
of nonperturbative observables.




\vskip 1.0cm
{\bf Acknowledgement}

We thank T.W. Chiu, S. Narison and Q. Zhao for helpful discussions.
This work was supported in part by the Ministry of Science and Technology of R.O.C. under Grant No.
MOST-110-2811-M-001-540-MY3.



\begin{thebibliography}{99}

\bibitem{SVZ} M.~A.~Shifman, A.~I.~Vainshtein and V.~I.~Zakharov, Nucl. Phys. B {\bf 147}, 385 (1979);
B {\bf 147}, 448 (1979).

\bibitem{Li:2020ejs}
H.~n.~Li and H.~Umeeda,
Phys. Rev. D \textbf{102}, 114014 (2020).

\bibitem{Coriano:1993yx} 
  C.~Coriano and H.~n.~Li,
  Phys.\ Lett.\ B {\bf 324}, 98 (1994).

\bibitem{Coriano:1998ge}
C.~Coriano, H.~n.~Li and C.~Savkli,
JHEP \textbf{07}, 008 (1998).

\bibitem{Leinweber:1995fn}
D.~B.~Leinweber,
Annals Phys. \textbf{254}, 328-396 (1997).

\bibitem{Gubler:2010cf}
P.~Gubler and M.~Oka,
Prog. Theor. Phys. \textbf{124}, 995 (2010).

\bibitem{Li:2020xrz} 
  H.~n.~Li, H.~Umeeda, F.~Xu and F.~S.~Yu,
Phys. Lett. B \textbf{810}, 135802 (2020).

\bibitem{Li:2020fiz}
H.~n.~Li and H.~Umeeda,
Phys. Rev. D \textbf{102}, no.9, 094003 (2020).

\bibitem{Ohtani:2012ps}
K.~Ohtani, P.~Gubler and M.~Oka,
Phys. Rev. D \textbf{87}, 034027 (2013).

\bibitem{Huang:1998wj}
T.~Huang, H.~Y.~Jin and A.~L.~Zhang,
Phys. Rev. D \textbf{59}, 034026 (1999).

\bibitem{Harnett:2000fy}
D.~Harnett and T.~G.~Steele,
Nucl. Phys. A \textbf{695}, 205-236 (2001).

\bibitem{PDG}
P.~A.~Zyla et al. (Particle Data Group), Prog. Theor. Exp. Phys. 2020, 083C01 (2020).

\bibitem{SN98} 
S.~Narison,
Nucl. Phys. B \textbf{509}, 312-356 (1998).

\bibitem{NSVZ} V.~A.~Novikov, M.~A.~Shifman, A.~I.~Vainshtein and V.~I.~Zakharov, 
Nucl. Phys. B {\bf 191}, 301 (1981); M.~A.~Shifman, Phys. Rep. {\bf 209}, 341 (1991).

\bibitem{Forkel:2000fd}
H.~Forkel,
Phys. Rev. D \textbf{64}, 034015 (2001).

\bibitem{Forkel:2003mk}
H.~Forkel,
Phys. Rev. D \textbf{71}, 054008 (2005).

\bibitem{Wen:2010as}
S.~Wen, Z.~Zhang and J.~Liu,
J. Phys. G \textbf{38}, 015005 (2011).

\bibitem{Wen:2010qoe}
S.~Wen, Z.~Zhang and J.~Liu,
Phys. Rev. D \textbf{82}, 016003 (2010).

\bibitem{Wang:2015mla}
F.~Wang, J.~Chen and J.~Liu,
Eur. Phys. J. C \textbf{75}, no.9, 460 (2015).

\bibitem{Yuan:2009vs}
X.~H.~Yuan and L.~Tang,
Commun. Theor. Phys. \textbf{54}, 495-498 (2010)

\bibitem{Mathieu:2008me}
V.~Mathieu, N.~Kochelev and V.~Vento,
Int. J. Mod. Phys. E \textbf{18}, 1-49 (2009).

\bibitem{NS80} V.~A.~Novikov, M.~A.~Shifman, A.~I.~Vainshtein and V.~I.~Zakharov, Nucl. Phys. B {\bf 165}, 67 (1980).

\bibitem{RRY} L.~J.~Reinders, H.~Rubenstein and S.~Yazaki, Phys. Rep. {\bf 127}, 1 (1985).

\bibitem{PV90} H.~Panagopoulos and E.~Vicari, Nucl. Phys. {\bf B332}, 261 (1990); 
A.~DiGiacomo, K.~Fabricius and G.~Paffuti, Phys. Lett. B {\bf 118}, 129 (1982).

\bibitem{SN10}  S.~Narison, Phys. Lett. B {\bf 693}, 559 (2010); erratum ibid, Phys. Lett. B {\bf 705}, 544 (2011);
Phys. Lett. B {\bf 706}, 412 (2012); Phys. Lett. B {\bf 707}, 259 (2012).


\bibitem{SVZ80} M.~A.~Shifman, A.~I.~Vainshtein and V.~I.~Zakharov, Nucl. Phys. 
B {\bf 166}, 493 (1980); H.~Leutwyler and A.~Smilga, Phys.
Rev. D {\bf 46}, 5607(1992); for two degenerate flavors see also R.~J.~Crewther, 
Phys. Lett. B {\bf 70}, 349 (1977).

\bibitem{LS92} H.~Leutwyler and A.~V.~Smilga,
Phys. Rev. D {\bf 46}, 5607 (1992).


\bibitem{NV89} S.~Narison and G.~Veneziano, Int. J. Mod. Phys. {\bf A} 4, 2751 (1989).

\bibitem{Close} C.~Amsler and F.~E.~Close, Phys. Lett. B {\bf 353}, 385 (1995); 
F~.E.~Close and A.~Kirk, Phys. Lett. B {\bf 483}, 345 (2000);
F.~E.~Close and Q.~Zhao, Phys. Rev. D {\bf 71}, 094022 (2005).

\bibitem{Giacosa:2005zt}
F.~Giacosa, T.~Gutsche, V.~E.~Lyubovitskij and A.~Faessler,
Phys. Rev. D \textbf{72}, 094006 (2005).

\bibitem{Vento:2004xx}
V.~Vento,
Phys. Rev. D \textbf{73}, 054006 (2006).

\bibitem{Fariborz:2006xq}
A.~H.~Fariborz,
Phys. Rev. D \textbf{74}, 054030 (2006).

\bibitem{Cheng:2015iaa}
H.~Y.~Cheng, C.~K.~Chua and K.~F.~Liu,
Phys. Rev. D \textbf{92}, no.9, 094006 (2015).

\bibitem{Noshad:2018afw}
H.~Noshad, S.~Mohammad Zebarjad and S.~Zarepour,
Nucl. Phys. B \textbf{934}, 408-436 (2018).

\bibitem{Guo:2020akt}
X.~D.~Guo, H.~W.~Ke, M.~G.~Zhao, L.~Tang and X.~Q.~Li,
Chin. Phys. C \textbf{45}, no.2, 023104 (2021).

\bibitem{Narison:2021xhc}
S.~Narison, Lupm and iHEPMAD,
[arXiv:2108.13089 [hep-ph]].

\bibitem{Chen:2021bck}
H.~X.~Chen, W.~Chen and S.~L.~Zhu,
[arXiv:2107.05271 [hep-ph]].

\bibitem{Bali} G.~S.~Bali et al., Phys. Lett. B {\bf 309}, 379 (1993).

\bibitem{Morningstar:1999rf}
C.~J.~Morningstar and M.~J.~Peardon,
Phys. Rev. D \textbf{60}, 034509 (1999).

\bibitem{Chen:2005mg}
Y.~Chen, A.~Alexandru, S.~J.~Dong, T.~Draper, I.~Horvath, F.~X.~Lee, K.~F.~Liu, N.~Mathur, C.~Morningstar and M.~Peardon, \textit{et al.}
Phys. Rev. D \textbf{73}, 014516 (2006).

\bibitem{Athenodorou:2020ani}
A.~Athenodorou and M.~Teper,
JHEP \textbf{11}, 172 (2020).

\bibitem{Zhang:2021itx}
L.~Zhang, C.~Chen, Y.~Chen and M.~Huang,
[arXiv:2106.10748 [hep-ph]].

\bibitem{Lee:1999kv}
W.~J.~Lee and D.~Weingarten,
Phys. Rev. D \textbf{61}, 014015 (2000).

\bibitem{Kataev:1981aw}
A.~L.~Kataev, N.~V.~Krasnikov and A.~A.~Pivovarov,
Phys. Lett. B \textbf{107} (1981), 115-118.


\bibitem{Huber:2020ngt}
M.~Q.~Huber, C.~S.~Fischer and H.~Sanchis-Alepuz,
Eur. Phys. J. C \textbf{80}, no.11, 1077 (2020).

\bibitem{Kaptari:2020qlt}
L.~P.~Kaptari and B.~K\"ampfer,
Few Body Syst. \textbf{61}, no.3, 28 (2020).

\bibitem{Page:1996ss}
P.~R.~Page and X.~Q.~Li,
Eur. Phys. J. C \textbf{1}, 579-583 (1998).

\bibitem{Wu:2000yt}
N.~Wu, T.~N.~Ruan and Z.~P.~Zheng,
Chin. Phys. \textbf{10}, 611-612 (2001).

\bibitem{BES:2006nqh}
M.~Ablikim \textit{et al.} [BES],
Phys. Rev. D \textbf{73}, 112007 (2006).

\bibitem {MARK} D. Coffman et al. (MARK-III Collaboration), Phys. Rev. D {\bf 41}, 1410 (1990);
J. E. Augustin et al. (DM2 Collaboration), Phys. Rev. D {\bf 42}, 10 (1990);
M. Ablikim et al. (BES Collaboration), Phys. Lett. B {\bf 594}, 47 (2004).

\bibitem{Cheng:2008ss}
H.~Y.~Cheng, H.~n.~Li and K.~F.~Liu,
Phys. Rev. D \textbf{79}, 014024 (2009).

\bibitem{He:2009sb}
S.~He, M.~Huang and Q.~S.~Yan,
Phys. Rev. D \textbf{81}, 014003 (2010).

\bibitem{Tsai:2011dp}
Y.~D.~Tsai, H.~n.~Li and Q.~Zhao,
Phys. Rev. D \textbf{85}, 034002 (2012).

\bibitem{Gutsche:2009jh}
T.~Gutsche, V.~E.~Lyubovitskij and M.~C.~Tichy,
Phys. Rev. D \textbf{80}, 014014 (2009).

\bibitem{Qin:2017qes}
W.~Qin, Q.~Zhao and X.~H.~Zhong,
Phys. Rev. D \textbf{97}, no.9, 096002 (2018).

\bibitem{SN83} S.~Narison,  Phys. Lett. B {\bf 125}, 501 (1983); Z. Phys. C {\bf 26}, 209 (1984).

\bibitem{Bonati:2015vqz}
C.~Bonati, M.~D'Elia, M.~Mariti, G.~Martinelli, M.~Mesiti, F.~Negro, F.~Sanfilippo and G.~Villadoro,
JHEP \textbf{03}, 155 (2016).

\bibitem{Aoki:2017paw}
S.~Aoki \textit{et al.} [JLQCD],
PTEP \textbf{2018}, no.4, 043B07 (2018).

\bibitem{Alexandrou:2017bzk}
C.~Alexandrou, A.~Athenodorou, K.~Cichy, M.~Constantinou, D.~P.~Horkel, K.~Jansen, G.~Koutsou and C.~Larkin,
Phys. Rev. D \textbf{97}, no.7, 074503 (2018).

\bibitem{Dimopoulos:2018xkm}
P.~Dimopoulos, C.~Helmes, C.~Jost, B.~Knippschild, B.~Kostrzewa, L.~Liu, K.~Ottnad, M.~Petschlies, C.~Urbach and U.~Wenger, \textit{et al.}
Phys. Rev. D \textbf{99}, no.3, 034511 (2019).

\bibitem{Chiu:2020ppa}
T.~W.~Chiu [TWQCD],
PoS \textbf{LATTICE2019}, 133 (2020).

\bibitem{Bhattacharya:2021lol}
T.~Bhattacharya, V.~Cirigliano, R.~Gupta, E.~Mereghetti and B.~Yoon,
Phys. Rev. D \textbf{103}, no.11, 114507 (2021).

\bibitem{GrillidiCortona:2015jxo}
G.~Grilli di Cortona, E.~Hardy, J.~Pardo Vega and G.~Villadoro,
JHEP \textbf{01}, 034 (2016).

\bibitem{Luciano:2018pbj}
F.~Luciano and E.~Meggiolaro,
Phys. Rev. D \textbf{98}, no.7, 074001 (2018).

\bibitem{Gorghetto:2018ocs}
M.~Gorghetto and G.~Villadoro,
JHEP \textbf{03}, 033 (2019).

\bibitem{Lucini:2004my}
B.~Lucini, M.~Teper and U.~Wenger,
JHEP \textbf{06}, 012 (2004).

\bibitem{Bennett:2020hqd}
E.~Bennett, J.~Holligan, D.~K.~Hong, J.~W.~Lee, C.~J.~D.~Lin, B.~Lucini, M.~Piai and D.~Vadacchino,
Phys. Rev. D \textbf{102}, no.1, 011501 (2020).

\bibitem{KL97}
K.~H.~Kwon and L.~Littlejohn, J. Korean Math. Soc. {\bf 34}, 973 (1997).

\bibitem{CDK} Y. Chung, H. G. Dosch, M. Kremer and D. Schall, Z. Phys. C {\bf 25}, 151 (1984).

\bibitem{SN95} S. Narison, Phys. Lett. B {\bf 361}, 121 (1995). 

\bibitem{SN09} S. Narison, Phys. Lett. B {\bf 673}, 30 (2009). 

\bibitem{Kallen:1955fb} 
  A.~O.~G.~K\"{a}ll\'en and A.~Sabry,
  Kong.\ Dan.\ Vid.\ Sel.\ Mat.\ Fys.\ Med.\  {\bf 29}, no. 17, 1 (1955).

\bibitem{Kwon:2008vq} 
  Y.~Kwon, M.~Procura and W.~Weise,
  Phys.\ Rev.\ C {\bf 78}, 055203 (2008).
  
\bibitem{Wang:2016sdt}
Q.~N.~Wang, Z.~F.~Zhang, T.~Steele, H.~Y.~Jin and Z.~R.~Huang,
Chin. Phys. C \textbf{41}, 074107 (2017).

\bibitem{Narison:2014wqa}
S.~Narison,
Nucl. Part. Phys. Proc. \textbf{258-259}, 189 (2015).
 
\bibitem{ST90} K.~G.~Chetyrkin, V.~P.~Spiridonov and S.~G.~Gorishnii, Phys. Lett. 
B {\bf 160}, 149 (1985); L.~R.~Surguladze and F.~V.~Tkachov, Nucl. Phys. B {\bf 331}, 35 (1990).

\bibitem{Yamazaki:2001er}
T.~Yamazaki \textit{et al.} [CP-PACS],
Phys. Rev. D \textbf{65}, 014501 (2002).

\bibitem{SW67} S. Weinberg, Phys. Rev. Lett. {\bf 18}, 507 (1967).


\bibitem{Donoghue:1993xb}
J.~F.~Donoghue and E.~Golowich,
Phys. Rev. D \textbf{49}, 1513-1525 (1994).


\bibitem{Kapusta:1993hq}
J.~I.~Kapusta and E.~V.~Shuryak,
Phys. Rev. D \textbf{49}, 4694-4704 (1994).

\bibitem{NSVZ79} V.A. Novikov, M.A. Shifman, A.I. Vainsthein, and V.I. Zakharov, 
Phys. Lett. B {\bf 86}, 347 (1979); Nucl. Phys. B {\bf 165}, 55 (1980).

\bibitem{BLP} E.~Bagan, J.~I.~Latorre, P.~Pascual and T.~Tarrach, Nucl. Phys. B {\bf 254}, 555 (1985).

\bibitem{Kataev:1981gr}
A.~L.~Kataev, N.~V.~Krasnikov and A.~A.~Pivovarov,
Nucl. Phys. B \textbf{198}, 508-518 (1982)
[erratum: Nucl. Phys. B \textbf{490}, 505-507 (1997)].

\bibitem{BS90} E.~Bagan and T.~G.~Steele, Phys. Lett. B {\bf 243}, 413 (1990).

\bibitem{CKS97} K.G. Chetyrkin, B.A. Kniehl, and M. Steinhauser, Phys. Rev. Lett. 
{\bf 79}, 2184 (1997).

\bibitem{HS01} D. Harnett and T.G. Steele, Nucl. Phys. A {\bf 695}, 205 (2001).

\bibitem{ZS03} A. Zhang and T.G. Steele, hep-ph/0304208.

\bibitem{AMM92} D. Asner, R.B. Mann, J.L. Murison and T.G. Steele, Phys. 
Lett. B {\bf 296}, 171 (1992).

\bibitem{NS81} V.~A.~Novikov and M.~A.~Shifman, Z. Phys. C {\bf 8}, 43 (1981).

\bibitem{MS81} M.~A.~Shifman, Z. Phys. C {\bf 9}, 347 (1981).

\bibitem{Narison:2018dcr}
S.~Narison,
Int. J. Mod. Phys. A \textbf{33}, no.10, 1850045 (2018); arXiv:1812.09360.

\bibitem{PTN} P.~Pascual and R.~Tarrach, Phys. Lett. B {\bf 113}, 495 (1982);
S.~Narison, Z. Phys. C {\bf 26}, 209 (1984);
J.~Bordes, V.~Gimenez and J.~A.~Penarrocha, Phys. Lett. B {\bf 223}, 251 (1989).

\bibitem{BLR} R.~A.~Bertlmann, G.~Launer and E.~de Rafael, Nucl. Phys. B {\bf 250}, 61 (1985).

\bibitem{Krasnikov:1982ea}
N.~V.~Krasnikov, A.~A.~Pivovarov and N.~N.~Tavkhelidze,
Z. Phys. C \textbf{19}, 301 (1983).


\bibitem{Sarantsev:2021ein}
A.~V.~Sarantsev, I.~Denisenko, U.~Thoma and E.~Klempt,
Phys. Lett. B \textbf{816}, 136227 (2021).

\bibitem{Klempt:2021ope}
E.~Klempt,
[arXiv:2108.12819 [hep-ph]].

\bibitem{Achasov:2020aun}
N.~N.~Achasov, J.~V.~Bennett, A.~V.~Kiselev, E.~A.~Kozyrev and G.~N.~Shestakov,
Phys. Rev. D \textbf{103}, no.1, 014010 (2021).

\bibitem{Fariborz:2003uj}
A.~H.~Fariborz,
Int. J. Mod. Phys. A \textbf{19}, 2095-2112 (2004).

\bibitem{Faddeev:2003aw}
L.~Faddeev, A.~J.~Niemi and U.~Wiedner,
Phys. Rev. D \textbf{70}, 114033 (2004).

\bibitem{Sun:2017ipk}
W.~Sun, L.~C.~Gui, Y.~Chen, M.~Gong, C.~Liu, Y.~B.~Liu, Z.~Liu, J.~P.~Ma and J.~B.~Zhang,
Chin. Phys. C \textbf{42}, no.9, 093103 (2018).


\bibitem{BESIII:2018dim}
M.~Ablikim \textit{et al.} [BESIII],
Phys. Rev. D \textbf{97}, no.5, 051101 (2018).

\bibitem{MCU} A.~Masoni, C.~Cicalo and G.~L.~Usai, J. Phys. G {\bf 32}, R293 (2006).


\bibitem{Wu:2011yx}
J.~J.~Wu, X.~H.~Liu, Q.~Zhao and B.~S.~Zou,
Phys. Rev. Lett. \textbf{108}, 081803 (2012).

\bibitem{Wu:2012pg}
X.~G.~Wu, J.~J.~Wu, Q.~Zhao and B.~S.~Zou,
Phys. Rev. D \textbf{87}, no.1, 014023 (2013).

\bibitem{Du:2019idk}
M.~C.~Du and Q.~Zhao,
Phys. Rev. D \textbf{100}, no.3, 036005 (2019).


\bibitem{Blok:1997yd}
B.~Blok and M.~Lublinsky,
Phys. Rev. D \textbf{57}, 2676 (1998).

\bibitem{GPP} M. Gonzalez-Alonso, A. Pich and J. Prades, 
Phys. Rev. D {\bf 81}, 074007 (2010); D {\bf 82}, 04019 (2010).

\bibitem{BGJ} D. Boito, M. Golterman, M. Jamin, K. Maltman and S. Peris, 
Phys. Rev. D {\bf 87}, 094008 (2013).

\bibitem{Dominguez:2014fua}
C.~Dominguez, L.~Hernandez, K.~Schilcher and H.~Spiesberger,
JHEP \textbf{03}, 053 (2015).

\bibitem{NPP} S.~Narison, N.~Pak and N.~Paver, Phys. Lett. B {\bf 147}, 162 (1984).
\bibitem{BBN} E.~Bagan, A.~Bramon and S.~Narison, Phys. Lett. B {\bf 196}, 203 (1987).

\bibitem{SVVZ} M.~A.~Shifman, A.~I.~Vainshtein, M.~B.~Voloshin and V.~I.~ 
Zakharov, Phys. Lett. B {\bf 77}, 80 (1978). 

\bibitem{NOS} V.~A.~Novikov, L.~B.~Okun, M.~A.~Shifman, A.~I.~Vainshtein, 
M.B. Voloshin and V.I. Zakharov, Phys. Rep. {\bf 41}, 1 (1978); 
Phys. Lett. B {\bf 67}, 409 (1977).

\bibitem{Krasnikov:1981vw}
N.~V.~Krasnikov and A.~A.~Pivovarov,
Phys. Lett. B {\bf 112}, 397 (1982); 
Sov. J. Nucl. Phys. \textbf{35}, 744 (1982);  Yad. Fiz. {\bf 35}, 1270 (1982).

\bibitem{Bakulev:1998pf}
A.~Bakulev and S.~Mikhailov,
Phys. Lett. B \textbf{436}, 351 (1998).

\bibitem{Pimikov:2013usa}
A.~Pimikov, S.~Mikhailov and N.~Stefanis,
Few Body Syst. \textbf{55}, 401 (2014).

\bibitem{MaiordeSousa:2012vv}
M.~Maior de Sousa and R.~da Silva,
Braz. J. Phys. \textbf{46}, 730 (2016).

\bibitem{LPN} J.~I.~Latorre, S.~Paban and S.~Narison, Phys. Lett. B {\bf 191}, 437 (1987).










































\end{thebibliography}
\end{document}